\documentclass[a4paper,twocolumn,11pt,accepted=2025-06-06]{quantumarticle}
\pdfoutput=1
\usepackage[utf8]{inputenc}
\usepackage[english]{babel}
\usepackage[T1]{fontenc}
\usepackage{amsmath}
\usepackage{amssymb}
\usepackage{graphicx}
\usepackage[colorlinks=true, allcolors=blue]{hyperref}
\usepackage{mathtools}
\usepackage{tikz}
\usetikzlibrary{decorations.pathreplacing}
\usepackage{dsfont}
\usepackage{bbm}
\usepackage{physics}
\usepackage{float}
\usepackage{caption,subcaption}
\usepackage{comment}
\usepackage{xcolor}
\usepackage[bottom]{footmisc}
\usepackage{blkarray}
\usepackage{breqn}
\usepackage[numbers,sort&compress]{natbib}
\usepackage[normalem]{ulem}

\usepackage{ifthen}

\newcommand{\Andrea}[2][]{%
  \textcolor{orange}{#2}%
  \ifthenelse{\equal{#1}{}}{}{%
    \textcolor{orange}{\textbf{[Andrea: #1]}}%
  }%
}
\newcommand{\Nastya}[2][]{%
  \textcolor{teal}{#2}%
  \ifthenelse{\equal{#1}{}}{}{%
    \textcolor{teal}{\textbf{[Nastya: #1]}}%
  }%
}

\def\Tr{\operatorname{Tr}}

\newcommand\mon{\textnormal{mon}}
\newcommand\uni{\textnormal{uni}}
\def\mmu{\boldsymbol \mu}
\def\mnu{\boldsymbol \nu}
\def\tmmu{\boldsymbol{\tilde{\mu}}}
\def\tmnu{\boldsymbol{\tilde{\nu}}}

\def\hchi{\hat \chi}
\def\hsigma{\hat \sigma}
\def\hgamma{\hat \gamma}

\def\hhc{\mathfrak{h}}
\def\eec{\mathfrak{e}}

\begin{document}

\title{Field theory for monitored Brownian SYK clusters}

\author{Anastasiia Tiutiakina}
\affiliation{Laboratoire de Physique Th\'eorique et Mod\'elisation, CNRS UMR 8089,
	CY Cergy Paris Universit\'e, 95302 Cergy-Pontoise Cedex, France}

\author{Hugo Lóio}
\affiliation{Laboratoire de Physique Th\'eorique et Mod\'elisation, CNRS UMR 8089,
	CY Cergy Paris Universit\'e, 95302 Cergy-Pontoise Cedex, France}

\author{Guido Giachetti}
\affiliation{Laboratoire de Physique Th\'eorique et Mod\'elisation, CNRS UMR 8089,
	CY Cergy Paris Universit\'e, 95302 Cergy-Pontoise Cedex, France}

\author{Jacopo De Nardis}
\affiliation{Laboratoire de Physique Th\'eorique et Mod\'elisation, CNRS UMR 8089,
	CY Cergy Paris Universit\'e, 95302 Cergy-Pontoise Cedex, France}

 \author{Andrea De Luca}
 \affiliation{Laboratoire de Physique Th\'eorique et Mod\'elisation, CNRS UMR 8089,
	CY Cergy Paris Universit\'e, 95302 Cergy-Pontoise Cedex, France}

\maketitle

\begin{abstract}
  We consider the time evolution of multiple clusters of Brownian Sachdev-Ye-Kitaev (SYK), i.e. systems of $N$ Majorana fermions with a noisy interaction term. In addition to the unitary evolution, we introduce two-fermion monitorings. We construct a coherent states path integral of the dynamics by generalizing spin coherent states for higher symmetry groups. We then demonstrate that the evolution of the replicated density matrix can be described by an effective field theory for the "light" degrees of freedom, i.e. the quantum fluctuations generated by the unitary evolution. This method is applied to both quadratic, where the field theory reduces to the nonlinear sigma model (NLSM), and also to interacting SYK clusters.  We show that in the stationary regime, two monitored clusters exhibit linear-in-$N$ entanglement, with a proportionality factor dependent on the strength of the unitary coupling.
\end{abstract}
 
Quantum systems out-of-equilibrium present a tremendous challenge for modern theoretical physics \cite{Calabrese,Polkovnikov2011,Huse2014}. From contemporary quantum machines to superconductors and black holes, understanding the dynamics of many-body quantum systems requires simple models that encapsulate their universal features. Over the past few decades, the Sachdev-Ye-Kitaev (SYK) model for interacting fermions  \cite{SachdevYe,Kitaevtalk}  has emerged as an optimal candidate to describe the dynamics of generic quantum systems that exhibit rapid scrambling behavior. The SYK model has provided significant insight into the chaotic properties and thermalization processes in such systems \cite{Maldacena:2015waa,Polchinski:2016xgd,Bagrets:2016cdf,PhysRevB.95.155131,Gu:2016oyy,Bagrets:2017pwq,Sunderhauf:2019djv,Agarwal:2020yky,Gu:2017njx,Gu:2017ohj}. Brownian SYK without measurements has been studied in several contexts, including its connection to gravity via holography \cite{Saad:2018bqo}, its role in quantum information theory  \cite{Jian:2022pvj,Tiutiakina:2023ilu} and quantum chaos \cite{Sunderhauf:2019djv}. The entropy dynamics of coupled Brownian SYK clusters were investigated in \cite{Jian:2020krd}, where it was found that, in the absence of measurements, the second Rényi entropy grows linearly with time before saturating at a coarse-grained value. 
An important new area of research in non-equilibrium quantum dynamics is the study of monitored quantum systems. These systems involve quantum degrees of freedom undergoing unitary many-body evolution, combined with local, projective or weak, measurements. This hybrid process can lead to unique dynamical phase transitions, known as measurement-induced phase transitions (MIPT) \cite{PhysRevLett.125.030505,PhysRevX.10.041020,PhysRevLett.125.070606,PhysRevB.101.104302,PhysRevB.101.060301,PhysRevB.102.224311,PhysRevB.102.014315,PhysRevB.101.235104,PhysRevLett.125.210602,PRXQuantum.2.010352,PhysRevB.106.L220304,PhysRevB.105.094303,PhysRevX.11.011030,PhysRevLett.126.060501,Lavasani:2020xea,PhysRevResearch.3.023200,Fisher:2022qey,PhysRevLett.128.010604,Sharma:2021guu,PhysRevX.12.041002,PhysRevLett.129.120604,PhysRevB.103.224210,PhysRevB.105.L241114}. MIPTs are characterized by a critical behavior that arises from the competition between unitary evolution, which tends to scramble and entangle distant degrees of freedom, and local measurements, which tend to project the quantum state into simpler product states.

Monitored fermions are a prime example of systems that might exhibit MIPT \cite{DeLuca:2019lfw,PhysRevLett.126.170602,PhysRevResearch.2.033017,PhysRevB.103.174303,PhysRevLett.127.140601,Jian:2021tli,PhysRevB.105.094303,PhysRevResearch.4.033001,PhysRevB.106.L220304,Buchhold:2022vyf,PhysRevResearch.5.033174,PhysRevB.108.165126,PhysRevX.11.041004,PhysRevLett.126.123604,LeGal:2022rwf,PhysRevB.108.L020306,PhysRevB.106.024304,Kells:2021nns,Poboiko:2023koc,Swann:2023vpg}, even when the coupling and measured operators are purely quadratic. This scenario bears strong similarities with Anderson localization and is effectively described, using the replica formalism, by a non-linear Sigma model (NLSM) field theory in the \( n \to 1 \) replica limit \cite{Poboiko:2023koc,PhysRevB.101.104302,Poboiko:2023yid}. A recent work introduced such a field theory via the equation of motion approach \cite{Fava:2023tgg}. However, a derivation from first principles using the coherent states formalism has not been attempted until now.
 
In this paper, we address this gap by deriving the emergent NLSM directly from the coherent states path integral representation \cite{RevModPhys.62.867,Nishiyama:1981uz} of the replicated dynamics. We consider interacting SYK clusters each composed of $N_F$ Majorana fermions undergoing continuous monitoring of quadratic (two-fermion) idempotent operators, as modeled in the framework of the stochastic Sch\"odinger equation~\cite{Wiseman1993}. We show how the NLSM emerges in a controlled way in the large limit $N_F$, by integrating out heavy degrees of freedom induced by the measurements. Our approach offers several advantages: it clearly elucidates the mechanism behind the emergence of the NLSM and provides a method to derive an effective field theory for any system of interacting electrons, regardless of whether the unitary part is quadratic or incorporates any global symmetry.
 
As a tractable example, we also focus on the specific case of two SYK clusters, where the path integral takes the form of two coupled $0+1$ dimensional field theories. There, we demonstrate how the entanglement of the large-time stationary state is given by the quadratic quantum fluctuations around the classical minima. The integration of such fluctuations gives an entanglement scaling linearly in $N_F$, thus proportionally to the volume of the cluster, with a prefactor dependent on the coupling strength between the clusters. 

The paper is organised as follows: in Sec.~\ref{sec:model}, we introduce the model, the replicated non-hermitian Hamiltonian, and the replica method for monitored fermionic systems; we explain how the path integral is expressed in terms of the fields associated with the coset $SO(2n)/U(n)$ and express the corresponding action in the semi-classical large-$N_F$ limit. In Sec.~\ref{sec:n2q2}, we show how, for $n = 2$, 
this representation reduces to standard $SU(2)$ spins and how spin-coherent states can be used to write the evolution of a generic initial state in path integral formulation. In Sec.~\ref{sec:coherent} we extend the coherent states approach for a generic number of replicas $n$. We derive the path-integral action, which includes the expectation value of the Hamiltonian on the path-integral states and the correspondent Berry phase. Subsequently, we show how the massive modes can be integrated out, to obtain the field theory description of the quantum fluctuating modes around the semi-classical minima, which are controlled at large $N_F$ and which correspond to the NLSM in the quadratic case. Finally, in the two last sections, we apply the method to two examples: i) the computation of the stationary purity of two monitored SYK clusters in terms of the quantum fluctuations over the minima of the action Sec.~\ref{sec:purity2}; ii) the explicit derivation of the NLSM for fermions without unitary interaction but competing measurements between two clusters Sec.~\ref{sec:measurementdynamics}. 

\section{The model and the replicated Hamiltonian description}\label{sec:model}
We consider systems under weak continuous monitoring and stochastic unitary evolutions. Generally, the evolution is described by a stochastic Schrödinger equation (SSE) \cite{Gross2018,Attal2010,Attal2006,Wiseman1993,Oksendal2003}, which updates the state of the system $|\psi\rangle $ as:
\begin{equation} \label{eq:SSE}
\begin{aligned}
    d \left| \psi \right>   = & -i dt  H^{\rm uni} \left|\psi \right>  
    +  \sqrt{\Gamma} \Bigg[ (M - \langle M \rangle ) dW  \\
    & -  \frac{\Gamma dt}{2} (M - \langle M \rangle )^2 \Bigg] | \psi \rangle , 
\end{aligned}
\end{equation}
where $H^{\rm uni}$ represents the unitary part of the evolution, $M$ is the operator being monitored, $dW$ is a standard Wiener process in the Ito convention and $\Gamma$ is the measurement rate. Typically, SSE emerges from repeated and frequent interactions of the system for a short time interval $dt$ with a sequence of ancillas, each of them measured projectively. By appropriately scaling the system/ancilla interaction in the limit $dt\to0$, the inherent randomness of the measurement outcomes leads to the Wiener process $dW$ and the nonlinearity, associated with the expectation value $\left<M\right> = \braket{\psi|M}{\psi}$, is a consequence of the Born rule~\cite{breuer2002theory}.

In this study, we consider a system composed of different clusters of Majorana fermions $\{\hchi_{i \nu}\}$, where $i \in \{1,\dots ,L\}$ is a cluster index and $\nu \in \{1,...,N_F\}$ is a "flavour" index within each cluster. These operators satisfy the usual anticommutation relations $\{\hchi_{i \nu}, \hchi_{j \mu}\} = 2 \delta_{ij} \delta_{\nu \mu}$. 
Eq.~\eqref{eq:SSE} is easily generalized to the case of monitoring of multiple operators, introducing independent Wiener processes for each of those. We choose to monitor $q$-body operators inside a cluster: 
introducing a corresponding set of independent Wiener processes $\{dW_{i, \boldsymbol{\tilde \nu}}\}$, we choose $\hat{M} := \hat{M}_{i, \boldsymbol{\tilde \nu}}$ with
\begin{equation}
    \hat{M}_{i, \boldsymbol{\tilde \nu}} = i^{q/2}\prod_{l=1}^{q}\hchi_{i\tilde \nu_l} .
\end{equation}
 Here, we use bold indices to represent an ordered set of flavours
and we primarily focus on quadratic operators, i.e. $q=2$. It is important to stress that, in such case, $\hat{M}_{i,\boldsymbol{\tilde \nu}}^2 = \hat{M}_{i,\boldsymbol{\tilde \nu}}$ and thus the non-hermitian part in Eq.~\eqref{eq:SSE} only involves quadratic operators in the $\hat{\chi}$'s.

Now, we introduce the unitary interaction. We allow fermions within each cluster and between clusters to interact via a generic white-noise coupling, which we refer to as \textit{Brownian}. Specifically, we consider interactions between different clusters indexed by $j, \ell$ through the interaction of an even number $q_J$ of
Majorana operators
\begin{equation}
\label{eq:Hunijj}
  H^{\uni}_{j\ell} = i^{\frac{q_J}{2}} \sum_{{\boldsymbol{\nu}}} \sum_{{\boldsymbol{\mu}}}  h^{j,\ell}_{ \boldsymbol{\mu}\boldsymbol{\nu}}(t) \prod^{q_J/2}_{k=1} \hchi_{j \mu_{k}} \prod_{l=1}^{q_J/2} \hchi_{\ell \nu_{l}} , 
\end{equation}
while the interactions within the same cluster are given by
\begin{equation}
\label{eq:Hunij}
  H^{\uni}_{j} = i^{\frac{q_J}{2}} \sum_{{\boldsymbol{\tilde \mu}}}  h^j_{ \boldsymbol{\tilde \mu}}(t) \prod^{q_J}_{k=1} \hchi_{i \tilde \mu_{k}} .
\end{equation}
\begin{figure}
    \centering
    \includegraphics[trim=3cm 6cm 7cm 6cm, clip, width=\columnwidth]{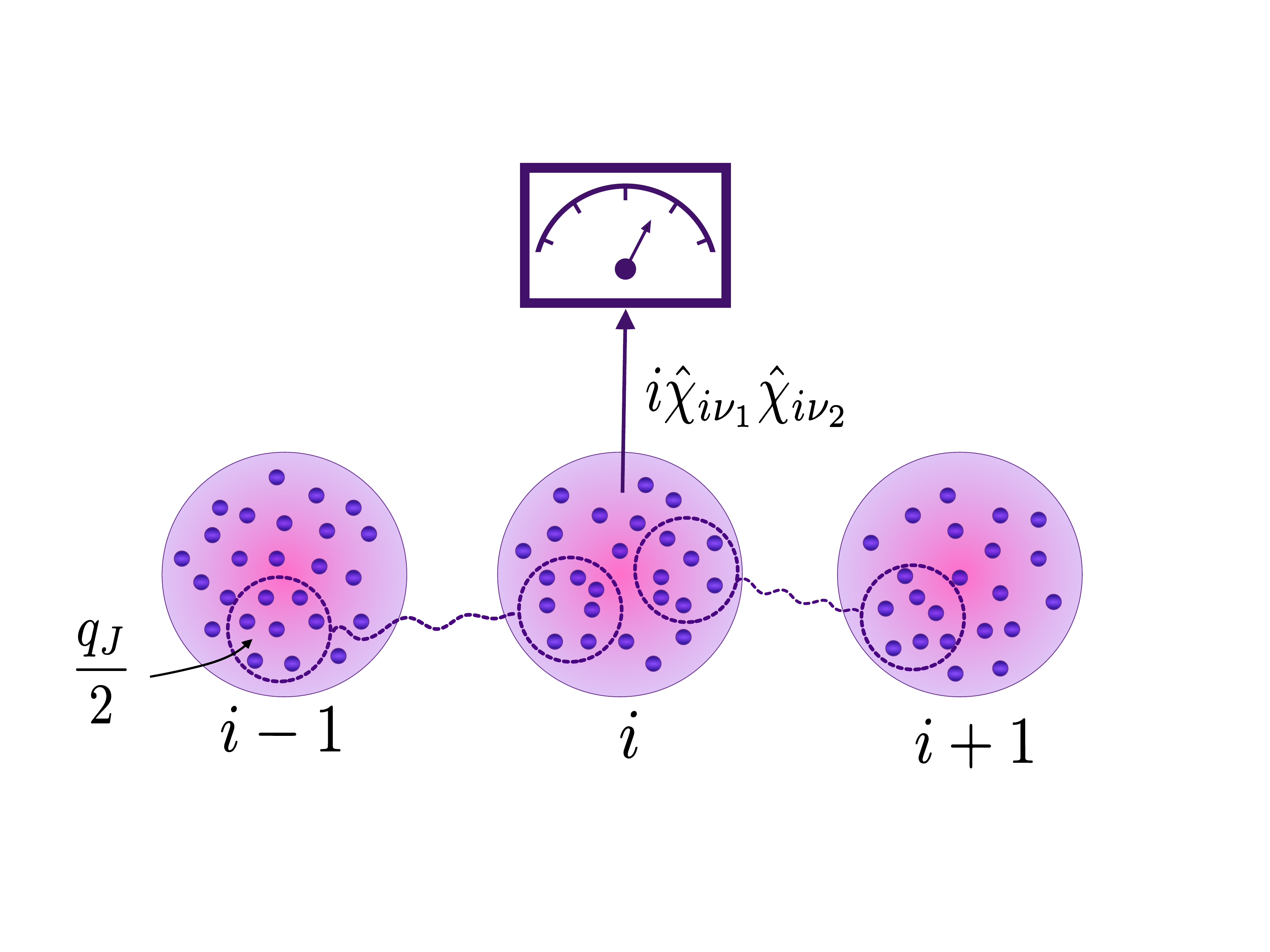}
    \caption{1D chain of Majorana clusters under continuous monitoring of bilinears inside each cluster, here $q_J$ is the number of interacting Majorana fermions.}
    \label{fig_model_scheme}
\end{figure}
Consistently, sums over bold indices are a shortcut for $\sum_{{\boldsymbol{\mu}}} = \sum_{1 \leqslant \mu_1 < \mu_2 < \dots < \mu_{q_J/2}\leqslant N_F} $ and $\sum_{{\boldsymbol{\tilde \mu}}} = \sum_{1 \leqslant \tilde \mu_1 < \tilde \mu_2 < \dots < \tilde \mu_{q_J}\leqslant N_F}$. The noisy couplings $h^{j,\ell}_{ \boldsymbol{\mu}\boldsymbol{\nu}}(t)$ and $h^j_{ \boldsymbol{\mu}}(t)$ are standard $\delta$ correlated white noises with unit variance for each value of their indices. 
We assume the clusters are arranged in a $1D$ chain, allowing interactions only between nearest neighbors clusters. Thus, combining these interaction terms, we obtain the Hamiltonian contribution to Eq.~\eqref{eq:SSE}
\begin{equation} \label{eq_H_uni}
  H^{\rm uni}(t) = \sqrt{J}\sum_i \left(H_{j j+1}^{\uni}(t)+H_{j}^{\uni}(t) \right),
\end{equation}
where, anticipating the average over the replicated system, we conveniently parameterised the strength of unitary interactions with $\sqrt{J}$. 
Note that in contrast with the Wiener process in Eq.~\eqref{eq:SSE}, the white noise entering Eq.~\eqref{eq_H_uni} are expressed in Stratonovich convention.

We will call a given realization of Wiener processes (associated with measurement outcomes) and noise in unitary terms a “trajectory”.
Given a trajectory, the time evolution $\rho(t)$ of an initial state specified by the matrix $\rho(0)$, can, in principle, be computed from Eq.~\eqref{eq:SSE}. 
We are interested in computing averages of scalar functionals $F[\rho(t)]$ of $\rho(t)$ at a given time $t$, over the trajectories. We will refer to this averaging as $\overline{F[\rho]}_{\rm SSE}$. However, 
except for the case of linear functionals, the nonlinearity of Eq.~\eqref{eq:SSE} prevents obtaining closed equations for this kind of quantities, thus hindering an analytical treatment. Let us notice that such a nonlinearity is linked to the fact that Eq.~\eqref{eq:SSE} preserves the normalization $ \Tr \rho(t) =1$ on each trajectory. An alternative approach~\cite{Fava:2023tgg, PhysRevB.101.104302,giachetti2023elusive} is thus to introduce a non-normalized version of $\rho(t)$, $\check{\rho}(t)$, such that $\rho(t) = \frac{\check{\rho}(t)}{\Tr \check \rho (t)}$, which evolves in time through a linear (non-hermitian) Hamiltonian dynamics. In particular,
\begin{equation}
  \check{\rho}(t) = K(t) \rho(0) K^\dagger(t) , 
\end{equation}
where the evolution operator $K(t)$ is linked via a time-ordered exponential 
\begin{equation}
  K(t) \equiv \mathcal{T} \exp(-i \int_0^t H(s) ds ) . 
\end{equation}
to the corresponding Hamiltonian 
\begin{align} \label{eq_H_uni_plus_mon}
  H(t) & = H^{\rm uni}(t) + i \sqrt{\Gamma} \sum_j H_{j}^{\mon}(t) , \\
  H^{\mon}_j & = i^{\frac{q}{2}} \sum_{\boldsymbol{\tilde \nu}} w^j_{\boldsymbol{\tilde \nu}}(t) \prod_{l=1}^{q}\hchi_{i\tilde \nu_l} .
\end{align}
In this formulation, the complexity of the monitoring dynamics has been hidden in the distribution of the couplings, $w^j_{\boldsymbol{\tilde \nu}}(t)$ which shall not be confused with the Wiener processes $dW$ in Eq.~\eqref{eq:SSE}. The connection between two approaches can be established by coupling the Hamiltonian to ancillas degrees of freedom, performing projective measurements on them and normalising the state after each measurement for more details see Appendix $A$ of \cite{giachetti2023elusive}. 
The couplings $\{w^j_{\boldsymbol{\tilde \nu}}(t)\}$ reflect the outcomes of the measurements on the ancillas and thus their distribution depends non-trivially on the state itself as dictated by Born rule. 
According to it, the probability of observing the outcomes $\{w^j_{\boldsymbol{\tilde \nu}}(t)\}$ is proportional to $\Tr \check\rho(t)$~\footnote{The proportionality constant can be explicitly fixed~(see for instance Suppl. Mat. in \cite{giachetti2023elusive}) but, in our case, it will not be necessary since there is always a natural way to account for it.}.
In turn, we can circumvent this non-linear averaging by explicitly including the Born's weight $\Tr \check\rho(t)$ as a factor in the quantity to be averaged. As a consequence, the residual distribution for the couplings $w^j_{\boldsymbol{\tilde \nu}}(t)$ is unbiased, i.e. it amounts to uncorrelated white noises. We denote as $\mathbb{E}_{\rm G}$ the resulting average, with $\mathbb{E}_{\rm G}[w^j_{\boldsymbol{\tilde \nu}}(t) w^{j'}_{\boldsymbol{\tilde \nu'}}(t')] = \delta_{jj'} \delta_{\boldsymbol{\tilde \nu}\boldsymbol{\tilde \nu'}} \delta(t-t')$. Therefore, the final quantum trajectory-averaged expectation values are expressed in terms of the Gaussian averages as,
\begin{equation}
\label{eq:replicatrickF}
  \overline{F[\rho]}_{\rm SSE} \equiv 
  \frac{\mathbb{E}_{\rm G}\left[F[\rho] \Tr {\check\rho}(t) \right]}{\mathbb{E}_{\rm G}[\Tr {\check\rho}(t) ]} ,
\end{equation}
where the denominator in the right-hand side accounts for the proportionality constant in the Born probabilities.
A notable example is the trajectory-averaged purity of a subsystem $A$, corresponding to the choice $F[\rho] = \Tr[\rho_A^2]$, with $\rho_A = \Tr_{\bar{A}} \rho$, where $\bar{A}$ is the complementary of $A$. Applying Eq.~\eqref{eq:replicatrickF}, we can express it as
\begin{equation}
\label{eq:purityreplica}
\begin{split}
  & \overline{\Tr\rho_A(t)^2}  = \frac{\mathbb{E}_{\rm G} \left[ \Tr  \check\rho_A(t)^2 (\Tr  \check\rho(t))^{-1} \right] }{\mathbb{E}_{\rm G}\left[\Tr  \check\rho(t) \right]}\\
 & =  \lim_{n \rightarrow 1} \frac{\mathbb{E}_{\rm G} \left[ \Tr  \check\rho_A(t)^2 (\Tr  \check\rho(t))^{n-2} \right]}{\mathbb{E}_{\rm G}[\left(\Tr  \check\rho(t)\right)^n]} ,
\end{split}
\end{equation}
with $\check\rho_A = \Tr_{\bar{A}} \check\rho$. In the last step, we applied the replica trick to facilitate the Gaussian averaging of $\mathbb{E}_{\rm G} \left[ \Tr \check\rho_A(t)^2 \left(\Tr \check\rho(t)\right)^{-1} \right]$, by treating $n$ as a parameter.
In particular, {for integer $n \geqslant 2$}, we can replace the powers in Eq.~\eqref{eq:purityreplica} with a tensor product, arriving at
\begin{equation} \label{eq_purity_final}
  \overline{\Tr\rho_A(t)^2} =  
  \lim_{n \rightarrow 1} \frac{\Tr \left[ \rho^{(n)}(t) \left(\mathcal{C}_{A,2} \otimes \mathbb{I} \right)\right]}
  {\Tr \left[ \rho^{(n)}(t) \right]} ,
\end{equation}
where $\rho^{(n)}(t) = \mathbb{E}_{\rm G}\left[ \check\rho(t)^{\otimes n} \right]$
and $\mathcal{C}_{A,2}=\sum_{j,j'} \left| j \right>\left<j' \right|_A \otimes \left|j' \right>\left< j \right|_{A} \otimes \mathbb{I}$ is a swap operator that exchanges, within the region $A$, the first two replicas among the total $n$, while acting as the identity elsewhere. Here  $\left|j\right>$ is a basis vector on a subspace $A$.
More generally, we can introduce permutation operators $\mathcal{C}_{A,\sigma}$ for any permutation $\sigma$ in the symmetric group with $n$ elements. Denoting a permutation $P = (2^{k_2} 3^{k_3} \ldots )$ as its cycle decomposition, with $k_m$ the number of cycles of length $m$, we can express generic functionals of the density matrix. For instance,
considering the $\alpha$-Renyi entropy of the subsystem $A$, defined as,
\begin{equation}
S_A^{(\alpha)} = \frac{1}{1-\alpha} \log \Tr[\rho_A^\alpha]  ,  
\end{equation}
we can express its average over trajectory within the replica formalism as
\begin{equation} \label{eq_renyi_limit}
\begin{split}
   &  \overline{S^{(\alpha)}_A(t)} =     \frac{1}{1-\alpha} \lim_{k \rightarrow 0}  \lim_{n \rightarrow 1} \\
   & \frac{1}{k}\left(\frac{\Tr \left[ \rho^{(n)}(t) \left(\mathcal{C}_{A,\alpha^k}\otimes \mathbb{I} \right)\right]}{\Tr \left[ \rho^{(n)}(t) \right]} - 1\right) ,
\end{split}
\end{equation}
where $P = (\alpha^k)$ is any permutation with $k$ cycles of length $\alpha$.

\subsection{Density matrix vectorization and replicated Hamiltonian} \label{sec:vect}
In order to simplify the averaging procedure, we perform a standard operator-to-state mapping. We vectorize the operators in a Hilbert space using a \textit{folding} procedure:
\begin{equation}
  \ket{ \hat O} = \sum_{kl} \bra{k} \hat O \ket{l} \ket{k} \otimes \ket{l} , 
\end{equation}
where $\{\ket{k}\}, \{\ket{l}\}$, are a fixed basis of the Hilbert space.
Then the folded replicated state is given by:
\begin{equation} \label{eq_folded_replicated_rho}
  \ket{\rho^{(n)}(t)} = \mathbb{E}_{\rm G} \left[ \left( K(t) \otimes K^*(t) \right)^{\otimes n} \right] \ket{\rho^{(n)}(0)} .
\end{equation}
According to this mapping, expectation values of operators are mapped onto overlaps:
\begin{equation} \label{eq_overlap_rho}
  \Tr \left[ \rho^{(n)} (t) \hat O \right] = \left< \hat O|\rho^{(n)}(t)\right> .
\end{equation}For further consideration, let us introduce the  indices $\sigma = \pm$ . More specifically, we use the notation $+$ for operators multiplying on the left and $-$ for operators multiplying on the right, i.e.,
$$
  A \rho B \to  A^+ (B^-)^t \ket{\rho} ,
$$
where $t$ denotes the transpose. 
Therefore, the term inside the average in Eq.~\eqref{eq_folded_replicated_rho} is given by:
\begin{equation} \label{eq:evol}
  \left( K(t) \otimes K^*(t) \right)^{\otimes n} = \mathcal{T} \exp \left(-i \int_0^t H^{(n)}(s) ds \right)\;  ,
\end{equation}
with the replicated Hamiltonian (see Appendix \ref{App:rep}):
\begin{equation} \label{eq_H_rep_f}
  \begin{split}
    H^{(n)}(t) & = \sum_{\sigma,a,j} \left[ (i)^{\frac{q_J}{2}}f_\sigma^{\rm uni} \sqrt{J} \left[\sum_{{\boldsymbol{\tilde \mu}}}  h^j_{ \boldsymbol{\tilde \mu}}(t) \prod^{q_J}_{k=1} \hchi^{(\sigma, a)}_{j \tilde \mu_{k}} \right. \right. \\ 
    &+ \left. \sum_{{\boldsymbol{\nu}}} \sum_{{\boldsymbol{\mu}}}  h^j_{ \boldsymbol{\mu}\boldsymbol{\nu}}(t) \prod^{q_J/2}_{k=1} \hchi^{(\sigma, a)}_{j \mu_{k}} \prod_{l=1}^{q_J/2} \hchi^{(\sigma, a)}_{j+1 \nu_{l}} \right]\\
    &+\left. (i)^{\frac{q}{2} + 1}\sigma f_\sigma^{\rm mon} \sqrt{\Gamma} \sum_{\boldsymbol{\tilde \nu}} w^j_{\boldsymbol{\tilde \nu}} \prod_{l=1}^{q}\hchi^{(\sigma, a)}_{j\tilde \nu_l}  \right]  ,
\end{split}
\end{equation}
where $f^{\rm uni}_+ = i^{q_J+ 2}$ and $f_-^{\rm uni} = 1$,  and $f^{\rm mon}_+ = -i^{q+2}$, $f^{\rm mon}_- = -1$. And $(\sigma,a)$ are replica indexes $\sigma\in \{+,-\}$ and $a \in {1,...,n}$.  We now perform averaging over the Gaussian distributed parameters $w$ and $h$ in order to find the evolution of the replicated state Eq. (\ref{eq_folded_replicated_rho}) (see Appendix \ref{App:aver}) :
\begin{equation}
    \mathbb{E}_G\left[\left( K(t) \otimes K^*(t) \right)^{\otimes n} \right]=e^{-t \mathcal{H}^n},
\end{equation}and obtain the effective Hamiltonian of the evolution 
\begin{equation} \label{repHam}
\begin{split}
  \mathcal{H}^{(n)}  = \frac{1}{2} & \left( J \sum_{j{\boldsymbol{\mu}} {\boldsymbol{\nu}}} \mathcal{H}^\uni_{j j+1, \boldsymbol{\mu}{\boldsymbol{\nu}}} \right. \\ 
  &\left.+ J \sum_{j{\boldsymbol{\tilde \mu}}}\mathcal{H}^\uni_{j, \boldsymbol{\tilde \mu}} - \Gamma \sum_{j \boldsymbol{\tilde \nu} } \mathcal{H}^\mon_{j,\boldsymbol{\tilde \nu} }\right),
\end{split}
\end{equation}
with
\begin{equation}
\begin{split}
  & \mathcal{H}^\uni_{jj+1,{\boldsymbol{\mu}} {\boldsymbol{\nu}}} = \left(i^{\frac{q_J}{2}} \sum_{\sigma, a}  f_\sigma^{\rm uni} \prod^{q_J/2}_{k=1} \hchi^{(\sigma, a)}_{i \mu_{k}} \prod_{l=1}^{q_J/2} \hchi^{(\sigma ,a)}_{i+1 \nu_{l}} \right)^2 ,
\end{split}
\end{equation}
\begin{equation}
\begin{split}
  & \mathcal{H}^\uni_{j,{\boldsymbol{\tilde \mu}}} = \left(i^{\frac{q_J}{2}} \sum_{\sigma, a}  f_\sigma^{\rm uni} \prod^{q_J}_{k=1} \hchi^{(\sigma ,a)}_{j{\tilde \mu}_{k}}\right)^2 ,
\end{split}
\end{equation}
and
\begin{equation}
    \mathcal{H}^\mon_{j,{\boldsymbol{\tilde \nu}}} = \left(i^{\frac{q}{2}} \sum_{\sigma, a} \sigma f_\sigma^{\rm mon} \prod_{l=1}^{q}\hchi^{(\sigma, a)}_{j{\tilde \nu}_l} \right)^2 ,
\end{equation}
where again $\sigma \in \{+,-\}$ and the summations are defined as  $\sum_{{\boldsymbol{\mu}}} = \sum_{1 \leqslant \mu_1 < \mu_2 < \dots < \mu_{q_J/2}\leqslant N_F}$ , $\sum_{{\boldsymbol{\tilde \mu}}} = \sum_{1 \leqslant {\tilde \mu}_1 < {\tilde \mu}_2 < \dots < {\tilde \mu}_{q_J}\leqslant N_F}$ and $\sum_{{\boldsymbol{\tilde \nu}}} = \sum_{1 \leqslant \nu_1 < \nu_2 < \dots < \nu_{q}\leqslant N_F}$. Notice that for replicated Majorana fermions, we have the commutation relations $\{ \hchi_{i \mu}^{(\sigma,a)}, 
    \hchi_{j \nu}^{(\sigma',a')}\} = \delta_{\sigma\sigma'} \delta_{a a'} \delta_{i j} \delta_{\mu \nu} $ that are distinct by the factor $2$ from the original ones. Finally, $n$-replica purity we are interested in can be expressed as a ratio of overlaps
\begin{equation} \label{eq:pur}
\frac{{\rm Tr}\left[\rho^{(n)}(t) \left(\mathcal{C}_{A,2} \otimes \mathbb{I}\right) \right]}
{{\rm Tr}\left[\rho^{(n)}(t) \right]}
=\frac{\left< \mathcal{C}_{A,2} |\rho^{(n)}(t) \right>}
{\left< \mathbb{I} |\rho^{(n)}(t) \right>}
.
\end{equation}
The folded replicated state can be expressed as an imaginary time Schr\"odinger evolution with the Hamiltonian Eq.~\eqref{repHam},
\begin{equation} \label{eq_folded_replicated_rho_v2}
  \ket{\rho^{(n)}(t)} = e^{-t \mathcal{H}^{(n)}} \ket{\rho^{(n)}(0)} \
;.
\end{equation}
More general Renyi entropies can be represented similarly via Eq.~\eqref{eq_renyi_limit}, replacing $\bra{\mathcal{C}_{A,2}}$ with the appropriate boundary state $\ket{B}$.

At late time, the imaginary time evolution will act as a projector onto the groundstate $\ket{\rm GS}$ of $\mathcal{H}^{(n)}$, thus we simplify the overlap as
\begin{equation}\label{eq:transitAmpl}
    \lim_{t \rightarrow \infty }
    \frac{\left< 
    B
    |\rho^{(n)}(t) \right>}
{\left< \mathbb{I} |\rho^{(n)}(t) \right>} = 
\frac{\braket{
B
}{\rm GS}}{\braket{\mathbb{I}}{\rm GS}}.
\end{equation}
In the next section, we discuss how to write this Hamiltonian in a compact way in terms of generators of the $SO(2n)$ algebra. Based on these generators, we will build a representation of the coherent states and write the path integral expression for the overlap Eq.~\eqref{eq:transitAmpl}.
\subsection{The fermionic \texorpdfstring{$SO(2n)$}{SO(2n} fields}\label{sec:so(2n)}

The replicated Hamiltonian can be expressed in terms of the following field operators
\begin{equation}\label{eq:fieldoperators}
  \hat{\Phi}_j^{(\sigma,a),(\sigma',a')} = \frac{i}{2 N_F} \sum_{\nu} [\hchi_{j\nu}^{(\sigma,a)}, \hchi_{j\nu}^{{(\sigma',a')}}] . 
\end{equation}
It is simple to show that they satisfy the same commutation relations as the elements of the $SO(2n)$ algebra~\cite{Fava:2023tgg}. 
Representing the replica indices with a single flattened index, the commutation relations of the new operators can be simplified as
\begin{multline}\label{eq:commPhi}
   \left[\hat{\Phi}_i^{\alpha \beta}, \hat{\Phi}_j^{\lambda \xi}\right]  =   \delta_{ij} \frac{i}{N_F} \times \\ (\delta_{\beta, \lambda} \hat\Phi_{i}^{\alpha \xi}
   -\delta_{\beta, \xi} \hat\Phi_{i}^{\alpha \lambda}
   +\delta_{\alpha, \xi} \hat\Phi_{i}^{\beta \lambda}
   -\delta_{\alpha, \lambda} \hat\Phi_{i}^{\beta \xi}
   ).
\end{multline}
To help clarify this algebra, we
recall that defining for $1\leq \alpha < \beta \leq 2n$, the $2n\times 2n$ matrices 
\begin{equation}\label{eq:Erepmajo}
[E^{\alpha\beta}]_{\alpha' \beta'} = \delta_{\alpha'}^\alpha \delta_{\beta'}^\beta   
\end{equation}
we see that this is the same algebra obtained replacing at each site $i$: $\hat{\Phi}^{\alpha\beta}_i \to W^{\alpha \beta} = \frac{i}{N_F} (E^{\alpha\beta} - E^{\beta \alpha})$, i.e. the antisymmetric matrices that generate $so(2n)$.

The replicated Hamiltonian expressed in terms of these fields therefore takes the form
\begin{equation}\label{eq:Htotal}
    \mathcal{H} =   \mathcal{H}_{\rm uni} -   \mathcal{H}_{\rm mon} ,
\end{equation}
with
\begin{equation}\label{eq:Huni}
\begin{split}
      \mathcal{H}_{\rm uni} =  
      \frac{J N_F^{q_J }}{2 (q_J/2)!^2} \sum_{j\alpha \alpha'} \Bigg[ f^{\rm uni}_{\sigma'} f^{\rm uni}_\sigma (\Phi^{\alpha \alpha'}_j)^{\frac{q_J}{2}}(\Phi^{\alpha \alpha'}_{j+1})^{\frac{q_J}{2}} \\
      + \frac{(q_J/2)!^2}{ q_J !}f^{\rm uni}_{\sigma'} f^{\rm uni}_\sigma (\Phi^{\alpha \alpha'}_j)^{q_J}\Bigg],
\end{split}
\end{equation}
and
\begin{equation}
    \mathcal{H}_{\rm mon} = \Gamma \frac{N_F^q}{ 2 q!} \sum_{j\alpha \alpha'}\sigma \sigma' f^{\rm mon}_\sigma    f^{\rm mon}_{\sigma'} (\hat \Phi^{\alpha \alpha'}_j)^q .
\end{equation}From now on, we will focus on the case where the monitoring is quadratic $q=2$, which allows for the simplification
\begin{equation}\label{eq:Hmeas}
\begin{split}
   \mathcal{H}_{\rm mon}\Big|_{q=2}=  & \Gamma  \frac{N_F^2}{4} \sum_{j \alpha \alpha'} \sigma \sigma' \Big(\Phi^{\alpha \alpha'}_j\Big)^2,
    \end{split}
\end{equation}
where $\{\alpha, \alpha'\}=\{(\sigma, a), (\sigma',a')\}$ are multi-indices.
The operators in Eq.~\eqref{eq:fieldoperators} are the main building blocks of the theory, as they allow defining a coherent states representation for any number of replicas.

\section{The case with \texorpdfstring{$n=2$}{n=2} replicas and spin coherent states}\label{sec:n2q2}

We first consider $n=2$ as an instructive and more familiar example and derive a nonlinear sigma model in the case $q_J=2$. Afterwards, we calculate the replicated purity for two clusters for both $q_J=2$ and $q_J=4$ and show its scaling. The case $n=2$ is particularly enlightening due to its mapping to a $SU(2)$ spin system~\cite{BAO2021168618}. 
We can explicitly construct this representation in terms of the Dirac's fermions \begin{equation}
\begin{split}
   & c_{a j \nu}^{\dagger} = \frac{\chi_{j\nu}^{(+,a)} - i \chi_{j \nu}^{(-,a)}}{\sqrt{2}} , \\
   & c_{a j \nu} =  \frac{\chi_{i\nu}^{(+,a)} + i \chi_{j \nu}^{(-,a)}}{\sqrt{2}}\;.
\end{split}
\end{equation}To do so, we firstly introduce two sets of spin $1/2$ operators, one for each replica $a = 1,2$, us the Jordan-Wigner transformation. To simplify the notation, we omit cluster index $j$. Then we combine $\sigma = \pm1$ replicas introducing $c^{\dag}_{a \nu} c_{\nu a}=\frac{1}{2}(S^z_{a\nu}+\mathbb{I})$. Then, we define raising/lowering spin operators as
\begin{equation}
\begin{split}
    c_{a\nu}^{\dag}=e^{i \pi \sum_{k=1}^{a-1} S_{k\nu}^+S_{k\nu}^-} S_{a\nu}^+,\\~c_{a\nu}=e^{-i \pi \sum_{k=1}^{a-1} S_{k\nu}^+S_{k\nu}^-} S_{a\nu}^-\;.
    \end{split}
\end{equation}
 We thus obtain six generators in total for each flavour $\nu$,  namely $\{S_{1, \nu}^z, S_{1, \nu}^{\pm}, S_{2 }^z, S_{2, \nu}^{\pm} \}$, consistent with the decomposition $so(4) \simeq su(2) \oplus su(2)$ with a representation of dimension $4$ (two spins $1/2$). We can further reduce the relevant Hilbert space using the conservation of parity. Labelling the states as the eigenvectors of $S^z_1, S^z_2$, we have that the two states $\{\ket{\uparrow \uparrow}, \ket{\downarrow \downarrow}\}$ are decoupled in the dynamics from $\{\ket{\uparrow \downarrow}, \ket{\downarrow \uparrow}\}$.
Since the identity matrix we choose as initial condition always lies in the former subspace, we can introduce a 
pseudospin basis with only two states: $\ket{\uparrow \uparrow}, \ket{\downarrow \downarrow}$. 
So, we can further reduce the operators for $a=1,2$ to a single spin $1/2$ in the space $\ket{\Uparrow} := \ket{\uparrow\uparrow}, \ket{\Downarrow} := \ket{\downarrow \downarrow}$. Let us denote simply as $S^x_{\nu},S^y_{\nu},S^z_{\nu}$ the spin operators acting on this $2$-dimensional space. Then,
\begin{equation}
\begin{split}
&S^x_{1\nu} S^x_{2\nu} = 
- S^y_{1\nu} S^y_{2\nu} 
= \frac{S^y_{\nu}}{2} , ~
S^x_{1\nu} S^y_{2\nu} = 
 S^y_{1\nu} S^x_{2\nu} 
= \frac{S^x_{\nu}}{2} , \\
&S^{z}_{1\nu}  = 
S^{z}_{2\nu} =   
 \frac{S^z_{\nu}}{2}  .
\end{split}
\end{equation}
It is useful to write explicitly the matrix $\Phi$ in Eq.~\eqref{eq:fieldoperators} in terms of this $su(2)$ algebra. Introducing the total spin generators $ \sum_{\nu} \vec{S}_{\nu}=\vec{S}$ and ordering the basis $(\sigma, a)$ as $((+,1), (+,2), (-,1),(-,2))$, it reads
\begin{equation}\label{eq:paramPhi} 
  \hat{\Phi} = \frac{1}{N_F} \begin{pmatrix}
    0 &  -S^y & S^z & S^x \\ S^y & 0 & -S^x & S^z  \\ -S^z & S^x & 0 & S^y \\ -S^x & - S^z & -S^y & 0 
  \end{pmatrix}.
\end{equation} And the cluster index can be recovered simply by replacement $\hat \Phi \rightarrow \hat \Phi_j$.
We stress how the original number of parameters expected from the $so(2n = 4)$ algebra, which simply required $\hat{\Phi}$ to be an antisymmetric matrix of generators, has been reduced from 6 to just $3$. We will discuss the nature of this reduction for generic replicas in section \ref{sec:coherent}.

\subsection{The case with \texorpdfstring{$q_J = 2$}{q_J = 2}}

By mapping to the spin operators Eq.~\eqref{eq:paramPhi}, the $n=2$ Hamiltonian Eq.~\eqref{eq:Htotal} can be written as
\begin{equation}\label{eq:spinHam}
  \mathcal{H} = - 2 J  \sum_j \vec{S}_j \cdot \vec{S}_{j+1} + 2 \Gamma \sum_j  (S_j^y)^2 .
\end{equation}
Some remarks about this expression is needed. Formally, each flavour contributes with a $2$-dimensional representation of $su(2)$, so that for each cluster, we have a $2^{N_F}$-dimensional representation. This representation is not irreducible. In particular, we are interested in quantities always involving states, which are completely symmetric under the exchange of spins within each clusters. An example is the late-time purity of the first cluster, which we identify with $A$ and we denote with $B := \bar{A}$. Then from Eq.~\eqref{eq:pur}, this quantity can be written as
\begin{equation}\label{eq:puroverlSpin}
 \lim_{t \rightarrow \infty} \frac{\Tr \left[ \rho^{(2)}(t) \left(\mathcal{C}_{2,A} \otimes \mathbb{I} \right)\right]}{\Tr \rho^{(2)}(t)}= \frac{1}{Z}\langle X_A, \otimes_{i \in B} Z_i  | GS\rangle  ,
\end{equation}which follows from Eq.~\eqref{boundstate}.
Normalization is given by the overlap of the ground state with the identity $Z=\langle \mathbb{I} |  GS\rangle$. The states $\ket{X_A}$ and $\ket{Z_i}$ correspond to the fully polarized product states in the $x$ and $z$ positive directions respectively, e.g. $\ket{Z_i} = \otimes_{\mu=1}^{N_F}\ket{\Uparrow_{i,\mu}}$ and similarly for $\ket{X_A}$. Being completely symmetric states, they belong to the representation with maximal spin $S := N_F/2$ and dimension $N_F+1$. Also, $\left| GS \right>$ is the ground state of the Hamiltonian Eq.~\eqref{eq:spinHam} reached starting from the identity initial state. So, all states appearing in Eq.~\eqref{eq:puroverlSpin} belong to this representation for each site $i= 1,\ldots, L$. Eq.~\eqref{eq:spinHam} was shown to exhibit a Kosterlitz-Thouless transition~\cite{BAO2021168618}, but the behavior of this transition is modified in the limit $n\to 1$~\cite{Fava:2023tgg}.

In the case of large $N_F$, the basis of spin coherent states offer a useful framework to deal with this computation. We will see that for $L=2$ clusters, this formalism leads to an accurate prediction for the behavior of the purity as a function of $N_F$. Instead, in the continuum space limit, this leads to the NLSM action. 
\begin{figure}
    \centering
    \includegraphics[trim={0 0.2cm 0 0.2cm}, clip,width=6.5cm]{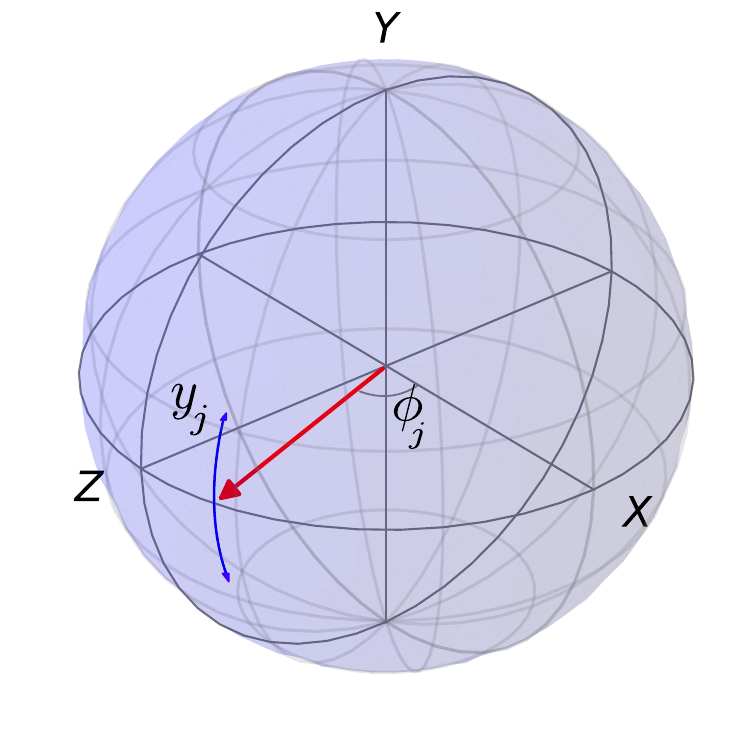}
    \caption{Representation of $j$-th spin vector on the Bloch sphere (the red line). Monitoring part of the evolution forces the vector to be in $XZ$ plane, however coupling of $y_j$ with $\phi_j$ through the kinetic term allows fluctuations in $Y$ direction (the blue line) see Eq.~\eqref{eq:act2}. }
    \label{fig_bloch_sphere}
\end{figure}
Given the form of Eq.~\eqref{eq:spinHam}, it is convenient to introduce a coherent states path integral using the $y$-axis as the reference direction. We thus define
\begin{equation}
\label{eq:su2cohe}
\ket{\Omega_{\hat{n}}} := e^{-i \frac{{\rm arccos} y}{ \sqrt{1-y^2}} ( \mathbf{\hat{n}} \cross \hat{\mathbf{y}})\cdot \mathbf{S}} \ket{\Uparrow^y} ~ .
\end{equation}
where $\hat{n}$ is a unit vector with $n_y = y$, $n_x = \sqrt{1-y^2} \sin \phi$, $n_z = \sqrt{1-y^2} \cos \phi$ (see Appendix \ref{App:SpinCoherent} for details) and $\ket{\Uparrow^y}$ is the fully polarized state in the positive $y$-direction. Then the path integral takes the form
\begin{equation}
\label{eq:overlapn2}
    \langle GS | X_A, Z_B \rangle \sim \lim_{T \rightarrow \infty} \prod_{j=1}^N \int^{1}_{-1} dy_j \int^{2 \pi}_0 d \phi_j  e^{-\mathcal{S}} ,
\end{equation}
with the action
\begin{equation}\label{eq:act}
    \mathcal{S} = \int_0^t dt\left( \sum^N_{j=1} K_j + \sum^N_{j=1} H_{j,j+1} \right) ,
\end{equation}
where the kinetic term is given by 
\begin{equation}
    K_j = -i S \dot \phi_j(1-y_j) ,
\end{equation}
and the Hamiltonian part is
\begin{equation}\label{eq:Huni2}
\begin{split}
    &H_{j,j+1} = \\
    & - 2 J S^2 \sum_j \bigg[ \sqrt{(1-y_j^2)(1-y_{j+1}^2)} \cos (\phi_j - \phi_{j+1})  \\
    & +  y_j y_{j+1} \bigg] + \Gamma \sum_j 2S^2 (y_j^2 + y_{j+1}^2).
\end{split}
\end{equation}
{We now take the limit of large $S$ (namely $N_F \gg 1 $ in each cluster) and redefine $y \rightarrow y/S$, so that $y \in [-S, S]$.}
We can then expand with respect to $y$ to obtain a quadratic action in these variables,
\begin{equation}\label{eq:act2}
\begin{split}
    \mathcal{S} & = \int^T_0 dt \Bigg[   -2 J S^2 \sum_{j=1}^N \cos (\phi_j- \phi_{j+1})   \\
    & + 2 \Gamma \sum_{j=1}^N y_j^2 - i \sum_{j=1}^N  y_j \dot{\phi}_j + O(y/S) \Bigg],
\end{split}
\end{equation}
where we neglected the quadratic order in $y$ from the unitary part, assuming the scaling of the coupling $J\sim S^{\alpha}$, with $\alpha \in (-2,0)$. Here we can integrate over $y$ to find the effective action
\begin{equation}
    \mathcal{S} = \int^T_0 dt \ \left[ \sum_{j=1}^N \frac{\dot{\phi}_j ^2}{8 \Gamma} -2 JS^2\sum_{j=1}^N \cos (\phi_j - \phi_{j+1}) \right] .
\end{equation}

\paragraph{Continuum space limit. -- }
To take the continuous limit and derive the NLSM action, we assume $\phi_{j+1} - \phi_{j} = \Delta x \partial_x \phi_j$, where $\Delta x$ is the lattice distance, and expand the action with respect to $\Delta x$. Taking the limit $\Delta x \rightarrow 0$, $L \rightarrow \infty$ one obtains
\begin{equation}
    \mathcal{S} = \int d^2x \ \left[\frac{(\dot{\phi}(x)) ^2}{8 \Delta x \Gamma} + JS^2  \Delta x (\partial_x \phi(x))^2 \right] ,
\end{equation}
corresponding to the celebrated NLSM action on the $U(1)$ group for the compactified field $\phi$.

\paragraph{Two clusters. --}
Next we consider the case $L=2$ of just two clusters and calculate the purity.
In this case, we introduce the relative coordinate and the centre of mass coordinate $\tilde \phi = (\phi_2 - \phi_1)/2$ and $\tilde \theta = (\phi_1 + \phi_2)/2$ so that
\begin{equation}
\begin{split}
\label{eq:actionn2q2}
     \mathcal{S} & = \mathcal{S}[\theta] + \mathcal{S}[\phi]  \\ 
     & = \int^T_0 dt \left[ \frac{\dot{\tilde \theta}^2}{4 \Gamma} + \frac{\dot{\tilde \phi}^2}{4 \Gamma} -2 JS^2 \cos 2 \tilde \phi \right].
\end{split}
\end{equation}
Notice that, we can compute the overlap exactly, taking into account the boundary state $\left|X_A,Z_B \right> = \left|X_1,Z_2 \right>$. 
First, for the large times we can interpret the overlap as the transition amplitude between the boundary state and the ground state of a Hamiltonian of a particle on a circle with a pendulum potential, namely
\begin{equation}
    \mathcal{H} =  \frac{\dot{\tilde{\phi}}^2}{4 \Gamma} - 2 J S^2 \cos 2 \tilde \phi .
\end{equation}
For large $S$, the ground state can be approximated to be same as the ground state of a harmonic oscillator with $\tilde \phi = 0$,
\begin{equation}
    \mathcal{H} = \frac{p^2}{2m} + 4 J S^2\tilde \phi^2 ,
\end{equation}
with $m \omega^2/2 = 4 J S^2$, explicitly
\begin{equation}
    \Psi_{\overline{GS}} (\tilde \phi) \sim e^{-m \omega \tilde \phi^2/2}.
\end{equation}
So for the overlap we have
\begin{equation}
    \langle X_A, Z_B |GS  \rangle = \int d \tilde \phi d \tilde \theta e^{-m \omega \tilde \phi^2/2} \left<X_A,Z_B|\tilde \phi,\tilde \theta \right> .
\end{equation} Here the boundary state $\left|X_A,Z_B \right>$ corresponds to the coherent state with $\tilde \phi_b=\frac{\pi}{2}$ and $\tilde \theta_b=0$.
Using the overlaps between two coherent states (see Appendix \ref{App:SpinCoherent}) we obtain 
the following integral
\begin{equation}
\begin{split}
     \langle X_A, Z_B   & |GS \rangle= \frac{1}{2^{4 S}} \int_{- \pi}^{ \pi} d \tilde \phi \int_{0}^{2 \pi} d \tilde \theta \\
     & e^{-m \omega \tilde \phi^2/2} \left(e^{-i(\frac{\pi}{2} - \tilde \phi)} + 1\right)^{2S} \left(e^{i \tilde \theta} + 1\right)^{2S},
\end{split}
\end{equation}
with $m \omega = 2 S \sqrt{\frac{ J}{ \Gamma}}$. The leading order of this expression is given by $\tilde \phi =\frac{\pi}{2}$ and $\tilde \theta = 0$ or $\tilde \theta = 2\pi$, and the latter gives the replicated purity as
\begin{equation}\begin{split}
    \lim_{t \rightarrow \infty} \frac{\Tr \left[ \rho_A^{(2)}(t) \left(\mathcal{C}_2 \otimes \mathbb{I} \right)\right]}{\Tr \rho^{(2)}(t)} \sim e^{-2 S \sqrt{\frac{ J}{\Gamma}}\frac{\pi^2}{8}} ,
    \end{split}
\end{equation}
where $J\sim S^{\alpha}$, with $\alpha \in (-2,0)$.
In the next subsection, we repeat the analysis for the interacting case $q_J=4$. We show that the replicated purity can be again approximated as an overlap of the ground state of an oscillator with the boundary state. 
\subsection{The case with \texorpdfstring{$q_J = 4$}{q_J = 4}}

We now consider the case where the unitary part is interacting, but measurements remain quadratic. In this subsection, we derive a field theory action and calculate the purity in the case of two clusters. Using the mapping between the operator $\hat \Phi$ and the spin operator in Eq.~\eqref{eq:paramPhi}, we can write the replicated Hamiltonian:
\begin{equation}
\begin{split}
    & \mathcal{H}  = \frac{J}{2} \sum_j  \left[ (S^y_j)^{2} (S^y_{j+1})^{2} - ((S^x_j)^{2} (S^x_{j+1})^{2} \right. \\ 
    & \left.+(S^z_j)^{2} (S^z_{j+1})^{2}) \right]  + \frac{J }{12}\sum_j\left[ (S^y_j)^{4} -((S^x_j)^4 \right.\\
    & \left. + (S^z_j)^4) \right] +  2 { \Gamma }\sum_j (S^y_j)^2.
\end{split}
\end{equation}
We can then repeat the procedure of the previous sections, calculating the effective action for the path integral and integrating out the massive degrees of freedom as we did in the $q_J=2$ case, we find the following effective action 
\begin{equation}
\begin{split}
    \mathcal{S} &  = \int^t_0 d \tau \ \left[ \frac{\dot{\tilde \theta}_j^2}{4 \Gamma}  + \frac{\dot{\tilde \phi}_j^2}{4 \Gamma} \right. \\ & \left.- \frac{J S^4}{24}(3+\cos  4 \tilde\phi_j)(3+ \cos 4 \tilde  \theta_j) \right],
\end{split}
\end{equation}
with $\tilde \phi_j = (\phi_{j+1}-\phi_j)/2$ and $\tilde \theta_j =(\phi_{j+1}+\phi_j)/2$. In the case of two clusters, for the large times, we can again compute the overlap explicitly, mapping the partition function into the Hamiltonian evolution.
This problem reduces to finding the ground state of the following Hamiltonian
\begin{equation}
\begin{split}
   H  = \frac{p_{\tilde \phi}^2}{2m} + \frac{p_{\tilde \theta}^2}{2m} - \frac{J S^4}{24}(3+\cos 4\tilde \phi)(3+ \cos  4\tilde \theta),
\end{split}
\end{equation}
where $m=\frac{1}{2 \Gamma}$ and $p_{\sigma}=\dot\sigma/(2\Gamma)$.
In order to find the ground states, we maximize the potential, applying the standard variational method
\begin{equation}
    \partial_{\tilde \phi} V = 0 \quad , \quad \partial_{\tilde \theta} V = 0 ,
\end{equation}
which corresponds to the condition:
\begin{gather}
     \sin 4\tilde \phi = 0 , \\ 
     \sin  4\tilde \theta= 0 ,
\end{gather}
where $\tilde \theta \in (0, 2 \pi)$ and $\tilde \phi \in (- \pi, \pi)$. This condition gives us a set of solutions for maximas
\begin{equation}\label{eq:min}
   \tilde \phi = n \frac{\pi}{2} \quad , \quad \tilde \theta = m \frac{\pi}{2} ,
\end{equation}
with $ n \in \{-2,-1,...,1,2\}$, $ m \in \{0,..,4\}$. We again expand around these minima (the consideration is the same for each of them), 
\begin{equation}
    \mathcal{H} = \frac{p^2_{\tilde \phi}}{2m} + \frac{p^2_{\tilde \theta}}{2m} + \frac{4 J S^4}{3}(\tilde \phi^2 + \tilde \theta^2) ,
\end{equation}
which is just a sum of two uncoupled oscillators. Thus, for the $\overline{GS}$ of the Hamiltonian we can write
\begin{equation}
    \Psi_{\overline{GS}}(\phi,\theta) \sim e^{- m \omega (\tilde \phi^2 + \tilde \theta^2)/2} ,
\end{equation}
with $m = \frac{1}{2 \Gamma}$, and $m \omega^2 = \frac{8 J S^4}{3}$. Finally, the overlap takes the form
\begin{equation}
\begin{split}
   &  \left<X_A,Z_B|GS\right> \sim \left< X_A, Z_B \right| \int d \tilde \phi d \tilde \theta \\
   & e^{- m \omega( \tilde \phi^2 + \tilde \theta^2)/2} \sum_{\rm min} \left|\tilde \phi - \tilde \phi_{\rm min}, \tilde \theta - \tilde \theta_{\rm min}\right>.
\end{split}
\end{equation}
Once again, using the overlaps of coherent states (Appendix \ref{App:SpinCoherent}) and the connection between the replicated purity and the overlap Eq.~\eqref{eq:puroverlSpin}, for $1/S^4<J \leq 1/S^2$ and dominating maximas Eq.~\eqref{eq:min} we obtain the result for the late-time purity 

\begin{equation}
    \lim_{t \rightarrow \infty} \frac{\Tr \left[ \rho_A^{(2)}(t) \left(\mathcal{C}_2 \otimes \mathbb{I} \right)\right]}{\Tr \rho^{(2)}(t)} \sim e^{- S^2 \sqrt{\frac{J}{3 \Gamma}} \left(\frac{\pi}{2}\right)^2},
\end{equation}
while in the case $1/S^2<J $, we can neglect the overlap, so we just have the Gaussian integral,
\begin{equation}
\begin{split}
  &   \lim_{t \rightarrow \infty} \frac{\Tr \left[ \rho_A^{(2)}(t) \left(\mathcal{C}_2 \otimes \mathbb{I} \right)\right]}{\Tr\rho^{(2)}(t)} 
  \\& = \frac{1}{2^{4 S} Z} \int d \tilde \phi d \tilde \theta e^{- m \omega (\tilde\phi^2 + \tilde\theta^2)/2} \sim e^{-4 S \log(2)},
\end{split}
\end{equation}
which gives a volume law scaling of the entanglement with respect to $S$. Here, $S=\frac{N_F}{2}$ effective spin and $N_F$ number of Majorana fermions in each cluster. Since we consider a system of two clusters, the entanglement entropy scales with cluster size $N_F$.
Thus, we have calculated the replicated purity for $n=2$ for two clusters in both cases $q_J=2$ and $q_J=4$ and derived the NLSM for $q_J=2$. 
In the next sections, we repeat this calculation for the generic number of replicas using the formalism of the coherent states on the initial $SO(n)$ group. We also discuss the scaling of the entanglement with the number of clusters $L$ in Sec(\ref{ssec:NLSM}). 
Next, we take the limit $n \rightarrow 1$ and arrive at the desired expression for the purity in the correct replica limit.

\section{Coherent states representation for \texorpdfstring{$n$}{n} replicas and the NLSM}\label{sec:coherent}

\subsection{General procedure for coherent states}
Before discussing the details of the fermionic coherent states for the $SO(2n)$ Lie group, it is useful to briefly remark about the general construction and how it applies to the familiar example of $SU(2)$ that we discussed above. 
In order to construct the coherent state representation of a given Lie group $\mathcal{G}$, we assume that $\{\hhc_a, \eec_{\vec\rho}, \eec_{-\vec\rho}\}$ denote the generators of its Lie algebra in the standard Cartan-Weyl basis and $\vec\rho \in \mathfrak{R}^+$, the set of positive roots. For a specific description of this basis in the case of $so(2n)$ using fermionic bilinears, see Appendix~\ref{sec:cartan}. 
Then, the coherent states can be written as
\begin{equation}
\label{eq:naivecoh}
\ket{g} = \Omega(g) \ket{\mathbb{I}}   , 
\end{equation}
where $g \in \mathcal{G}$ and $\Omega$ is an irreducible representation $\mathcal{G}$. The state $\ket{\mathbb{I}}$ is a reference state within the chosen representation that we take as normalised $\braket{\mathbb{I}}{\mathbb{I}} = 1$. Since we are dealing with unitary representations, this definition ensures that the coherent state $\ket{g}$ is normalised 
\begin{equation}
\label{eq:cohenorm}
    \braket{g}{g} = 1\;.
\end{equation}
There are multiple choices for the state $\ket{\mathbb{I}}$, but a essential requirement is that it is a highest-weight of the representation~\cite{STONE1989399}.
As it is well known, the coherent states are a largely overcomplete basis. However, 
one can use Haar measure over $\mathcal{G}$ to define a resolution of the identity in the form
\begin{equation}
\label{eq:resolid}
    \int_{\operatorname{Haar}} dg \ket{g} \bra{g} = \mathbb{I},
\end{equation}
which simply follows from an application of Schur's lemma. By repeatedly inserting the resolution of the identity Eq.~\eqref{eq:resolid} in between small time steps of the evolution Eq.~\eqref{eq_folded_replicated_rho_v2}, one obtains the prototypical path-integral representation
\begin{multline}
\label{eq:pathintgen}
    \left<B \right|e^{-t \mathcal{H}^{(n)}}\left|\rho^{(n)}(0)\right>\\ = 
    \lim\limits_{\substack{N_t \to \infty \\ \delta t \to 0}} \int \prod_{j=1}^{N_t} 
    d g(t_j) 
    \braket{g(t_j) |e^{-\delta t \mathcal{H}}}{g(t_{j+1})} \\:=
    \int \mathcal{D}[g] e^{-S} ,
\end{multline}
where the boundary states determine the boundary conditions for $g$ within the path integral. The action $S := \int_0^t dt L[g]$ and the Lagrangian is obtaining by expanding $\braket{g(t_j) |e^{-\delta t \mathcal{H}}}{g(t_j)}$ at first order in $\delta t$ and reads
\begin{equation}
   L[g] := \braket{\dot g}{g} + \braket{g|\mathcal{H}^{(n)}}{g}  \;. \label{eq:Lgeneral}
\end{equation}
Note that $\bra{\dot g} = \lim_{\delta t \to 0} (\bra{g(t_{j+1})} - \bra{g(t_{j})})/\delta t$ is nothing more than a formal expression at this stage. In fact, beyond being overcomplete, the definition Eq.~\eqref{eq:naivecoh} provides multiple representatives of the \textit{same} quantum state and this leads to a problem in defining two states that are close one to the other. 
In particular, this redundancy precludes a proper definition of the Berry phase, the first term of Eq.~\eqref{eq:Lgeneral}. To overcome this difficulty, we denote as $\mathcal{H} \subset \mathcal{G}$ the stability subgroup of the reference state, defined by
\begin{equation}
    \Omega(h) \ket{\mathbb{I}} = e^{i \phi_h} \ket{\mathbb{I}},
\end{equation}
i.e. the elements of $\mathcal{G}$ which leave the reference state unchanged, up to a phase with $\phi_h \in [0,2\pi)$. An appropriate evaluation of the overlap $\braket{\dot g}{g}$ requires removing this redundancy. Formally, this is done by reducing the volume of integration from the whole group to the coset $\mathcal{G} \to \mathcal{G}/\mathcal{H}$, so that each coherent state is associated with a unique element $\tilde{g} \in \mathcal{G}/\mathcal{H}$.

This construction is familiar in the case of the standard $SU(2)$ that we used explicitly in Eq.~\eqref{eq:su2cohe}. In such a case, the stabilizer group $\mathcal{H} \simeq U(1)$, corresponding to the transformations generated by $S_y$. It follows that the coherent states are in one-to-one correspondence with $SU(2)/U(1) \simeq S^2$, i.e. the 2-dimensional sphere, parameterised in Eq.~\eqref{eq:su2cohe} with cylindrical coordinates $(y, \phi)$.

The Cartan-Weyl basis provides an effective way to express the elements of the coset. Indeed, the requirement that $\ket{\mathbb{I}}$ is a highest-weight ensures that it is a common eigenvector for all elementns of the Cartan subalgebra $\hhc_a$ annihilated by $\eec_{\vec\rho}$ for all positive roots
\begin{align}
    &\mathfrak{h}_a \ket{\mathbb{I}} = \lambda_a \ket{\mathbb{I}} , \quad \lambda_a \in \mathbb{C} 
    ,
    \label{eq:cartaneigen}
    \\ &\mathfrak{e}_{\vec\rho} \ket{\mathbb{I}} = 0 , \quad \forall \vec\rho\in\mathfrak{R}^+ .\label{eq:annihilateplus}
\end{align}
Beyond this general properties, in many relevant cases, the reference state $\ket{\mathbb{I}}$ might also be annihilated by a subset of the negative roots. So, decomposing the negative roots $\mathfrak{R}^- = \mathfrak{R}^-_0 \cup \mathfrak{R}^-_{\mathcal{H}}$, we have additionally
\begin{equation}
\eec_{\vec{\rho}} \ket{\mathbb{I}} = 0 , \quad \forall \vec{\rho} \in \mathfrak{R}^{-}_0 .
\label{eq:annihilateminus}
\end{equation}
By means of the exponential map, one has that the generators of the stability group $\mathcal{H}$ are precisely those in eqs.~(\ref{eq:cartaneigen}, \ref{eq:annihilateplus}, \ref{eq:annihilateminus}). Thus, this provides an effective representation for the elements of the coset in terms of the residual elements that do not annihilate the reference state, i.e.
\begin{equation}
\label{eq:cohstEtaGen}
    \Omega(\eta) \ket{\mathbb{I}} := \exp\left[{\sum\limits_{\vec \rho \in \mathfrak{R}^-_{\mathcal{H}}}
    \eta_{\vec\rho} \eec_{\vec\rho} - \text{c.c.}
    }\right] \ket{\mathbb{I}},
\end{equation}
with $\eta_{\vec\rho} \in \mathbb{C}$ and the minus sign ensures unitarity of the representation. Note that the normalisation condition Eq.~\eqref{eq:cohenorm} remains true, so we will be dealing with normalised coherent states.
In the following, we will see how this general construction applies to the case of the fermionic representation built in Sec.~\ref{sec:cartan}.

\subsection{Fermionic coherent states for \texorpdfstring{$SO(2n)$}{SO(2n)}}

We momentarily focus on a single cluster. As we have shown in Sec.~\ref{sec:so(2n)}, the fields $\hat \Phi^{\alpha \beta}$ span the $so(2n)$ Lie algebra and represent a complete set of generators. 
As we have already seen, the Hamiltonian can be expressed solely in terms of these operators. As clarified in the previous section, another ingredient in the definition of the coherent states is the reference state, which we will act on with the group elements generated by the operators $\hat \Phi$. 
Following \cite{Fava:2023tgg}, we define the reference state $\ket{\mathbb{I}}$ to be Gaussian and satisfy the condition
\begin{equation}\label{eq:ref}
  \bra{\mathbb{I}} \hat \Phi \ket{\mathbb{I}} = \frac{\Sigma}{2},
\end{equation}
with the usual symplectic matrix given by
\begin{equation}
\Sigma = \begin{pmatrix}
0 & \mathbb{I}_n \\ - \mathbb{I}_n & 0 
\end{pmatrix}
.
\end{equation}
Such a state is also a highest-weight for the $so(2n)$ algebra. We
provide a summary of the Cartan-Weyl basis for $so(2n)$ in Appendix~\ref{sec:cartan}. More directly, we set $\ket{\mathbb{I}}$ as the fully occupied state
\begin{equation}
\label{eq:cdacI}
c^\dag_{a \nu} c_{a \nu} \ket{\mathbb{I}} = \ket{\mathbb{I}}  , 
\end{equation}
where $a = 1,\ldots, n$ and $\mu = 1,\ldots, N_F$. Note that for the case $n = 2$, this state reduces to $\ket{\mathbb{I}} = \ket{\Uparrow}$, so there is a discrepancy with the convention we took in Eq.~\eqref{eq:su2cohe}, where the $y$ direction was used.
The state defined by Eq.~\eqref{eq:cdacI} is also consistent with  Eq.~\eqref{eq:ref}.
As explained in Appendix~\ref{sec:cartan}, the generators $\hhc_a = c^\dag_a c_a - 1/2$ (see Eq.~\eqref{eq:cartangen}), so that Eq.~\eqref{eq:cartaneigen} holds with $\lambda_a = 1/2$. Instead, $\eec_{\vec\rho}$ associated to the positive roots $\vec\rho \in \mathfrak{R}^+$ split into two kinds: $\{c^\dag_a c^\dag_{a'}\}_{a<a'}$ and $\{c^\dag_a c_{a'}\}_{a<a'}$ and both annihilate the reference state in agreement with Eq.~\eqref{eq:annihilateplus}.

Taking the complex conjugate, we can write the generators associated with negative roots as two families: $\{c_a c_a'\}_{a<a'}$ and $\{c_a^\dag c_{a'}\}_{a>a'}$. Consistently with Eq.~\eqref{eq:annihilateminus}, we see that the latter also annihilate the reference state.
We can thus simply characterize the stability group as generated by the set of $n^2$ operators $\{\frac{1}{N_F} \sum_\nu c^{\dagger}_{a \nu} c_{a'\nu} - \frac{1}{2} \delta_{aa'}\}_{a,a'=1}^n $. As these are the generators of the unitary group $U(n) \simeq \mathcal{H}$, we deduce that the coherent states are isomorphic to $SO(2n)/U(n)$. Following the general prescription in Eq.~\eqref{eq:cohstEtaGen}, we can represent the coherent states using the generators that do not annihilate the references state
\begin{equation}\label{eq:cohstEta}
\begin{split}
   & \left| O_\eta\right> = \exp\left[{\sum\limits_{\nu,1 \leq a < a' \leq n} \eta_{a a'} c^{\dagger}_{a\nu} c^{\dagger}_{a'\nu} - \text{h.c.}} \right] \left| \mathbb{I} \right> 
    \\& \equiv T_{\eta}  \left| \mathbb{I} \right> .
\end{split}
\end{equation}
As we can see, the presence of the stability group explains the reduction in the effective dimensionality of the coherent states: While the group $SO(2n)$ is a manifold of size $n(2n-1)$, $SO(2n)/U(n)$ has dimension $n(n-1)$. In the case $n=2$, this reduces to $2$, consistent with the coordinates $(y, \phi)$ parameterizing the sphere in Eq.~\eqref{eq:overlapn2}.

Notice that we easily recover the cluster index and write the generalisation as a tensor product over multiple clusters
\begin{equation}
\begin{split}
    \bigotimes_{j=1}^L & \left| O_{\eta^j}\right> \\ & = \exp\left[{\sum\limits_{j,\nu,1 \leq a< a' \leq n} \eta^j_{a a'} c^{\dagger}_{a\nu j} c^{\dagger}_{a'\nu j} - \text{h.c.}}\right] \left| \mathbb{I} \right> .
\end{split}
\end{equation}

In the following, we will discuss how this formalism can be used to compute the relevant quantities. For instance, following the derivation in Eq.~\eqref{eq:pathintgen}, but with the replacement $\ket{g} \to \ket{O_\eta}$, we arrive at the expression of the purity as a path integral
\begin{equation}
    I:=\left< \mathcal{C}_{A,2} \right|e^{-t \mathcal{H}^{(n)}}\left|\rho^{(n)}(0)\right>=
    \int \mathcal{D}[\eta] e^{-S}\;.
\end{equation}
As explained in Eq.~\eqref{eq:Lgeneral}, 
this formal expression involves three ingredients: 
a resolution of the identity in terms of the Haar measure on the coset space, the overlap between two close coherent states at two consequent times $\braket{O_\eta(t_j)}{ O_\eta(t_{j+1})}$, leading to the Berry phase and the expectation value of the Hamiltonian on the coherent states $\braket{O_\eta | \mathcal{H}^{(n)}}{ O_\eta }$. We will start deriving an expression for the latter. Then, we will discuss the Berry phase and the geometric structure of the coset manifold, thus leading to the Haar measure and the resolution of the identity.

\subsection{The Hamiltonian in the coherent states basis}
As we saw, one of the ingredients of the path integral is the calculation of the expectation value of the Hamiltonian on the coherent states (see Appendix \ref{App:matel}). One can notice that our Hamiltonian depends on $\hat \Phi_i$, so we calculate the expectation values of the powers of $\hat \Phi_i$. Here, the large number of flavors becomes useful: the operator $\hat\Phi$ in Eq.~\eqref{eq:fieldoperators} is expressed as an average over a large number of independent operators. Thus, quantum fluctuations in higher-order moments are subleading and one has
\begin{equation}
\begin{aligned}
   & \left<O_{\eta}\right| (\hat\Phi^{\alpha \alpha'}_i)^{m} \left|O_{\eta} \right> \\
   = & \left<O_{\eta}\right| \hat\Phi^{\alpha \alpha'}_i \left|O_{\eta} \right>^m \left(1+O\left(\frac{1}{N_F}\right)\right).
\end{aligned}
\end{equation}
Therefore we need to determine the expectation values of the operators $\hat \Phi$. Moreover, we can express the operators $\hat \Phi$ in terms of the Dirac fermions, as was discussed in the previous subsection. 
As a result, we reduce the problem to finding the expectation values of the quadratic operators consisting of $c^{\dagger}$ and $c$. Let us consider one of these quadratic operators,
\begin{equation}
\begin{split}
    &\left<O_\eta \right| c^{\dagger}_{a\nu}c_{a' \nu}\left| O_\eta \right> = \left<\mathbb{I} \right| T^{-1}_{\eta} c^{\dagger}_{a\nu} T_{\eta }T^{-1}_{\eta} c_{a' \nu} T_{\eta}\left| \mathbb{I} \right> ,
\end{split}
\end{equation}
where $T_{\eta}$ is defined in Eq.~\eqref{eq:cohstEta}.
So we are interested in finding the transformation of each fermion under the operator $T_{\eta}$. As the transformation has to preserve the fermion algebra, it must take the form of a Bogoliubov transformation, that we parameterise as
\begin{equation}\label{Ttr}
\begin{bmatrix}
    T_{\eta} c_{a\nu} T^{-1}_{\eta} \\
    T_{\eta} c^{\dagger}_{a\nu} T^{-1}_{\eta}
\end{bmatrix} = \mathbb{T}^{\dagger} \begin{bmatrix}
    c_{a \nu} \\
    c^{\dagger}_{a \nu}
\end{bmatrix}
= \begin{bmatrix}
    U^{\dagger}_{ab} c_{b \nu} + V^{\dagger}_{ab} c^{\dagger}_{b\nu} \\
    V^T_{ab} c_{b \nu} + U^T_{ab} c^{\dagger}_{b \nu}
\end{bmatrix},
\end{equation}
with
\begin{equation}
    \mathbb{T}^{\dagger} = \begin{pmatrix}
        U^{\dagger} & V^{\dagger} \\
        V^t & U^t
    \end{pmatrix}.
\end{equation}
Here the summation over repeating replica index $b$ is implicit. 
We can explictly connect these matrices $U$ and $V$ with the matrix $\eta$ (see Appendix \ref{App:matel}) as 
\begin{equation}
\label{eq:UVeta}
    U^{\dagger} = U = \cos(\sqrt{ \eta \eta^{\dagger}})\ , \ V^{\dagger} =- \frac{\sin(\sqrt{ \eta \eta^{\dagger}})}{\sqrt{ \eta \eta^{\dagger}}} \eta .
\end{equation}
Finally, we express the expectation value $\left<O_\eta\right|\hat \Phi^{\alpha \alpha'} \left|O_\eta \right> = \Phi^{\alpha \alpha'}$ in terms of these variables. 
We define
\begin{equation}
\hat\Psi^{a,b} = \frac{1}{N_F}\sum_\mu\left( \begin{array}{c|c}
       c_{a,\mu}^\dag c_{b,\mu}  & c_{a,\mu}^\dag c_{b,\mu}^\dag  \\ \hline
       c_{a,\mu} c_{b,\mu}  & 
      c_{a,\mu} c_{b,\mu}^\dag  
     \end{array} \right) \;.
\end{equation}
Then, on the reference state we simply get
\begin{equation}
\label{eq:Psi0def}
    \Psi_0 := \braket{\mathbb{I} | \hat\Psi}{\mathbb{I}} = \left(\begin{array}{c|c}
        \mathbb{I} & 0 \\ \hline
        0 & 0
    \end{array} \right)
    ,
\end{equation}
and on any coherent state
\begin{equation}
\label{eq:PsiTT}
\begin{split}
    &\Psi := 
    \left<O_\eta \right| \hat\Psi \left| O_\eta \right>=\\
    &= \mathbb{T}^* \Psi_0 \mathbb{T}^{t} = \begin{pmatrix}
        U^t U^t & U^t V^t \\
        -V^{\dagger} U^t & -V^{\dagger} V^t
    \end{pmatrix} \;.
    \end{split}
\end{equation}
Finally, we can go back to expressing the expectation value of the Majorana operators. This is easily done by introducing the transformation to go from Dirac to Majorana fermions. We can write compactly
\begin{equation}
    \begin{pmatrix}
        c^\dag\\
        c
    \end{pmatrix} = R^\dag 
        \begin{pmatrix}
        \chi^+ \\
        \chi^-
    \end{pmatrix}\;,
\end{equation}
where $(c^\dag, c)$ is a shortcut for $(c^\dag_1,\ldots, c_n^\dag, c_1,\ldots, c_n)$ and similarly for $(\chi^+,\chi^-)$. The matrix $R$ takes the form of a $2n\times 2n$ matrix
\begin{equation}
    R = \frac{1}{\sqrt{2}} \begin{pmatrix}
        1 & 1 \\
        i & -i
    \end{pmatrix} \otimes \mathds{1}_n
    ,
\end{equation}
where the first matrix acts in the $(+,-)$ index space and the identity acts in the $a=1,\ldots,n$ replica space. We finally arrive at
\begin{equation}
    \Phi = \begin{pmatrix}
        \Phi^{++} & \Phi^{+-} \\
        \Phi^{-+} & \Phi^{--}
    \end{pmatrix} 
    = i R \Psi R^{\dagger} - i \frac{\mathds{1}}{2}
    \;.
    \label{eq:PhiFromPsi}
\end{equation} 
Combining Eq.~(\ref{eq:UVeta},\ref{eq:PsiTT},\ref{eq:PhiFromPsi}), each of the components of the matrix $\Phi$ is now expressed in terms of the parameters $\eta$. In the next subsection, we proceed similarly for the kinetic term of the path integral.

\subsection{The kinetic term of the action}\label{sec:kin}
We now focus on expressing the kinetic term (or Berry phase) in the parameterization Eq.~\eqref{eq:cohstEta}. The fundamental ingredient is the overlap 
$\braket{O_\eta}{O_{\eta'}}
$ and we now show how it can be obtained by a particular form of Baker-Campbell-Hausdorff (BCH)~\cite{RevModPhys.62.867}. 
To simplify the notation, we momentarily work at $N_F=1$ avoiding the sum over the greek flavour index. General formulas can be recovered at the end by tensor product over the flavour degrees of freedom. 
We have the identity
\begin{multline}\label{eq:BCH1}
\exp\left[{\sum_{a< a'} \eta_{a a'} c^{\dagger}_{a} c^{\dagger}_{a'} - \text{h.c.}} \right] = \\
   = W_{-\tau}^\dag 
    \exp[\sum_{a,b = 1}^n \xi_{a,b} (c^\dag_a c_{b}- \delta_{ab}/2)]
    W_\tau    \;,
\end{multline}
with $W_\tau = \exp[\sum_{a<a'} \tau_{a,a'} c_a^\dag c_{a'}^\dag]$ and where $\tau$ and $\gamma$ are two complex $n\times n$ matrices. Clearly, one can choose $\tau^t = - \tau$. As this relation only depends on the commutation relations within the algebra, its proof and the explicit relation between the matrices $\gamma$ and $\tau$ with $\eta$ can be derived by using the fundamental representation (see Appendix \ref{App:matel}). We obtain
\begin{subequations}
\label{eq:etatau}
    \begin{align}
    &z = \eta \frac{\sin(\sqrt{\eta^{\dagger} \eta})}{\sqrt{\eta^{\dagger} \eta}}, \\
    &e^\xi = (1 - z z^\dag )^{1/2} \;, \quad \tau = z(1 - z^{\dagger} z)^{-1/2}. 
\end{align}
\end{subequations}
Eq.~\eqref{eq:BCH1} is particularly useful when acting on the reference state $\ket{\mathbb{I}}$. Indeed, as this state is completely filled
$W_{\tau} \ket{\mathbb{I}} = \ket{\mathbb{I}}$. Additionally, expanding the exponential, one clearly has
\begin{multline}
\label{eq:Agammasimpl}
     \exp[\sum_{a,b} \xi_{a,b} (c^\dag_a c_{b}-1/2 \delta_{ab})] \ket{\mathbb{I}} = \\ =
     \exp[\sum_{a} \xi_{a,a} (c^\dag_a c_{a}-1/2)] \ket{\mathbb{I}} = A_\gamma \ket{\mathbb{I}}\;,
\end{multline}
with $A_\gamma = \prod_a e^{\gamma_{a,a}/2}$.
Thus, we simply have the representation
\begin{align}
    \ket{O_\eta} = A_\gamma  W^\dag_{-\tau}\ket{\mathbb{I}} \;.
\end{align}
Note that $A_\gamma$ is implicitly a function of the matrix $\eta$, so we write $A_\gamma = A[\eta]$. Let us use this expression to find the overlap between two arbitrary coherent states,
\begin{equation}
\label{eq:OAAN}
    \braket{O_\eta}{O_{\eta'}} = A[\eta] A[\eta'] \mathcal{N}[\tau, \tau^{\prime}] ,
\end{equation}
where we have set
\begin{equation}
\mathcal{N}[\tau, \tau^{\prime}] = \bra{\mathbb{I}}  W_{-\tau'} W_{-\tau}^\dag \ket{\mathbb{I}} .
\end{equation}
The explicit form of this factor can be found using direct manipulations of fermion operators~(see for instance~\cite{PhysRevB.87.245107}). One has
\begin{equation}
\label{eq:dettau}
    \mathcal{N}[\tau, \tau'] = \det(\mathbb{I} + \tau^{\dagger} \tau')^{N_F/2} \;,
\end{equation}
where in this expression we reintroduced an arbitrary number of flavors, simply noting that the overlap is factorised over them so $N_F$ just appears in the exponent. Finally, the expression of $A[\eta]$ could be obtained by direct manipulations. However,  it is more convenient to use the fact that 
coherent states $\ket{O_\eta}$ are normalised (see Eq.~\eqref{eq:cohenorm}), so that setting $\eta' = \eta$ (and thus $\tau' = \tau$), we must have
\begin{equation}
    A[\eta]^{-2} = \mathcal{N}[\tau, \tau] \;.
\end{equation}
Using the mapping Eq.~\eqref{eq:etatau}, we can always relate the matrix $\tau$ to $\eta$ parameterising the coherent state. Also, we observe that $\mathcal{N}[\tau, \tau']$ depends on the complex variables $\tau, \tau'$ but it is a real function when $\tau' = \tau$. So, it is convenient to use the standard notation for complex functions $\mathcal{N}[\tau, \tau'] \to \mathcal{N}[\tau, \tau^\ast]$. Thus, 
\begin{equation}
\label{eq:overlapN}
    \braket{O_\eta}{O_{\eta'}} = 
    \frac{\mathcal{N}[\tau, \tau^{\ast \prime}]}{\sqrt{\mathcal{N}[\tau, \tau^\ast]\mathcal{N}[\tau', \tau^{\ast \prime}]}} \;.
\end{equation}
As a side note, we observe that the square modulus of the overlap can be obtained more directly by noting that $\rho_\tau := \ket{O_\tau}\bra{O_\tau}$ defines a density matrix and the overlap $\Tr[\rho_\tau \rho_{\tau'}]$ between two density matrices has a simple form in terms of their correlation matrices~\cite{PhysRevB.87.245107}, thus
\begin{equation}
\label{eq:overlapPhi}
|\braket{O_\tau}{O_{\tau'}}|^2  = \det\left(\frac{\mathbb{I} - 4 \Phi_\tau \Phi_{\tau'}}{2}\right)^{N_F/2} \;.
\end{equation}
and using the explicit relation between $\Phi_\tau$ and $\tau$, which follows by Eqs.~(\ref{eq:PhiFromPsi}, \ref{eq:etatau}), one can verify that Eqs.~(\ref{eq:overlapN},\ref{eq:overlapPhi}) are consistent.

Finally, setting $\eta' = \eta + d \eta$, we can  expand with respect to $d \eta$ to find
\begin{equation}
\begin{split}
    &\ln \braket{O_\eta}{O_{\eta + d\eta}} = -\frac{1}{2} \sum_{a<b} \\
    &  \left[\partial_{\eta_{ab}}(\ln \mathcal{N}) d\eta_{ab} - \partial_{\eta^\ast_{ab}}(\ln \mathcal{N}) d\eta^\ast_{ab} \right] .
\end{split}
\end{equation}
We see that this expression is invariant under reparameterizations of the coherent states. So, it is also convenient to use Eq.~\eqref{eq:etatau} to express the coherent states directly in terms of $\tau$, i.e. $\ket{O_\eta} \to \ket{O_\tau}$. Then, we can express the kinetic term as
\begin{equation}\label{eq:kin}
\begin{split}
    &K = -\lim_{dt \to 0}\frac{\ln \braket{O_\tau}{O_{\tau + \dot{\tau} dt}}}{dt} =  \frac{1}{2} \sum_{a<b}  \\
    & \left[\partial_{\tau_{ab}} (\ln \mathcal{N}) \dot{\tau}_{ab} - \partial_{\tau_{ab}^\ast} (\ln \mathcal{N}) \dot{\tau}^\ast_{ab} \right].
\end{split}
\end{equation}
and using the explicit expression Eq.~\eqref{eq:dettau}
\begin{equation}
\label{eq:kintau}
    K = \frac{N_F}{4} \Tr \Big[( \mathbb{I} + \tau^{\dagger} \tau)^{-1} (\tau^{\dagger} \dot \tau - \dot \tau^\dagger \tau)\Big] .
\end{equation}
Here let us make the notation more intuitive and compare the calculation with the case $n=2$. First, we can parameterized $\tau$ in terms of the coordinates on the Bloch sphere:
\begin{equation}
    \tau=e^{-i \phi}\begin{pmatrix} 0& \tan \frac{\theta}{2}\\
    -\tan \frac{\theta}{2}&0
 \end{pmatrix}.
\end{equation}
Then the kinetic term takes the form
\begin{equation}\label{eq:kin_tau_spin}
    K = -i \frac{N_F}{2} \int dt \, (1 - \cos \theta) \, \dot{\phi}\;,
\end{equation}
which corresponds to the kinetic term we derived in Section~\ref{sec:n2q2} for the spin chain. Note that this parameterization uses the $z$-axis as a reference axis. Therefore, to obtain a complete mapping for $n = 2$, one must perform a coordinate transformation, $(x, y, z) \rightarrow (z, x, y)$. In this case, the kinetic term obtained will differ from Eq.~\eqref{eq:kin_tau_spin} by a total derivative.
\subsection{Geometric considerations}
A few observations are in order. The manifold defined by the coset $SO(2n)/U(n)$ (and more generally by the Lie group over the stability group $\mathcal{G}/\mathcal{H}$) naturally has the structure of a K\"ahler manifold~\cite{WIEGMANN1989311}. It has real dimensions $2n(n-1)$, which we can parameterise in terms of the  $n(n-1)$  complex components of the antisymmetric matrix $\tau$. Then, the K\"ahler potential and the metric $\mathtt{g}$ are provided by
\begin{equation}
\label{eq:kahler}
    F[\tau, \tau^\ast] = \log \mathcal{N}[\tau, \tau^\ast] , \; \mathtt{g}_{ab, cd} = \frac{\partial^2 F}{\partial \tau_{ab} \partial \tau^\ast_{cd}}\;,
\end{equation}
where $g_{ab, cd}$ is a $n(n-1)/2 \times n(n-1)/2$ matrix. The metric allows us to express explicitly the Haar measure in terms of the current parametrization in the usual form
\begin{equation}
\label{eq:haarmeas}
    \int_{\rm Haar} dO_\tau := \int \left[\prod_{a<b} d\tau_{ab} d\tau_{ab}^\ast\right] |\det \mathtt{g}|\;.
\end{equation}
From Eq.~\eqref{eq:dettau} and Eq.~\eqref{eq:kahler}, we obtain the expression
\begin{equation}
    \det \mathtt{g} \propto \det(\mathbb{I} + \tau^\dag \tau)^{-(n-1)}\;,
\end{equation}
where we neglect normalisation factors that are irrelevant in the ratio Eq.~\eqref{eq:transitAmpl}.

Also, when considering closed paths $\tau(t) = \tau(0)$, it is well known that the total contribution of the kinetic term
has a geometric interpretation as the 2-dimensional integral of a 2-form inside the coset manifold $SO(2n)/U(n)$. This is seen easily adding a fictitious extra coordinate $s \in [0,1]$ which connects the orbit to the identity element: choosing any smooth $\tau(s,t)$ such that $\tau(s = 0, t) = 0$ and $\tau(s = 1, t) = \tau(t)$, we can write
\begin{multline}
    \int_0^t dt' K = \int_{\substack{s \in [0,1] \\ t' \in [0,t]}} ds dt' \sum_{\substack{a<b\\ c<d}} g_{ab,cd} \\ 
        (\partial_t \tau_{a,b} \partial_s \tau_{c,d}^\ast
        -\partial_s \tau_{a,b} \partial_t \tau_{c,d}^\ast
        )\;,
\end{multline}
which can be recognised as the Wess-Zumino-Witten term in the current parametrization~\cite{Witten:1983ar, Felder1988}.

We now have all the ingredients to write the path integral for our system.
In the next sections, we derive the NLSM action for $q_J=2$ using this formalism. 
To do so, we perform the integration over the massive modes coupled to the measurement rate, as we did in the case $n=2$. 

\subsection{Integration over the massive modes}\label{ssec:integ_tau}
In this subsection, we write the path integral and perform the integration over the heavy modes obtaining an effective description of the residual light ones. Parameterization in terms of the complex antisymmetric matrix $\tau$ provides a convenient set of variables for this analysis. In fact, writing $\tau = \tau_R + i \tau_I$ where $\tau_R, \tau_I$ are real antisymmetric matrices, we will see that the heavy modes are spanned by $\tau_I$. In terms of this parameterization, the measure in Eq.~\eqref{eq:haarmeas} takes the form
\begin{equation}
\label{eq:haarmeasri}
    \int_{\rm Haar} dO_\tau := \int \frac{d\tau_{R} d\tau_{I}}{\det(\mathbb{I} + \tau^\dag \tau)^{n-1}}\;,
\end{equation}
where will  collectively indicate $d\tau_R = \prod_{a<b} d\tau_{R,ab}$ and analogously for the imaginary part.
We momentarily drop the unitary part since the modes it consists of do not participate in the Gaussian integration to the leading order (see Appendix \ref{App:tauexp}), similarly to the case $n=2$. 
Let us first consider the path integral for a single cluster undergoing only local measurements,
\begin{equation}\label{action}
\begin{split}
    I & = \int \frac{d \tau_R d \tau_I}{[\det(\mathbb{I} + \tau^{\dagger} \tau)]^{(n-1)}}  \times  \\ 
    &  \exp\left(-\int_0^T dt [K(\tau_R, \tau_I) + \mathcal{H}_{\rm mon}(\tau_R, \tau_I)] \right) ,
\end{split}
\end{equation}
where the $\mathcal{H}_{\rm mon}$ is written in Eq.~\eqref{eq:Hmeas} for $q_J=2$ and for a single cluster and the kinetic term is given in Eq.~\eqref{eq:kintau}.
Proceeding analogously to what was done for the $n=2$ case, we wish to integrate the quantum fluctuations, represented by $\tau_I$, which become small due to the measurement part of the Hamiltonian. At the zero-th order in $\tau_I$, the field operator is given by 
\begin{equation}
\label{eq:PhitauR}
    \Phi = \frac{1}{2}\begin{pmatrix}
        O(\tau_I) & \frac{1 - \tau_R}{1 + \tau_R} +  O(\tau_I^2) \\
        -\frac{1 + \tau_R}{1 - \tau_R} +  O(\tau_I^2) &  O(\tau_I)
    \end{pmatrix}
    .
\end{equation}
By expanding to quadratic order for the measurement in a single cluster, we obtain
\begin{equation}
    \mathcal{H}_{\rm mon} = \Gamma N_F^2 \text{tr}\left( \frac{1}{1 - \tau_R^2} \tau_I \frac{1}{1 - \tau_R^2} \tau_I \right) \; .
\end{equation}
We can also expand the kinetic term to first order in $\tau_I$ (see Appendix \ref{App:tauexp}),
\begin{equation}
    K = i N_F \text{tr}\left(  \dot{\tau}_R \frac{1}{1 - \tau_R^2} \tau_I \frac{1}{1 - \tau_R^2}\right) + O(\tau_I^3).
\end{equation}

Finally, we arrive at the action written in terms of $\tau_R$ and $\tau_I$,
\begin{widetext}
\begin{equation}\label{eq:fullNLSM}
    I = \int \frac{d\tau_R}{[\det(1 - \tau_R^2)]^{(n-1)/2}} \int d\tau_I \ e^{-i N_F \int dt \ \text{tr}\left( \dot{\tau}_R \frac{1}{1 - \tau_R^2} \tau_I \frac{1}{1 - \tau_R^2}\right) +  \Gamma N_F^2 \int dt \ \text{tr}\left( \frac{1}{1 - \tau_R^2} \tau_I \frac{1}{1 - \tau_R^2} \tau_I \right)} .
\end{equation}
\end{widetext}
To perform the computation, we diagonalize the matrix $\tau^2_R = O D O^{T}$ (the matrix $\tau^2_R$ is symmetric so it can be diagonalized with an orthogonal transformation), where $D = \text{diag}(\lambda_{R,1}^2, ..., \lambda_{R,1}^2)$ and $\lambda_{R,i}$ are the eigenvalues of $\tau_R$. 
By applying the change of variables
\begin{equation}
    \tau_I' = O \tau_I O^t, \quad \tau_R' = \tau_R ,
\end{equation}
which has unit Jacobian, and by rescaling the variables as $\tau_I' \rightarrow \tau_I' / N_F$, the Gaussian integration over $(\tau_I)_{ij}$ gives
\begin{equation}\label{eq:smtau}
    \int \frac{d\tau_R}{[\det(1 - \tau_R^2)]^{\frac{(n-1)}{2}}} e^{-\frac{1}{4 \Gamma} \int dt \ \text{tr}\left( \frac{1}{1 - \tau_R^2} \dot{\tau}_R \frac{1}{1 - \tau_R^2} \dot{\tau}^T_R \right)} .
\end{equation}
One can easily verify that the matrix $Q = ({1 - \tau_R})/({1 + \tau_R})$ appearing in Eq.~\eqref{eq:PhitauR} is an orthogonal matrix in $SO(n)$; such a representation goes under the name of Cayley transform, see Appendix \ref{App:Caley}. The Jacobian of the change of variable is precisely the determinant appearing in Eq.~\eqref{eq:smtau}, namely 
\begin{equation}
     \int \frac{d\tau_R}{[\det(1 - \tau_R^2)]^{\frac{(n-1)}{2}}} = \int_{\rm Haar} dQ\;,
\end{equation}
where the integral in the right-hand side is over the Haar measure in $SO(n)$. Therefore by changing variables, the integral in Eq.~\eqref{eq:smtau} can be expressed in terms of $Q$, by noticing the following relation
\begin{equation}
    \text{tr} \left( \partial_t Q \partial_t Q^t \right) = 4 \text{tr} \left( \dot{\tau}_R \frac{1}{1 - \tau_R^2} \dot{\tau}^T_R \frac{1}{1 - \tau_R^2} \right),
\end{equation}
from which we can finally conclude that the integral of Eq.~\eqref{eq:fullNLSM} is the same as a free NLSM path integral, with the usual kinetic term 
\begin{equation}
    I = \int dQ e^{-\frac{1}{16 \Gamma} \int dt \ \text{tr} \left( \partial_t Q \partial_t Q^t \right)} .
\end{equation}

\subsection{Many-clusters continuous limit and NLSM }\label{ssec:NLSM}
Now we can recover the cluster index and express the action in the continuous limit. The generalization of the kinetic part can be achieved by adding an extra index,
\begin{equation}
    K=\frac{1}{16 \Gamma} \int dt \sum_{i}~~{\rm tr} \Big(\partial_t Q_i \partial_t Q_i^t \Big) .
\end{equation}
For the Hamiltonian part, we use the definition given in Eq.~\eqref{eq:Huni}. Note that we have already assumed that $\tau_I$ is small and we have integrated it out in the path integral. Therefore, in the unitary part, we only need the leading order, which we obtain by setting $\tau_I=0$,
\begin{equation}
\begin{aligned}
    \mathcal{H}_{\rm uni}\Big|_{q_J=2} & = \frac{J N_F^2}{4} \sum_{i} {\rm tr}(Q_i Q_{i+1}^t)+O(\tau^2_I)\\ 
    \mathcal{H}_{\rm uni}\Big|_{q_J=4} & = -\frac{J N_F^4}{64}\sum_{i,aa'}   \left( (Q^{aa'}_i)^2 (Q^{aa'}_{i+1})^2 \right.\\&\left.
        +\frac{1}{6}  (Q^{aa'}_i)^4\right)+O(\tau^2_I).
    \end{aligned}
\end{equation}
Expanding the unitary part to zeroth order in $\tau_I$ results in the same procedure as in the example with two replicas considered earlier. 
To take the continuous limit we set
\begin{equation}
\begin{split}
  &  Q_{i+1}+Q_{i}\approx 2  Q(x_i)\\&
  Q_{i+1}-Q_i\approx 2 \Delta x \partial_x Q(x_i),
     \end{split}
\end{equation}
where we assume that $x_i$ is a point in between the $i$-th and $(i+1)$-th positions in the chain.
The kinetic term then becomes
\begin{equation}
\begin{split}
    &\lim_{L \rightarrow \infty} K=\frac{1}{16 \Delta x \Gamma} \int d^2x~~{\rm tr} \Big(\partial_t  Q  \partial_t  Q^t  \Big).
    \end{split}
\end{equation}
In the $q_J=2$ case up to an additive constant ${\rm tr}(\Phi^t_i \Phi_i)=\frac{n}{2}$, we get the spatial part of the NLSM
\begin{equation}
\lim_{L \rightarrow \infty} \mathcal{H}_{\rm uni}\Big|_{q_J=2}\sim- \frac{JN_F^2}{2}\Delta x^2 {\rm tr}(\partial_x  Q^t \partial_x  Q)+O(\Delta x^3) .
\end{equation}
Performing the same type of expansion in the case $q=4$ and keeping the second order term in $\Delta x $ we get

\begin{equation}
\begin{split}
     &\lim_{L \rightarrow \infty} \mathcal{H}_{\rm uni}\Big|_{q_J=4}\sim 
     -\frac{JN_F^4}{64}\left( \frac{7 }{6}  \sum_{a a',i}  [Q^{aa'}(x_i)^4\right.\\&\left.-Q^{aa'}(x_i)^2 (\Delta x)^2 (\partial_x Q^{aa'}(x_i))^2\right.\\& \left.-\frac{2}{3} Q^{aa'}(x_i)^3 \Delta x \partial_x Q^{aa'}(x_i) +O(\Delta x)]\right )
  .
  \end{split}
\end{equation}
Finally, the actions for both models in the continuous limit are expressed as
\begin{widetext}
\begin{subequations}
\begin{align}
    \mathcal{S}\Big|_{q_J=2} & =\int d^2 x~~\left(\frac{1}{16 \Gamma \Delta x}{\rm tr} \Big(\partial_t  Q \partial_t  Q^t \Big)+ \frac{J N_F^2}{2} \Delta x {\rm tr}(\partial_x  Q^t \partial_x  Q)\right)\;, \\
    \mathcal{S}\Big|_{q_J=4} & =\int d^2 x~~\left(\frac{1}{16 \Gamma \Delta x}{\rm tr} \Big(\partial_t  Q \partial_t  Q^t \Big)+\frac{J N_F^4}{64}  \sum_{aa'}\left( \frac{1}{\Delta x }(Q^{aa'})^4-\right.\right.\\  &\left.\left. (Q^{aa'})^2 \Delta x (\partial_x Q^{aa'})^2-\frac{2}{3} (Q^{aa'})^3 \Delta x \partial_x Q^{aa'}  \right) \right)\; .\notag
\end{align}
\end{subequations}
\end{widetext}
In the case \(q=2\), we obtain the celebrated nonlinear sigma model for the \(SO(n)\) group manifold, and the explicit coefficients coincide with those derived in \cite{Fava:2023tgg} after the change of variables
\[
  J \;\longrightarrow\;\frac{2J}{\Delta x\,N_F}\,, 
  \qquad
  \Gamma \;\longrightarrow\;\frac{\Gamma}{\Delta x\,N_F}\,,
\]
which accounts for different conventions and the fact that here measurements act on a single cluster rather than coupling adjacent ones.

A renormalisation‐group analysis then reveals distinct phases \cite{Fava:2023tgg}. In the disentangling phase, the steady‐state entanglement entropy behaves as area-law.
In the stable non‐trivial phase, one finds logarithmic scaling of entanglement entropy. 
At \(t\gg\tau_P\), where $\tau_P$ is a purification time-scale, the system reaches a pure‐state trajectory whose bipartite entanglement across an interval of length \(L/2\) obeys
\[
  S_n \;\sim\;\,(\ln L)^2 \;.
\]At \(J=0\) and $N_F=1$ as we discuss in Sec. (\ref{sec:measurementdynamics}) the effective target manifold becomes $\frac{SO(2N)}{U(N)}$, and the $\theta$-term drives the theory to an unstable fixed point at \(\theta=\pi\).  This gives a genuine critical point on the measurement-only axis, at which scale invariance is restored and the entanglement law becomes purely logarithmic.  In particular,

\[
  S \sim \,\ln L \,.
\]
For general \(N_F\), there are \(N_F\) such  fixed points as the replica-symmetry parameter \(\Delta\) is tuned from \(-1\) to \(+1\), each exhibiting the same \(\ln L\) scaling.

Instead, in the non-gaussian case $q > 2$,  we do not have a continuous $SO(2n)$ symmetry, but the field theory we derived has the expected permutation symmetry $Q \to \Pi_1 Q \Pi_2$, with $\Pi_{1,2} \in S_n$, the $n$-elements symmetric group. 
In the next sections, we show other applications of the path integral approach. \footnote{Similar results were found in the recent works \cite{Guo:2024bhi,Poboiko:2024jzp} where the interacting case was as well considered.}
In particular, we calculate the replicated purity in the systems of two-clusters.

\section{Two-cluster stationary purity  }\label{sec:purity2}
In this section, we redo the two-cluster calculation of the purity that we performed in section \ref{sec:n2q2}, but for a generic number of replicas $n$. 
Consider the action for the sigma model in the $q_J=2$ case, 
\begin{equation}
\begin{split}
    \mathcal{S}=\int d t \tr\Big(\frac{1}{16 \Gamma  }\sum_{i=A,B} \partial_t  Q_i^t \partial_t  Q_i-\frac{J N_F^2}{4}  Q_A  Q_{B}^t \Big) ,
    \end{split}
    \end{equation}
where  $Q \in SO(n)$.  
In the generic $n$ case, we introduce new variables $Q_+$ and $Q_-$, playing the role of center of mass and relative coordinates respectively,
\begin{equation} \label{basischange}
     Q_A=Q_+Q_-^t \ , \ Q_B=Q_+ Q_- .
\end{equation}
and applying the general procedure in Eqs.~(\ref{eq:pur},\ref{eq:transitAmpl}), in order to compute the stationary purity we need
\begin{equation}
     \braket{\mathcal{C}_{A,2}}{ GS} = \int dQ_+ dQ_- e^{- \mathcal{S}} \braket{\mathcal{C}_{A,2}}{Q_+, Q_-}\;.
\end{equation}
Due to the large weight of the unitary evolution term, we can assume that $Q_-=e^{\delta Q}$, where the cluster-to-cluster fluctuations $\delta Q$ are small. 
Doing the expansion with respect to $\delta Q$ and rescaling $\delta Q \rightarrow N_f^{-1/4} \delta Q$ , $t \rightarrow \tau/\sqrt{N_f}$  one finds
\begin{equation}
\begin{split}
    &\mathcal{S}=  \int d \tau \left[ \frac{1}{8 \Gamma} \left( \sqrt{N_F} \tr\Big( \dot{Q}_+^t\dot{Q}_+ \Big)+ \tr \Big( \delta \dot{Q}^t \delta \dot{Q} \Big) \right.\right.\\&\left. \left.+ \tr \Big( [\delta \dot{Q}, \delta Q^t]  \dot{Q}^t_+ Q_+\right)+ \frac{J N^{3/2}_F}{2} \tr \Big( \delta  Q^t \delta  Q \Big) \right],
    \end{split}
\end{equation}
Note that in contrast with the $n=2$ case, the $Q_+$ does not decouple from the rest (compare with $\tilde\theta$ in Eq.~\eqref{eq:actionn2q2}). However, since $\dot{Q}_+$ has a large weight $\sqrt{N_F}$, we can use saddle point to perform the integral. Specifically, we write $Q_+ = Q_0 \exp{N_F^{-1/4} \delta \Theta}$, where $Q_0$ is chosen to maximize the overlap $\braket{\mathcal{C}_{A,2}}{Q_+ = Q_0, Q_-}$. We postpone the determination of the value of $Q_0$ and expand the fluctuations in $\delta \Theta$, which we can integrate out the 
\begin{equation}
  \begin{split}
    \mathcal{S} & =\frac{1}{8 \Gamma}  \int d t \left[ \tr  \Big(\delta \dot{  Q}^t \delta \dot{ Q} \Big)  + 4 J N^2_f \Gamma \tr \Big( \delta  Q^t \delta  Q \Big) \right]\\
    & = \frac{1}{4\Gamma}  \int d t \sum_{i>k} \left( \delta \dot{Q}^2_{ik} + 4 J N^{2}_f \Gamma \delta Q^2_{ik} \right) \;.
    \end{split}
\end{equation}
Thus, at large $N_F$, we obtained the action for $n(n-1)/2$ independent harmonic oscillators.
As in the case $n=2$, for large $t$, the dominating configuration comes from the ground state of the harmonic oscillators
\begin{equation}
     \mathcal{H} = \frac{1}{4 \Gamma}\sum_{i>k}\delta \dot{Q}^2_{ik} +  J N^{2}_F\sum_{i>k} \delta Q^2_{i k} ,
\end{equation}
with $m =  1/(2 \Gamma) $, $m \omega^2/2 =  J N^2_F$, whose wave function reads
\begin{equation}
    \Psi_{\overline{GS}} (\delta Q) \sim e^{-m \omega \sum_{i>k} \delta Q_{i j}^2/2} .
\end{equation}
Therefore the integral we want to compute is the overlap of the ground state of the Hamiltonian and the boundary state $\left< \mathcal{C}_{A, 2}  | GS \right>$, resulting in
\begin{equation}\label{eq:overlosc}
\begin{split}
    &\left<\mathcal{C}_{A,2} | GS \right>=\\&=\int d \delta Q e^{- m \omega \sum_{i>k} \delta Q^2_{ik}/2} \left<\mathcal{C}_{A,2} | Q_-,Q_0 \right>,
    \end{split}
\end{equation}
where $\left|Q_i\right>$ is a coherent state after taking the limit $ \tau_I \rightarrow 0$ .
The final ingredient of this calculation is understanding the overlap between an arbitrary coherent state and the boundary state $\left<\mathcal{C}_{A,2}| Q_-,Q_0\right>$.
This quantity can be calculated with the help of the formula Eq.~\eqref{eq:overlapPhi}, where we expressed the overlap of two arbitrary coherent states as a determinant of the corresponding $\Phi$'s. More explicitly, using $Q = (1 - \tau_R) / (1 + \tau_R)$ and Eq.~\eqref{eq:PhitauR}, we can relate $\Phi$ and $Q$ as
\begin{equation}
    \Phi =\frac{1}{2} \begin{pmatrix}
    0 & Q \\
    -Q^t & 0 
    \end{pmatrix} \;,
\end{equation}
where we ignore the sub-leading fluctuations in $\tau_I$.
Using the expression in Eq.~\eqref{eq:overlapPhi} and expression of the boundary state in terms of orthogonal matrices (see Appendix \ref{App:bound_state}), we get
\begin{equation}
\begin{split}
    &\left<\mathcal{C}_{A,2}| Q_0,Q_-  \right> = \braket{\mathcal{C}_A}{Q_-} \braket{\mathbb{I}}{Q_0}=\\&=
    \det[\frac{\mathbb{I}-2\Sigma \Phi(Q_0)}{2} \frac{\mathbb{I}-2\Omega^t_A \Sigma \Omega_A \Phi(Q_-)}{2}]^{N_F/4}
    \\=&\frac{1}{2^{N_f n }} \det\left[(\mathbb{I}_{n \times  n}+Q_0)(\mathbb{I}_{n \times  n}+Q_{-,1\leftrightarrow{}  2})\right]^{N_F/2},
    \end{split}
\end{equation}
where we used 
\begin{equation}
    \det[\frac{1-2\Sigma \Phi(Q_0)}{2} ]=\det\frac{1}{2}\begin{pmatrix}
        \mathbb{I}+Q^t_0&0\\
       0 & \mathbb{I}+Q_0
    \end{pmatrix},
\end{equation}
\begin{equation}\begin{split}
   &\det[ \frac{1-2\Omega_A^t \Sigma \Omega_A \Phi(Q_-)}{2} ]\\&=\det \frac{1}{2}\begin{pmatrix}
        \mathbb{I}+Q^t_{-,1 \leftrightarrow{}2 }&0\\
       0 & \mathbb{I}+Q_{-,1 \leftrightarrow{}2 }
    \end{pmatrix},
    \end{split}
\end{equation}
and $Q_{-,1 \leftrightarrow{}2 }$ is $Q_-$ with the first and the second rows of the matrix swapped, and with the first row multiplied by $-1$. Therefore, the Eq.\eqref{eq:overlosc} can be rewritten as
\begin{equation}
    \begin{split}
    &\left<\mathcal{C}_{A,2} | GS \right>=\\&=\frac{1}{2^{N_F n}}\int d \delta Q e^{- N_F \sqrt{\frac{2 J}{\Gamma}} \sum_{i>k} \delta Q^2_{ik}/2} \\ &\times e^{\frac{N_F}{2}\log \det\left[(\mathbb{I}_{n \times  n}+Q_0)(\mathbb{I}_{n \times  n}+Q_{-,1\leftrightarrow{}  2})\right]},
    \end{split}
\end{equation}the leading order of this expression is given by $Q_0=\mathbb{I}$ and $Q_{-,1\leftrightarrow{}  2}= \mathbb{I}$. Now we find $\delta Q= \log(Q_{-})$:
\begin{equation}
    \delta Q=\begin{bmatrix}
        0 & \frac{\pi}{2} & 0 \\
        -\frac{\pi}{2} & 0 & 0 \\
        0 & 0 & \mathbb{O} \\
    \end{bmatrix},
\end{equation}
then the replicated purity takes the form
\begin{equation}
\begin{split}
     &\lim_{t \rightarrow \infty}\frac{\Tr \left[ \rho^{(n)}(t) \left(\mathcal{C}_{A,2} \otimes \mathbb{I} \right)\right]}{\Tr \rho^{(2)}(t)}=\\&=\frac{1}{Z}\left<\mathcal{C}_{2,A}| GS  \right>\sim e^{- \frac{N_F}{2} \sqrt{\frac{ J}{\Gamma}}\left(\frac{\pi}{2}\right)^2} . 
     \end{split}
\end{equation}
Therefore, the limit $n \rightarrow 1$ becomes trivial and gives purity
\begin{equation}\label{eq_purity_gaussian}
    \left.\overline{\Tr\rho_A(t)^2} \right|_{t \rightarrow \infty}\sim e^{-\frac{N_F}{2}\sqrt{\frac{J}{ \Gamma} } \left(\frac{\pi}{2}\right)^2 },~~~1/N_F^2 < J \leq 1 ,
\end{equation}
which interpolates between area-law for $J \leqslant 1/N_F^2$ and volume-law for $J \geqslant 1$. One can notice that this result is consistent with $n=2$ case.

\begin{figure*}[t]
  \centering
  \begin{subfigure}[t]{0.473\linewidth}
    \includegraphics[trim=0 0 0 0, clip, width=\columnwidth]{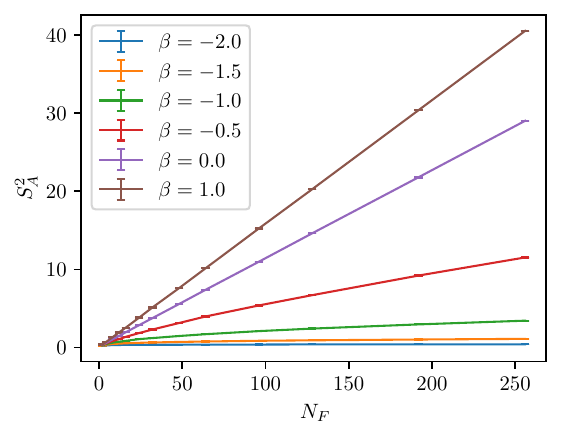}
    \caption{$S_A^2$ as a function of $N_F$, for different $\beta$.}
    \label{fig_2cluster_gaussian_entropies_NF_S2}
  \end{subfigure}
  \begin{subfigure}[t]{0.487\linewidth}
    \includegraphics[trim=0 0 0 0, clip, width=\columnwidth]{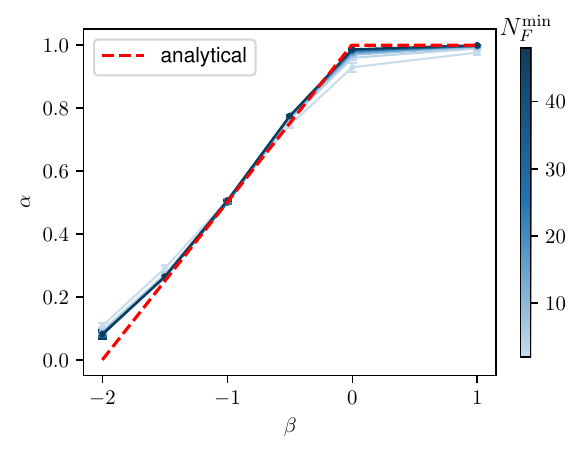}
    \caption{$\alpha$ from a fit with $N_F \geqslant N_F^{\textnormal{min}}$, as a function of $\beta$.}
    \label{fig_2cluster_gaussian_entropies_beta_alpha}
  \end{subfigure}
  \caption{ED results for the saturated ($t \rightarrow \infty$) second Rényi entropy $S_A^2$, with $q = q_J = 2$, $J = N_F^\beta$ and $\Gamma = 1$. 
  The entropy is expected to follow a law $S_A^2 \propto N_F^\alpha$, where the exponent $\alpha$ can be found with a fit.}
  \label{fig_2cluster_gaussian_entropies}
\end{figure*}

We compare our predictions for the purity with numerical exact diagonalization (ED) simulations (details in Appendix \ref{App:num}). 
We compute the saturated second Rényi entropy, given by $S_A^2 = - \log \Tr \rho_A^2$, and approximate $\overline{S_A^2} \approx - \log \overline{\Tr \rho_A^2}$, assuming that the fluctuations are suppressed at large $N_F$. 
Setting a power law dependency of $J$ with $N_F$, $J = N_F^\beta$, different entropy growths are observed in Fig. \ref{fig_2cluster_gaussian_entropies_NF_S2}.
Assuming a power law $S_A^2 \propto N_F^\alpha$, the exponent $\alpha$ can be found with a fit.
Our theretical prediction for $\alpha$ can be obtained from the expression of Eq.~\eqref{eq_purity_gaussian}, with $\alpha = 0$ corresponding to an area-law state and $\alpha = 1$ to a volume-law state, giving this way
\begin{equation}
    \alpha = 
    \left\{
    \begin{array}{lll}
        1 + \beta/2 & , &  -2 < \beta \leqslant 0 \\
        1 & , & \beta > 0
    \end{array}
    \right.
    \, .
\end{equation}
We find that the numerically obtained $\alpha$ agrees with the predictions of Eq.~\eqref{eq_purity_gaussian}, as revealed by the convergence to the dashed line in Fig. \ref{fig_2cluster_gaussian_entropies_beta_alpha}.
For $\beta = -2$, the numerics and the analysis agree less well, but that is precisely the point where we expect $\alpha = 0$ and the numerically obtained fit is affected by strong finite-size effects. 

Now we move to the case $q_J=4$. 
We will follow the same steps as for the case $q=2$. 
Let us again start with the action,
\begin{equation}
\begin{split}
   -\mathcal{S}=-\frac{1}{16 \Gamma} \int dt~~\sum_{j=A,B }{\rm tr} \Big( \partial_t Q_j \partial_t Q_j^t \Big) \\ +\frac{JN_f^4}{64}\int d t\left[  \sum_{\alpha \alpha'} (Q^{\alpha \alpha'}_A)^2 (Q^{\alpha \alpha'}_B)^2\right.  \\ \left. +\frac{1}{6} \sum_{j \alpha \alpha'} (Q^{\alpha \alpha'}_j)^4 \right] .
   \end{split}
\end{equation}
The minima of the action are given by the permutation matrices with determinant equal to one.
We then introduce new variables
\begin{equation} 
    Q_A \rightarrow D_A Q_A, ~~~Q_B \rightarrow D_B Q_B ,
\end{equation}
where $D_i$ is a permutation matrix, and $ Q_i=e^{\delta  Q_i}$ is a relative coordinate close to the identity matrix, with $\delta  Q$ a skew-symmetric matrix. 
We expand the exponent concerning $\delta Q$ and the expansion reveals once again the action of the quadratic oscillators,
\begin{equation}
\begin{split}
   &-\mathcal{S}=-\frac{1}{16 \Gamma} \sum_{j=A,B }\int dt~~{\rm tr} \left(\delta \dot Q_j ( \delta \dot Q_j)^t\right)\\&+\frac{JN_F^4}{48}\int dt ~({\rm tr}(\delta Q_A\delta Q_A)+{\rm tr}(\delta Q_B\delta Q_B)).
   \end{split}
  \end{equation}
Taking into account the boundary states we get the following expression for the overlap,
\begin{equation}
\begin{split}
    \left<  \mathcal{C}_{2,A}| GS\right>&=\frac{1}{2^{N_F n}}\int d \delta Q_A d \delta Q_B  \\
    \times & e^{- m \omega \sum_{\alpha \alpha' i} (\delta Q^{\alpha \alpha'}_{i})^2/2}\\
    \times & e^{N_F {\rm tr} \log(\mathbb{I}+e^{\delta Q_{B, 1 \leftrightarrow{}2}})+N_F{\rm tr} \log(\mathbb{I}+e^{\delta Q_{A}})}.
    \end{split}
\end{equation}
where $m\omega= \frac{N_F^2}{2} \sqrt{\frac{J}{3 \Gamma}}$. 
Once again, using the overlaps of coherent states  for $1/N_F^4<J \leq 1/N_F^2$ and dominating maximas $e^{\delta Q_A}=\mathbb{I}$ and $e^{\delta Q_{B,1\leftrightarrow{}  2}}= \mathbb{I}$  we obtain the result for the late-time purity 
\begin{equation}
    \left< \mathcal{C}_{2,A}|GS  \right>\sim e^{-\frac{N_F^2}{4} \sqrt{\frac{J}{3 \Gamma}} \left(\frac{\pi}{2}\right)^2} .
\end{equation}
Finally, in the case, $1/N_F^2<J<1$ we neglect the overlap and get the Gaussian integral
\begin{equation}
\begin{split}
   & \frac{1}{2^{N_F n}}\int d \delta Q_A d \delta Q_B e^{- m \omega \sum_{\alpha \alpha' i} (\delta Q^{\alpha \alpha'}_{i})^2/2} \\
   &\sim e^{-N_F n \log(2)}.
\end{split}
\end{equation}
In both cases, the replica limit $n \rightarrow 1$ can be obtained trivially, resulting in the infinite-time purities
\begin{equation}\label{eq_purity_qJ4}
    \begin{split}
        &\left.\overline{\Tr\rho_A(t)^2}\right|_{t\rightarrow \infty} \sim e^{-\frac{N_F^2}{4}\sqrt{\frac{J}{3\Gamma}} \left(\frac{\pi}{2}\right)^2},~\frac{1}{N_F^4} < J \leq \frac{1}{N_F^2}\\&
        \left.\overline{\Tr\rho_A(t)^2}\right|_{t \rightarrow \infty}\sim e^{-N_F \log(2)},~\frac{1}{N_F^2}<J<1 \,.
    \end{split}
\end{equation}

\begin{figure*}[t]
  \centering
  \begin{subfigure}[t]{0.48\linewidth}
    \includegraphics[trim=0 0 0 0, clip, width=\columnwidth]{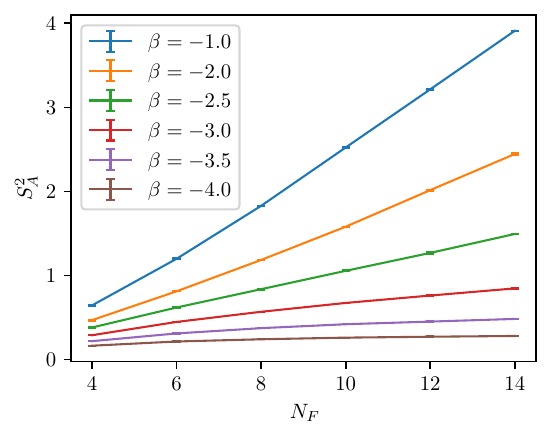}
    \caption{$S_A^2$ as a function of $N_F$, for different $\beta$.}
    \label{fig_2cluster_trotterv2_entropies_NF_S2}
  \end{subfigure}
  \begin{subfigure}[t]{0.50\linewidth}
    \includegraphics[trim=0 0 0 0, clip, width=\columnwidth]{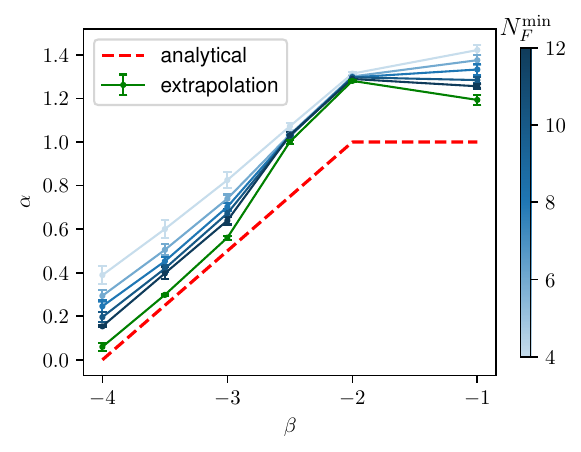}
    \caption{$\alpha$ from a fit with $N_F \geqslant N_F^{\textnormal{min}}$, as a function of $\beta$.}
    \label{fig_2cluster_trotterv2_entropies_beta_alpha}
  \end{subfigure}
  \caption{ED results for the saturated ($t \rightarrow \infty$) second Rényi entropy $S_A^2$, with $q = 2$, $q_J = 4$, $J = N_F^\beta$ and $\Gamma = 1$. 
  The entropy is expected to follow a law $S_A^2 \propto N_F^\alpha$, where the exponent $\alpha$ can be found with a fit.}
  \label{fig_2cluster_trotterv2_entropies}
\end{figure*}

Analogously to the Gaussian case, we compare the predictions to numerical ED simulations (details in Appendix \ref{App:num}).
However, now the system is interacting (exponentially hard to simulate in $N_F$) and there is a larger number of terms in the unitary part of the Hamiltonian, so results were only obtained for $N_F \leqslant 16$.
Setting $J = N_F^\beta$, different entropy growths are observed in Fig. \ref{fig_2cluster_trotterv2_entropies_NF_S2}.
The exponent $\alpha$ is obtained by fitting $S_A^2 \propto N_F^\alpha$ in Fig. \ref{fig_2cluster_trotterv2_entropies_beta_alpha}, progressively increasing the minimum number of flavours $N_F^\min$.
We observe a qualitative similarity of the dependence of $\alpha$ with $\beta$ between the ED results and the analytical prediction of Eq.~\eqref{eq_purity_qJ4}, which corresponds to
\begin{equation}
    \alpha = 
    \left\{
    \begin{array}{lll}
        2 + \beta/2 & , &  -4 < \beta \leqslant -2 \\
        1 & , & \beta > -2
    \end{array}
    \right.
    \, .
\end{equation}
Although this generally corroborates our analytical results, due to large finite-size effects, the two do not quantitively match, even though increasing $N_F^\min$ generally brings us closer to the analytical prediction. 
Finite-size effects are also evidently present, since regimes where $\alpha > 1$ are not possible in the thermodynamic limit (the maximum is set by the volume-law at $\alpha = 1$).
An extrapolation to $N_F^\min \rightarrow \infty$ is performed by linearly fitting $\alpha$ as a function of $1/N_F^\min$, which takes us even closer to the predicted values, but still suffers from finite-size effects.
{
Finite-size effects also progressively increase as we approach the volume-law regime in the range $-4 \leqslant \beta \leqslant -2$. 
Going deeper into the volume-law regime, finite-size effects seem to reduce, possibly since measurements become irrelevant, as Eq.~\eqref{eq_purity_qJ4} suggests.
}

In this section, we calculated the purity for a system with two clusters ($A$ and $B$) in the case $q_J=2$ and $q_J=4$. 
{
In both cases, we find that the purity interpolates between area-law and volume-law regimes, which we confirm numerically.
However, in the $q_J = 2$ case, this interpolation occurs when $J$ scales as $1/N_F^2 < J \leq 1$, whereas for $q_J = 4$, due to the larger strength of unitary evolution, it occurs for the scaling $1/N_F^4 < J \leq 1/N_F^2$.
}

\section{Field theory for the measurement-only dynamics}\label{sec:measurementdynamics}
In this section, we change the protocol of our model slightly, showing how the method based on coherent states can be flexible and used in different setups. In particular, we consider a model in which there is no unit dynamics but quadratic operator measurements are performed on all possible choices of operators between adjacent clusters. More explicitly, the model is described by the non-hermitian Hamiltonian Eq.~\eqref{eq_H_uni_plus_mon} where the unitary part is set to zero and where now the monitoring part acts on two consequential sites   
\begin{equation}
     H_{j,j+1}^{\rm mon} = i \sum_{\mu,\nu} M^{j}_{\mu \nu} \hat \chi_{j, \mu} \hat \chi_{j+1,\nu} ,
\end{equation}
where, as usual, the monitoring coefficients are white noises
\begin{equation}
    \mathbb{E}_{\rm G}[M^j_{\mu\nu}(t_1)M^i_{\alpha \beta}(t_1)]=\delta_{i j} \delta_{\mu \alpha} \delta_{\nu \beta} \delta(t_1-t_2) .
\end{equation}
When looking at this model on a $1D$ chain of $N$ clusters with $i,j = 1,\ldots, N$, the non-trivial dynamics results from the competition between measurements trying to project onto entangled dimers between 
even/odd sites ($2i, 2i +1$) and odd/even ones ($2i-1, 2i$). To lift the degeneracy between these two processes, we can introduce the staggering of the measurement strength $\Gamma$ as $\Gamma_j=[1+(-1)^j \Delta]\Gamma$. 
In  \cite{Fava:2023tgg}, it was argued, based on symmetry considerations, that the measurement-only phase can also be described by a NLSM.
It was conjectured that the sigma model for this system exhibits a larger symmetry, namely $SO(2n)/U(n)$ and have a topological term, which changes the phase diagram. 
In this context, we proceed at explicitly deriving this model conferming the conjectured form of the NLSM including the topological term. 
Performing the average on the replicated system as usual, we find the replicated Hamiltonian
\begin{equation}
\begin{split}
   \mathcal{H}^{\rm mon}=  &  \frac{N_F^2}{4}\sum_{\alpha,\alpha'} \Gamma_i \sigma \sigma'   \Phi^{\alpha \alpha'}_i \Phi^{\alpha \alpha'}_{i+1} .
    \end{split}
\end{equation}
First, we perform a canonical change of variables on odd sites that does not change commutation relations $\Phi^{+-}_{2i+1} \rightarrow -\Phi^{+-}_{2i+1}$ and $\Phi^{-+}_{2i+1} \rightarrow -\Phi^{-+}_{2i+1}$.
 Then for $n=2$, via the mapping to spins Eq.~\eqref{eq:paramPhi}, this model describes a $SU(2)$ antiferromagnet and was extensively studied in the literature, see for example \cite{Affleck:1985jy}. Note that antiferromagnetic chains can be divided into two interpenetrating subchains, which exhibit a staggered spin arrangement
 \begin{equation}
     S_{i} \rightarrow (-1)^{i+1} S_{i} .
 \end{equation}
 Then, by constructing the coherent states path integral, one can arrive at the celebrated $O(3)$ non-linear sigma model action, which here we shall generalise to the generic replica,  $SO(2n)/U(n)$ group.
Before presenting the explicit procedure, it is useful to describe the general outline. As before, we will obtain the NLSM first through expansion in coherent states and then by integrating fast modes, leading to an invariant field theory in the geometry of the manifold. In this case, we anticipate that the manifold on which the coherent states live is as before $SO(2n)/U(n)$ but it coincides with the manifold of the NLSM itself: the origin of this fact lies in the fact that we will integrate out half of the modes corresponding to fluctuations within a dimer, going from a system length $L$ to one length $L/2$. This differs from the procedure presented in Sect.~\ref{ssec:integ_tau} in which the system kept the original length, but part of the manifold (parametrized by the imaginary part of $\tau$) was integrated out.
We proceed therefore in an analogous way, and we perform the change of variable on even sites 
 \begin{equation}\label{eq:staggeringphi}
     \Phi_{2 i} \rightarrow - \Phi_{2 i}.
 \end{equation}
We then arbitrarily arrange the spins into different dimers containing even and odd sites $\{(12),(34),..,(N-1,N)\}$, as done also in \cite{Affleck:1985jy}. 
As explained in Sec.~\ref{sec:coherent}, the matrix $\Phi_i$ is parametrized in terms of the complex antisymmetric matrix $\tau_i$.
Aiming at a continuous space limit, we rewrite the space dependence of the $\tau_i$'s as follows
\begin{equation}
\begin{split}
   &\Delta x \delta \tau_{2 i-1}= \tau_{2i-1}-\tau_{2i}, \\&
   \Delta x  \partial_x \tau_{2 i}= \tau_{2i+1}-\tau_{2i}.
   \end{split}
\end{equation}
Note that at this stage this division is somewhat arbitrary, as the first field $\delta \tau$ gives the increment of the field $\tau$ inside each dimer, and the second field $\partial_x \tau $ the increment between two neigbouring dimers. 
Using the method we introduced in the previous sections, we compute the kinetic term and the matrix element of the Hamiltonian on the coherent states, keeping only the leading order in $N_F$. We therefore obtain the following action.  
\begin{equation}
\begin{split}
   - \mathcal{S}=\int d t~\sum_{i} & \bigg(K_{2i-1}-K_{2i}  \\
   + & H_{2i-1, 2i}+H_{2i,2i+1}\bigg),
   \end{split}
\end{equation}
where the kinetic terms $K_i$ are given by 
\begin{equation}
\begin{split}
     &K_i=\ln \braket{O_{\tau_i}}{O_{\tau  + \dot \tau_i d t}} =\\
     &  \frac{1}{2} \sum_{\alpha \beta} \left[\partial_{\tau_{\alpha \beta}}(\ln \mathcal{N}_i) \dot \tau_{i,\alpha \beta} - \partial_{\tau_{i,\alpha \beta}^\ast}(\ln \mathcal{N}_i) \dot \tau^\ast_{i,\alpha \beta} \right] d t.
    \end{split}
\end{equation}
We note that the kinetic terms of even sites have different signs inside the sum due to the change of variables in Eq.~\eqref{eq:staggeringphi}. 
We can now expand the difference of the kinetic terms within each dimer with respect to $\delta \tau_{2 i-1}$,
\begin{equation}
    K_{2i-1}-K_{2i}= 2 N_F \mathtt{g}_{\alpha \beta}(\tau^{2i-1}) \dot \tau^{2i-1}_{\alpha} (\delta\tau^{2i-1}_{\beta})^*-c.c. ,
\end{equation}
where $\mathtt{g}_{\alpha \beta}= {\partial^2 \ln \mathcal{N}}/{\partial \tau_{\alpha } \partial \tau^*_{\beta}}$ is the metric tensor introduced in Eq.~\eqref{eq:kahler}. Here, to simplify the notation we are using the index $\alpha$ indicating the $n(n-1)/2$ ordered pairs $(a,b)$. We also expand the Hamiltonian in these new variables. 
Finally, taking the continuous limit $N \rightarrow \infty$, $\Delta x \rightarrow 0$  results in the action
\begin{widetext}
\begin{equation}
\begin{split}
    -S & =  2 N_F \int d^2 x \mathtt{g}_{\alpha \beta}(\tau)  \dot \tau_{\alpha} \delta  \tau_{\beta}^*-c.c.-8 N_F^2\Delta x \Gamma \int d^2x  \mathtt{g}_{\alpha\beta}(\tau) \delta \tau_\alpha \delta  \tau_\beta^\ast\\
    & -4N_F^2 \Gamma(1+\Delta) \Delta x \int d^2x  \mathtt{g}_{\alpha \beta }(\tau) \delta  \tau_{\alpha } \partial_x \tau^*_\beta-4 N_F^2 \Gamma(1+\Delta) \Delta x \int d^2x  \mathtt{g}_{\alpha \beta }(\tau) \partial_x  \tau_{\alpha}\delta  \tau^*_{\beta } \\
    & -4 N_F^2\Gamma(1+\Delta) \Delta x \int d^2x \mathtt{g}_{\alpha \beta }(\tau) \partial_x  \tau_{\alpha} \partial_x \tau^*_{\beta } .
\end{split}
\end{equation}
\end{widetext}
Now we can perform the Gaussian field integral of $\delta \tau$ (see Appendix \ref{App:mon}) and finally obtain the NLSM action
\begin{widetext}
\begin{equation}
\begin{aligned}
    -S_{\rm NLSM } & =2 N_F^2 \Gamma(1-\Delta^2) \Delta x \int d^2x~ \partial_x \tau_{\alpha} \mathtt{g}_{\alpha \beta} \partial_x \tau^*_{\beta} \\
    & -\frac{1}{2 \Gamma \Delta x} \int d^2 x~ \dot \tau_{\alpha} \mathtt{g}_{\alpha \beta} \dot \tau^*_{\beta}-
    \frac{\Theta}{ \pi}  \int d^2x~ (\dot \tau_{\alpha} \mathtt{g}_{\alpha\beta} \partial_x \tau^*_{\beta}-\partial_x \tau_{\alpha} \mathtt{g}_{\alpha \beta} \dot \tau^*_{\beta}) ,
    \end{aligned}
\end{equation}
\end{widetext}
where we obtained the angle $\Theta=\pi N_F (1+\Delta)$, associated to the celebrated topological term of the action. 
We leave the study of the phase diagram via the renormalisation group analysis of this field theory to future work (see also the discussion in~\cite{Fava:2023tgg}).

\section{Conclusions}

In this paper, we have shown how to extend the formalism of spin coherent states to generic $SO(2n)$ symmetry for any number of replicas $n$. 
Borrowing from some previous works on coherent states  \cite{RevModPhys.62.867} we have extended and clarified many aspects of the formalism and used it to integrate out the heavy degrees of freedom and derive the effective field theories for monitored SYK clusters. 

Our work therefore reveals the path to rigorously deriving the relevant field theories for generic monitored fermionic systems, both Gaussian and interacting. 
As the first application of the method, we have derived the stationary purity of two SYK clusters, both interacting and not, in terms of the quantum fluctuations of the field theory. 
We have also demonstrated that when clusters interact only through a nearest-neighbor, quadratic,  monitoring Hamiltonian, the effective field theory is the $SO(2n)/U(n)$ NLSM, which extends the $O(3)$ field theory for antiferromagnetic spin chains \cite{Affleck:1985jy}. 
The case of intra-cluster monitoring and nearest-neighbor unitary interactions instead gives the $O(n)$ NLSM for orthogonal matrices, as previously found in \cite{Fava:2023tgg,2302.09094} with a different method.

Our approach is sufficiently general to allow for different applications and generalisations.
The entanglement structure of the monitored fermions could be investigated \cite{PhysRevE.104.014146,PhysRevB.106.L220304,PhysRevB.108.165126,Piccitto2024,PhysRevB.108.104313}, the role of the $U(1)$ or $SU(2)$ symmetry in interacting or Gaussian fermions \cite{2407.08045,Poboiko:2023koc,PhysRevLett.129.120604,PhysRevX.12.041002,PhysRevLett.129.200602}, and in cases with boundary driving \cite{PhysRevB.106.024304} or structured measurements. 
We leave these exciting questions for near future works.

\section*{Acknowledgements}
 We acknowledge inspiring discussions with Lorenzo Piroli, Adam Nahum and Denis Bernard. 
 J.D.N., A.T. and H.L. are founded by the ERC Starting Grant 101042293 (HEPIQ) and the ANR-22-CPJ1-0021-01. ADL and GG acknowledge the support of the ANR JCJC grant ANR-21-CE47-0003 (TamEnt).

\newpage
\onecolumn
\appendix

\section{More on the replicated Hamiltonian}\label{App:rep}

For a given realization of the noises $h, w$, the density matrix of the system will evolve as
\begin{equation}
    \rho(t) = K(t) \rho(0) K^{\dag}(t) , \qquad K(t) = T \exp (- i \int_0^t \; dt' H(t')) .
\end{equation}
We want to study the evolution of $n$ replicas of the density matrix $\rho(t)$. In order to do so, we represent the dynamics in the vectorised and replicated Hilbert space. 
We introduce the replica indices $\alpha = (\sigma, a)$ with $\sigma = \pm$ and $a = 1,\ldots,n$. 
More specifically, we recall that within the standard vectorization construction, we use the $+$ for operators multiplying on the left and the $-$ for operators multiplying on the right, i.e.
\begin{equation}
    A \rho B = A^+ (B^-)^t \ket{\rho},
\end{equation}
where $t$ denotes the transpose.
We can thus define the replicated Hamiltonian as
\begin{equation}
    H^{(n)} = \sum_{a}  H^{(+,a)} -
    (H^{(-,a)})^\ast ,
\end{equation}
where we have replicated the system with a standard tensor product of the original Hamiltonian with identity on the remaining replicas so that $[H^{\alpha}, H^{\alpha'}] = 0$ whenever $\alpha \neq \alpha'$. 
More explicitly, our Majorana operators are transformed as
\begin{equation}
    \hat{\gamma}_{k}^{(+,a)} = \mathds{1}^{2(a-1)} \otimes \hat\chi_k \otimes \mathds{1}^{2n - 2a + 1} , \qquad \hat{\gamma}_{k}^{(-,a)} = \mathds{1}^{2a-1} \otimes \hat\chi_k^\ast \otimes \mathds{1}^{2n - 2a} ,
\end{equation}
so that we ordered the replicas as $((+,1), (-,1), (+,2), (-,2), \ldots, (+,n), (-,n))$.
The replicated Hamiltonians are obtained by substituting the original Majorana operators with the replicated ones $ H^\alpha = H_{\hat{\chi}_k \rightarrow \hat{\gamma}^\alpha_k}$.
However, the $\hat{\gamma}$ operators are not genuine Majorana fields since they commute (instead of anticommuting) on different replicas. 
To adjust this, we can introduce the following Klein factors 
\begin{equation}
    F^{(\sigma, a)} = \prod_{i,\mu}
    \chi_{i,\mu}^{(\sigma, a)},
\end{equation}
which allows us to define the operators
\begin{equation}
\begin{aligned}
    & \hchi^{(+,a)}_{i,\mu} = i (\prod_{a'<a} F^{+,a'} F^{-,a'}) F^{+, a} \gamma_{i,\mu}^{+, a} = i \mathcal{S}^a \gamma_{i,\mu}^{+,a} 
    ,
    \\
    & \hchi^{(-,a)}_{i,\mu} = (\prod_{a'<a} F^{+,a'} F^{-,a'}) F^{+, a} \gamma_{i,\mu}^{-, a} =  \mathcal{S}^a \gamma_{i,\mu}^{-,a} 
    ,
\end{aligned}
\end{equation}
from which we recover the standard anticommutation relations, up to a constant normalization,
\begin{equation}
    \{ \hchi_{k}^{\alpha}, 
    \hchi_{k'}^{\alpha'}\} = \delta_{\alpha, \alpha'} \delta_{k,k'} .
\end{equation}
One can verify that, for even $M$,
\begin{equation}
\hgamma_{k_1}^{\alpha}\ldots\hgamma_{k_M}^{\alpha} = \hchi_{k_1}^{\alpha}\ldots\hchi_{k_M}^{\alpha} .
\end{equation}
Thus the replicated Hamiltonians are easily expressed in terms of the proper Majorana operators with a simple substitution $\hgamma_k^\alpha \rightarrow \hchi_k^\alpha$, resulting in
\begin{equation}
  \begin{split}
    H^{(n)}(t)   = \sum_{\sigma,a,i} \left\{\rule{0cm}{20pt}\right. & (i)^{\frac{q_J}{2}}f_\sigma^{\rm uni} \sqrt{J} \left[\sum_{{\boldsymbol{\tilde \mu}}}  h^i_{ \boldsymbol{\tilde \mu}}(t) \prod^{q_J}_{k=1} \hchi^{(\sigma, a)}_{i \tilde \mu_{k}} +    \sum_{{\boldsymbol{\nu}}} \sum_{{\boldsymbol{\mu}}}  h^i_{ \boldsymbol{\mu}\boldsymbol{\nu}}(t) \prod^{q_J/2}_{k=1} \hchi^{(\sigma, a)}_{i \mu_{k}} \prod_{j=1}^{q_J/2} \hchi^{(\sigma, a)}_{i+1 \nu_{j}} \right]  \\
    + \ &  (i)^{\frac{q}{2} + 1}\sigma f_\sigma^{\rm mon} \sqrt{\Gamma} \sum_{\boldsymbol{\tilde \nu}} w^i_{\boldsymbol{\tilde \nu}} \prod_{j=1}^{q}\hchi^{(\sigma, a)}_{i\tilde \nu_j}  \left.\rule{0cm}{20pt}\right\}
,
\end{split}
\end{equation}
where $f^{\rm uni}_{+}=(-1)^{q_J/2+1}$, $f^{\rm uni}_-=1$, $f^{\rm mon}_+=-(-1)^{q/2+1}$ and $f^{\rm mon}_-=-1$,
and extra signs (global for either the unitary or monitoring parts) have been absorbed in the definition of our random variables $h, w$.

\section{Averaging of replicated evolution operators}\label{App:aver}
In this appendix, we perform the averaging of the evolution operator in Eq.~\eqref{eq:evol} and find the averaged replicated Hamiltonian.
The Gaussian average of the evolution operator is given by
\begin{equation}\label{eq_avg_rep_ham_def}
   \mathbb{E}_{\rm G}\left[ \mathcal{T} \exp(-i \int_0^t H^{(n)}(s) ds ) \right] \cong \mathbb{E}_{\rm G}\left[ \prod_j \mathcal{T} \exp(-i \int_{t_j}^{t_j + \Delta t} H^{(n)}(s) ds ) \right] ,
\end{equation}
where in the last step we divided the evolution into segments of length $\Delta t \ll 1 $, with $t_j = j\Delta t$.
To perform the averaging, we expand the exponential into its Taylor series.
Since the white noise terms have zero mean, the leading term in Eq.~\eqref{eq_avg_rep_ham_def} is the second-order one, 
\begin{equation}
  \begin{aligned}
    & \mathbb{E}_{\rm G}\left[ \prod_j \mathcal{T} \exp(-i \int_{t_j}^{t_j + \Delta t} H^{(n)}(s) ds ) \right] \\
    = \ & \mathbb{E}_{\rm G}\left[\prod_j \left( 1 -i \int_{t_j}^{t_j + \Delta t} H^{(n)}(s) ds - \frac{1}{2} \iint_{t_j}^{t_j + \Delta t} \mathcal{T} H^{(n)}(s_1) H^{(n)}(s_2)ds_1 ds_2  + \dots \right) \right]  \\
    = \ & 1 - \frac{1}{2} \sum_j \int_{t_j}^{t_j + \Delta t} \int_{t_j}^{t_j + \Delta t} \mathcal{T} \mathbb{E}_{\rm G}\left[ H^{(n)}(s_1) H^{(n)} (s_2)\right]ds_1 ds_2  + \dots \\ 
    = \ & 1 - \sum_j \int_{t_j}^{t_j + \Delta t} \int_{t_j}^{t_j + \Delta t} \delta(s_1 - s_2) \mathcal{H}^{(n)} ds_1 ds_2  + \dots  = 1 - \mathcal{H}^{(n)}t + \dots = \exp(-t \mathcal{H}^{(n)}) 
  .
  \end{aligned} 
\end{equation}
Therefore we find that the averaged evolution operator gives the imaginary time evolution of the averaged replicated Hamiltonian 
\begin{equation}
  \mathcal{H}^{(n)} = \frac{1}{2} \left( J \sum_{i\mmu {\boldsymbol{\nu}}} \mathcal{H}^\uni_{i i+1, \boldsymbol{\mu}{\boldsymbol{\nu}}} \right. \\ \left.+J\sum_{i\tilde \mmu}\mathcal{H}^\uni_{i, \boldsymbol{\tilde \mu}}- \Gamma \sum_{i \boldsymbol{\tilde \nu} } \mathcal{H}^\mon_{i,\boldsymbol{\tilde \nu} }\right),
\end{equation}
with
\begin{equation}
\begin{split}
  & \mathcal{H}^\uni_{ii+1,\mmu {\boldsymbol{\nu}}} =\left(i^{\frac{q_J}{2}}  \sum_{\sigma, a}  f_\sigma^{\rm uni}\prod^{q_J/2}_{k=1} \hchi^{(\sigma, a)}_{i \mu_{k}} \prod_{j=1}^{q_J/2} \hchi^{(\sigma ,a)}_{i+1 \nu_{j}} \right)^2,
  \end{split}
\end{equation}
\begin{equation}
\begin{split}
  & \mathcal{H}^\uni_{i,\tilde \mmu} =\left(i^{\frac{q_J}{2}}  \sum_{\sigma, a}  f_\sigma^{\rm uni}\prod^{q_J}_{k=1} \hchi^{(\sigma ,a)}_{i \tilde \mu_{k}}\right)^2,
  \end{split}
\end{equation}
and
\begin{equation}
    \mathcal{H}^\mon_{\tilde \mnu} = \left(i^{\frac{q}{2}}  \sum_{\sigma, a} \sigma f_\sigma^{\rm mon}\prod_{j=1}^{q}\hchi^{(\sigma, a)}_{i\tilde \nu_j}   \right)^2 .
\end{equation}
Thus the replicated state $\rho^{(n)}(t)$ will evolve in imaginary time with the Hamiltonian $\mathcal{H}^{(n)}$, meaning that, at $t\rightarrow \infty$, it will act as a projector onto the groundstate $\ket{\mathrm{GS}}$ of $\mathcal{H}^{(n)}$.

\section{Boundary states}\label{App:bound_state}

The operators we calculate in Eq.~\eqref{eq_purity_final}, get vectorized into states according to Eq.~\eqref{eq_overlap_rho}.  
These are the boundary states, and it is important to understand how the replicated Majorana operators act on them.
We will start with the identity boundary state $\ket{\mathbb{I}}$. 
Before vectorization,
\begin{equation}
  \hgamma_{k}^{a} \mathbb{I} \hgamma_{k}^{a} = \mathbb{I} \, ,
\end{equation}
therefore 
\begin{equation}
  \hgamma_{k}^{(+,a)} \hgamma_{k}^{(-,a)} \ket{\mathbb{I}} = \ket{\mathbb{I}} \, .
\end{equation}
Now, we want to express this relation in terms of the new Majorana fields $\hchi^{(\pm,a)}_{k}$.
We have that
\begin{equation}
   i \hchi^{(+,a)}_{k} \hchi^{(-,a)}_{k} = - \mathcal{S}^a \hgamma_{k}^{(+,a)} \mathcal{S}^a \hgamma_{k}^{(-,a)} 
   = - (-1)^{M-1} \mathcal{S}^a \mathcal{S}^a \hgamma_{k}^{(+,a)} \hgamma_{k}^{(-,a)} ,
\end{equation}
where we anticommute $\hgamma_{k}^{(+,a)}$, $M-1$ times to move  through the string $F^{(+, a)}$ (here $M$ is the total number of Majorana operators $M = L N_F$ in each replica).
Computing the square of operator $\mathcal{S}^a$ results in $(\mathcal{S}^a)^2  = (-1)^{M(M-1)/2}$ such that, if we consider $M$ to be a multiple of four, then
\begin{equation}\label{eq_chi_boundary_id}
  i \hchi^{(+,a)}_{k} \hchi^{(-,a)}_{k} \ket{\mathbb{I}} = \ket{\mathbb{I}} .
\end{equation}
Next, we consider the state $\ket{\mathcal{C}_{A,2}}$, where the tensor product with identity is implicit. 
If $k \in \overline{A}$, 
\begin{equation}
  \hgamma_{k}^{a} \big( \mathcal{C}_{A,2} \otimes \mathbb{I} \big) \hgamma_{k}^{a} = \mathcal{C}_{A,2} \otimes \mathbb{I} 
  \ \Leftrightarrow \
  i \hchi^{(+,a)}_{k} \hchi^{(-,a)}_{k} \ket{\mathcal{C}_{A,2}} = \ket{\mathcal{C}_{A,2}} 
  ,
\end{equation}
acting just as the identity, as was previously derived. 
Instead for $k \in A$ we have 
\begin{align} \label{eq:bound}
  \begin{split}
    \hgamma_{k}^{1} \big( \mathcal{C}_{A,2} \otimes \mathbb{I} \big) \hgamma_{k}^{2} & = \mathcal{C}_{A,2} \otimes \mathbb{I} , \\
    \hgamma_{k}^{2} \big( \mathcal{C}_{A,2} \otimes \mathbb{I} \big) \hgamma_{k}^{1} & = \mathcal{C}_{A,2} \otimes \mathbb{I} .
  \end{split} 
\end{align}
With careful consideration of Majorana commutations and Klein factors, analogously to what was done for the identity boundary state, these relations can be vectorized and written in terms of $\hchi$ as
\begin{equation} \label{eq_cycle_bound_chi}
  \begin{aligned}
     i \hchi_{k}^{(+, 1)} \hchi_{k}^{(-, 2)} \ket{ \mathcal{C}_{A,2} } & = \ket{ \mathcal{C}_{A,2} } ,\\
     -i \hchi_{k}^{(+, 2)} \hchi_{k}^{(-, 1)} \ket{ \mathcal{C}_{A,2} }  & = \ket{ \mathcal{C}_{A,2}  } .
  \end{aligned} 
\end{equation}

Finally, we can calculate the expectation value of the fields $\hat \Phi$ on the boundary states,
\begin{equation}\label{eq_bound_state_phi_exp}
  \bra{\mathbb{I}} \Phi_i \ket{\mathbb{I}} = \bra{\mathcal{C}_{A,2} }\hat \Phi_{i \neq A} \ket{\mathcal{C}_{A,2}} = \frac{\Sigma}{2}  
  \ , \
  \bra{\mathcal{C}_{A,2} }\hat \Phi_A \ket{\mathcal{C}_{A,2}} = \frac{1}{2}
  \begin{bmatrix}
    0 & 0 & 0 & 0 & 1 & 0 \\
    0 & 0 & 0 & -1 & 0 & 0 \\
    0 & 0 & 0 & 0 & 0 & \mathbb{I}_{n-2} \\
    0 & 1 & 0 & 0 & 0 & 0 \\
    -1 & 0 & 0 & 0 & 0  & 0 \\
    0 & 0 & - \mathbb{I}_{n-2} & 0 & 0 & 0 \\
  \end{bmatrix}
  ,
\end{equation}
where $\Sigma$ is the symplectic matrix
\begin{equation}
\Sigma = \begin{pmatrix}
0 & \mathbb{I}_n \\ - \mathbb{I}_n & 0 
\end{pmatrix}
.
\end{equation}

Here we can also derive the form of the boundary state $\left|\mathcal{C}_{A,2} \right>$ as a coherent states in the case of two clusters. So we want to find the corresponding orthogonal matrices that define the boundary state. We suppose 
\begin{equation}
    \ket{\mathcal{C}_{A,2}}=e^{\frac{i N_F}{2} {\rm tr} (\omega^t_A \hat \Phi_A)}\left| \mathbb{I} \right>_A e^{\frac{i N_F}{2} {\rm tr} (\omega^t_B \hat \Phi_B)}\left| \mathbb{I} \right>_B,
\end{equation}
where $e^{\omega_A}=\Omega_A$ and $e^{\omega_B}=\Omega_B$ we are looking for. Finally, it is straightforward to confirm
\begin{equation}\begin{split}
    & \bra{\mathcal{C}_{A,2}} \hat \Phi_A  \ket{\mathcal{C}_{A,2}}=\frac{\Omega_A^t \Sigma \Omega_A}{2},\\
    & \bra{\mathcal{C}_{A,2}} \hat \Phi_B \ket{\mathcal{C}_{A,2}}=\frac{\Omega_B^t \Sigma \Omega_B}{2}.
     \end{split}
\end{equation}
By comparing these equations to  Eq.~\eqref{eq_bound_state_phi_exp} we find:
\begin{equation}\label{boundstate}
\begin{split}
    \ket{\mathcal{C}_{A,2}} = \ket{\Omega_A \Omega_B} \quad \Rightarrow \quad &\Omega_A = 
    \begin{bmatrix}
        0 & -1 & 0 \\
        1 & 0 & 0 \\
        0 & 0 & \mathds{1} \\
    \end{bmatrix}
    , \
    \Omega_B = \mathds{1}.
    \end{split}
\end{equation}
\section{Spin Coherent States}\label{App:SpinCoherent}

Here, we describe the path integral formalism that we use in the section \ref{sec:n2q2} in order to derive the replicated purity mapped to the spin system. The object of our interest is the overlap Eq.~\eqref{eq:transitAmpl},
which can be written as a path integral using the coherent states formalism. The coherent spin state associated with the direction $\mathbf{n} = \left( \sin \theta \sin \phi,\cos \theta ,\sin \theta \cos \phi \right)$, (notice that here we define $y$ as a reference direction) is defined as
\begin{equation}
\ket{\Omega_{\theta,\phi}} := e^{-i \frac{\theta}{ \sin \theta} ( \mathbf{\hat{n}} \cross \hat{\mathbf{y}})\cdot \mathbf{S}} \ket{s} \; .
\end{equation}
In the expression Eq.~\eqref{eq:transitAmpl} we can discretize time and perform usual procedure for the path integral formalism inserting the identity at each time step:
\begin{equation}
\begin{split}
   &\lim_{T \rightarrow \infty} \langle X_A, Z_B | e^{-T \mathcal{H}} | \rho^{(2)}(0) \rangle \sim \\&
  \lim_{N \rightarrow \infty, \delta t \rightarrow 0} \int \prod^N_{i=1} d \Omega_{\phi_i, \theta_i} \left<  X_A, Z_B \right|e^{-\delta t \mathcal{H}} \left|\Omega_{\phi_1, \theta_1}^{A,B} \right> \left< \Omega_{\phi_1, \theta_1}^{A,B}\right | e^{-\delta t \mathcal{H}} \left|\Omega^{A,B}_{\phi_2, \theta_2} \right>...\left< \Omega_{\phi_N, \theta_N}^{A,B}  \right|e^{-\delta t \mathcal{H}} \left|\rho^{(2)}(0)  \right>\;,
  \end{split}
\end{equation}
where $\left|\Omega^{A,B}_{\phi_1, \theta_1} \right>=\left|\Omega_{\phi_1, \theta_1} \right>_A \otimes \left|\Omega_{\phi_1, \theta_1} \right>_B$, $A$ is a subsystem and $B=\overline{A}$ defines the rest of the system, $N$ represents the number of time steps. Doing the expansion of the exponent up to linear order in $\delta t$ can get:
\begin{equation}
    \left< \Omega^{A,B}_{\phi_i, \theta_i}\right | e^{-\delta t \mathcal{H}} \left|\Omega^{A,B}_{\phi_{i+1}, \theta_{i+1}} \right>=\left< \Omega^{A,B}_{\phi_i, \theta_i}|\Omega^{A,B}_{\phi_{i+1}, \theta_{i+1}} \right>-\delta t\left< \Omega^{A,B}_{\phi_i, \theta_i}\right | \mathcal{H} \left|\Omega^{A,B}_{\phi_{i+1}, \theta_{i+1}} \right>+O(\delta t)\;,
\end{equation}and up to the same order we calculate the matrix element of the replicated Hamiltonian
\begin{equation}
    \frac{\left< \Omega^{A,B}_{\phi_i, \theta_i}\right | \mathcal{H} \left|\Omega^{A,B}_{\phi_{i+1}, \theta_{i+1}} \right>}{\left< \Omega^{A,B}_{\phi_i, \theta_i}|\Omega^{A,B}_{\phi_{i+1}, \theta_{i+1}} \right>}= \left< \Omega^{A,B}_{\phi_i, \theta_i}\right | \mathcal{H} \left|\Omega^{A,B}_{\phi_{i}, \theta_{i}} \right>+O(\delta t).
\end{equation}
So we can finally get the matrix element of the exponent
\begin{equation}
    \left< \Omega^{A,B}_{\phi_i, \theta_i}\right | e^{-\delta t \mathcal{H}} \left|\Omega^{A,B}_{\phi_{i+1}, \theta_{i+1}} \right>\sim \left< \Omega^{A,B}_{\phi_i, \theta_i}|\Omega^{A,B}_{\phi_{i+1}, \theta_{i+1}} \right> e^{-\delta t \left< \Omega^{A,B}_{\phi_i, \theta_i}\right | \mathcal{H} \left|\Omega^{A,B}_{\phi_{i}, \theta_{i}} \right>}\;,
\end{equation}where $\left< \Omega^{A,B}_{\phi_i, \theta_i}|\Omega^{A,B}_{\phi_{i+1}, \theta_{i+1}} \right>=\left< \Omega_{\phi_i, \theta_i}|\Omega_{\phi_{i+1}, \theta_{i+1}} \right>_A \left< \Omega_{\phi_i, \theta_i}|\Omega_{\phi_{i+1}, \theta_{i+1}} \right>_B $. For the kinetic term calculation, we also perform the expansion of the overlaps at subsequent time steps
\begin{equation}
\begin{split}
    &\left< \Omega_{\phi_i, \theta_i}|\Omega_{\phi_{i+1}, \theta_{i+1}} \right>=\left(e^{i \delta \phi_i} \sin \frac{\theta_i}{2}\sin \frac{\theta_{i+1}}{2}+\cos \frac{\theta_i}{2}\cos \frac{\theta_{i+1}}{2}\right)^{2S}\sim\\&\sim\left(e^{i \delta \phi_i} \sin \frac{\theta_i}{2}\left(\sin \frac{\theta_{i}}{2}+\frac{1}{2}\cos \frac{\theta_{i}}{2} \delta t \dot \theta_i\right)+\cos \frac{\theta_i}{2}\left(\cos \frac{\theta_{i}}{2}-\frac{1}{2}\sin \frac{\theta_{i}}{2} \delta t \dot \theta_i\right)\right)^{2S}\sim \\& \sim \left(1+i  \dot \phi_i \delta t \sin^2 \frac{\theta_i}{2} \right)^{2S}\sim  e^{i S \delta t \dot \phi_i (1-\cos \theta_i)}\;,
    \end{split}
\end{equation}where $ \dot \phi_i = (\phi_{i+1}-\phi_i)/\delta t$ and $\dot \theta_i = (\theta_{i+1}-\theta_i)/\delta t$. Now we move to cylindrical coordinates and set $\cos \theta_j=y_j$. Finally, in the continuum limit $\delta t\rightarrow 0$ we arrive at usual kinetic term:
\begin{equation}
    \int  d \Omega_{\phi, \theta} e^{i S \sum_{j\in A,B}\int dt \dot \phi_j (1-y_j )}\;.
\end{equation}
Together with the diagonal part of the Hamitlonian Eq.~\eqref{eq:spinHam} we get the path integral representation for the replicated purity of the spin system
\begin{equation}
    \lim_{T \rightarrow \infty} \langle X_A, Z_B | e^{-T \mathcal{H}} | \rho^{(2)}(0) \rangle \sim \int d \Omega_{\phi, \theta} e^{i S \sum_{j \in A,B}\int dt \dot \phi_j (1-y_j )-(J \vec n_A \vec n_B+\Gamma(n^2_{y,A}+n^2_{y,B}))}\;.
\end{equation}In this appendix we performed the derivation of the path integral used in the section \ref{sec:n2q2}.

\section{Cartan basis \label{sec:cartan}}
Here we explain how to move to the Cartan basis in the case of $so(2n)$ algebra, which we use to define the coherent states for the generic number of replicas.
Since $so(2n)$ is a semisimple Lie algebra, it is possible to transform it into the standard Cartan basis. Let us first recall that $so(2n)$ corresponds to a Dynkin diagram within the $D_n$ series with shape,
\begin{equation}
\scalebox{0.6}{
\begin{tikzpicture}
    \tikzset{every node/.style={circle, draw, minimum size=1cm, inner sep=0pt}} 
    \node (a1) at (0,0) [circle, draw] {$\vec\pi_1$};
    \node (a2) at (2,0) [circle, draw] {$\vec\pi_2$};
    \node (a3) at (4,0) [circle, draw] {$\vec\pi_3$};
    \node (an1) at (8,0) [circle, draw] {$\vec\pi_{n-2}$};
    \node (an2) at (10,0) [circle, draw] {$\vec\pi_{n-1}$};    
    \draw (a1) -- (a2);
    \draw (a2) -- (a3);
    \draw[dotted] (a3) -- (an1); 
    \draw (an1) -- (an2);

    \node (an) at (8,2) [circle, draw] {$\pi_n$};
    \draw (an1) -- (an); 
\end{tikzpicture}
}
\end{equation}
where the $\vec\pi_a$ with $a = 1,\ldots, n$ are the set of simple roots. All roots can be embedded in a $n$-dimensional real vector space endowed with a scalar product such that $\langle \vec\pi_a, \vec\pi_a \rangle = 2$ and $\langle \vec\pi_a, \vec\pi_{a'} \rangle = -1$ for all connected nodes in the diagram (and zero otherwise). Explicitly considering the orthogonal basis $[\vec{e}_a]^b = \delta_{a}^b $ in $\mathbb{R}^n$, we can set 
\begin{align}
    &\vec\pi_a = \vec{e}_a - \vec{e}_{a+1} , \quad a = 1,\ldots, n-1 , \\
    &\vec{\pi}_n = \vec{e}_{n-1} + \vec{e}_{n}.
\end{align}
Then, the full set of roots of $so(2n)$ can be obtained via the action of the Weyl group over the set of simple roots.
This leads to the full set of roots $\mathfrak{R}$ made of all vectors with two non-vanishing components of modulus $1$.
Explicitly, we decompose them into positive and negative roots as $\mathfrak{R} = \mathfrak{R}^+ \cup \mathfrak{R}^-$, with  $\mathfrak{R}^+ = \{ \vec{e}_{|a|} + \operatorname{sgn}(b) \vec{e}_{|b|} | a\in \mathbb{N}, b \in \mathbb{Z}, 1\leq a<|b|\leq n\}$ and $\mathfrak{R}^- = - \mathfrak{R}^+$.
We can then define the Cartan-Weyl basis of generators of $so(2n)$ as $\{\hhc_{a}, \eec_{\vec\rho}, \eec_{- \vec\rho} \}$.
The $\hhc_a$ are the commuting generators within the Cartan subalgebra, while the $\eec_{\vec{\rho}}$ are raising/lowering operators associated to each positive root $\vec\rho \in \mathfrak{R}^+$.
In this basis, the algebra of $so(2n)$ takes the form
\begin{equation}\label{eq:comm}
\begin{split}
    &[\hhc_a, \hhc_{a'}] = 0 ,  \\
    &[\hhc_a, \eec_{\vec{\rho}}] = [\vec\rho]_a  \eec_{\vec\rho} ,\\
    &[\eec_{\vec{\rho}}, \eec_{-\vec\rho}] = \sum_a [\vec\rho]_a \hhc_a ,\\
    &[\eec_{\vec\rho}, \eec_{\vec{\rho}^{\ \prime}}] = N_{\vec\rho, \vec{\rho}^{\ \prime}} \eec_{\vec\rho + \vec{\rho}^{\ \prime}} .
\end{split}
\end{equation}
where 
$a,a' = 1,\ldots, n$, $\vec\rho, \vec{\rho}^{\ \prime} \in \mathcal{R}$
the structure constants $N_{\vec\rho, \vec{\rho}^{\ \prime}}$ vanish if $\vec\rho + \vec{\rho}^{\ \prime} \notin \mathcal{R}$, while their explicit value depends on the normalisation convention and will be clarified afterward. 
We can now see how the Cartan-Weyl basis can be expressed realised in terms of fermionic operators. It is convenient to first move to the Dirac fermions representation,
\begin{equation}\label{eq:dirac}
\begin{split}
   & c_{a j \nu}^{\dagger} = \frac{\chi_{j\nu}^{(+,a)} - i \chi_{j \nu}^{(-,a)}}{\sqrt{2}} , \\
   & c_{a j \nu} =  \frac{\chi_{i\nu}^{(+,a)} + i \chi_{j \nu}^{(-,a)}}{\sqrt{2}},
\end{split}
\end{equation}
with the corresponding inverse relations
\begin{equation}
\begin{split}
&\chi_{j\nu}^{(+,a)} = \frac{c_{a j \nu}^{\dagger} + c_{a j \nu}}{\sqrt{2}},\\
&\chi_{j\nu}^{(-,a)} = i \frac{c_{a j \nu}^{\dagger} - c_{a j \nu}}{\sqrt{2}}.
\end{split}
\end{equation}
Since we have a $so(2n)$ algebra at each site, for simplicity of notation, we drop the chain index $j$ in the definition of the ladder operators and recover it later.
First, the generator within the Cartan subalgebra are direcly associated with occupation numbers setting 
\begin{equation}
\label{eq:cartangen}
\hhc_{a} = \frac1 {N_F} \sum_\nu (c^\dag_{a \nu} c_{a \nu} - 1/2)
=  \frac{i}{ N_F} \sum_{\nu} \chi^{(+,a)}_{\nu} \chi^{(-,a)}_{\nu}  =  \Phi^{(+,a),(-,a)} .
\end{equation}
Then the raising/lowering operators can then be defined for $\vec\rho \in \mathfrak{R}^+$ as
\begin{equation}
\label{eq:eecpos}
    \eec_{\vec\rho} = 
    \begin{cases}
    \frac 1 {N_F}\sum_\nu c^\dag_{a,\nu} c_{|b|,\nu} , & b<0, \\
    \frac 1 {N_F}\sum_{\nu} c^\dag_{a,\nu} c_{|b|,\nu}^\dag , & b>0, 
    \end{cases}
\end{equation}
and extended to the negative roots using $\eec_{-\vec\rho} = \eec_{\vec\rho}^\dag$.

\section{Generic n: Matrix elements of the Hamiltonian }\label{App:matel}
In this appendix, we provide calculation of the matrix elements of the Hamiltonian needed for the formulation of the path integral.
As we already discussed in the section \ref{sec:coherent} in order to find the expectation value of the Hamiltonian on the coherent states $\left< O_\eta \right|
\mathcal{H}^{(n)}\left| O_\eta \right>$, we need to find the expectation values of operators $\left< O_\eta \right|
\hat \Phi\left| O_\eta \right>$. Notice that \begin{equation}
    \left< O_\eta \right|
 (\hat\Phi^{\alpha \alpha'}_i)^{\frac{q_J}{2}} (\hat\Phi^{\alpha \alpha'}_{i+1})^{\frac{q_J}{2}}\left| O_\eta \right>= \left< O_\eta \right|
\hat \Phi^{\alpha \alpha'}_i\left| O_\eta \right>^{\frac{q_J}{2}}\left< O_\eta \right|
\hat \Phi^{\alpha \alpha'}_{i+1}\left| O_\eta \right>^{\frac{q_J}{2}}\left( 1+ O\left(\frac{1}{N_F}\right) \right),
\end{equation}
due to the fact that Majorana operators in the definition of the coherent states are uncorrelated for different flavours and clusters.
Then moving to the representation in terms of Dirac fermions, we arrive to the calculation of the  matrix elements of quadratic operators in $c^{\dagger},c$. 
For instance, let us consider $\hat \Phi^{++}$:
\begin{equation}\label{eq:Phipp}
\begin{split}
   \left<O_\eta \right| \hat \Phi^{++} \left| O_\eta \right>= \frac{i}{N_F}(1-\delta_{a a'}) \sum_\nu \left<O_\eta \right| \hat \chi^{+}_{a \nu} \hat \chi^{+}_{a' \nu}\left| O_\eta \right>\\
   = \frac{i}{2N_F}\left<O_\eta \right|  \sum_\nu (c^{\dagger}_{a \nu} + c_{a \nu})(c^{\dagger}_{a'\nu} + c_{a' \nu})\left| O_\eta \right>  - \frac{i}{2} \delta_{aa'}.
\end{split}
\end{equation}
 Using Eq.~\eqref{Ttr} for each quadratic operator in fermions we obtain:
\begin{equation}\label{ccd}
\begin{split}
    &\frac{1}{N_F} \sum_{\nu}\left<O_\eta \right| c^{\dagger}_{a\nu}c_{a' \nu}\left| O_\eta \right> =(U^{t}U^t)_{a a'},\\
    &\frac{1}{N_F} \sum_{\nu} \left<O_\eta \right| c^{\dagger}_{a\nu}c^{\dagger}_{a' \nu}\left| O_\eta \right>=(U^t V^t)_{aa'},\\
     &\frac{1}{N_F} \sum_{\nu}\left<O_\eta \right| c_{a\nu}c_{a' \nu}\left| O_\eta \right>=-(V^{\dagger} U^t)_{aa'},\\
     &\frac{1}{N_F} \sum_{\nu}\left<O_\eta \right| c_{a\nu}c^{\dagger}_{a' \nu}\left| O_\eta \right>=-( V^{\dagger} V^t)_{aa'}.
     \end{split}
\end{equation}
Now we need to specify the form of the matrices $U$ and $V$. To do so we consider the operator $T$ acting on the fermionic operators, in particular:
\begin{equation}
    T_{\eta} c_{a\nu} T_{\eta}^{-1}= e^{\sum_{\nu,1 \leq a< a' \leq n} \eta_{a a'} c^{\dagger}_{a\nu} c^{\dagger}_{a'\nu}- h.c.} c_{a\nu} e^{-\sum_{\nu,1 \leq a< a' \leq n} \eta_{a a'} c^{\dagger}_{a\nu} c^{\dagger}_{a'\nu}+ h.c.},
\end{equation}
where we assume the summation over repeating indexes. Expanding to the second order in $\eta$ we can find:
\begin{equation}
     T_{\eta} c_{a\nu} T_{\eta}^{-1}=c_{a\nu}-\frac{1}{2} \eta_{ab} \eta^{\dagger}_{ba'} c_{a' \nu}-\eta_{aa'} c^{\dagger}_{a'}+O(\eta^3),
\end{equation}
recovering the orders we can obtain
\begin{equation}
     T_{\eta} c_{a\nu} T_{\eta}^{-1}=[\cos \sqrt{\eta \eta^{\dagger}}]_{a a'} c_{a' \nu}-\left(\frac{\sin \sqrt{\eta^{\dagger} \eta}}{\sqrt{\eta \eta^{\dagger}}} \eta \right)_{aa'} c^{\dagger}_{a' \nu}.
\end{equation}
We will use these relations to find the matrix elements of $\hat \Phi$, but first let us find the connection between variables $\eta$ that we were working with before and variables $\tau$ that we introduced in Eq.\eqref{eq:BCH1}. To do so, it is more convenient to work in the fundamental representation of $so(2n)$. This amounts to replacing the fermionic bilinears with the $2n \times 2n$ matrices 
\begin{equation}
\label{eq:fundrepr}
\begin{aligned}
    & \sum_{\nu} c_{a\nu}^{\dagger} c_{a'\nu} - \frac{1}{2} \delta_{aa'} \longrightarrow E^{a,a'} - E^{n+a', n+a} \\
    & \sum_{\nu} c_{a\nu}^{\dagger} c^{\dagger}_{a'\nu} \longrightarrow E^{a, n+a'} - E^{a', n+a} \\
    & \sum_{\nu} c_{a\nu} c_{a'\nu} \longrightarrow E^{n+a, a'} - E^{n+a', a},
    \end{aligned}
\end{equation}
where $E^{\alpha,\beta}$ is a $2n \times 2n$ matrix with $+1$ in the $i$-th column and $j$-th row, as introduced in Eq.~\eqref{eq:Erepmajo}. To clarify the notation, by the arrow $\rightarrow$ we mean that we replace the generators on the left with their matrix representation on the right. We also stress that this representation is unitarily equivalent to the one introduced for $\hat\Phi^{\alpha\beta} \to W^{\alpha\beta}$ below Eq.~\eqref{eq:Erepmajo}.

A useful fact is that that for any antisymmetric matrix $M$, i.e. $M^t = -M$, also $M f(M^\dag M) = f(M M^\dag) M$ is antisymmetric. First, notice that in this representation, the operator that generates the coherent states can be expressed as
\begin{equation}\label{eq:matrepEta}
    e^{\sum_{\nu, 1 \leq a < a' \leq n} \eta_{a a'} c^{\dagger}_{a\nu} c^{\dagger}_{a'\nu} - \text{h.c.}}=
    \begin{pmatrix}
      \sqrt{\mathbb{I}_n-z z^{\dagger}}& z\\
      -z^{\dagger}&\sqrt{\mathbb{I}_n- z^{\dagger}z}
    \end{pmatrix}, \quad 
    z:=\eta \frac{\sin{\sqrt{\eta^{\dagger}\eta}}}{\sqrt{\eta^{\dagger}\eta}}
\end{equation}
On the other hand, in this representation, the right-hand side of Eq.~\eqref{eq:BCH1} is
\begin{equation}\label{eq:matrepTau}
    \begin{pmatrix}
      \mathbb{I}_n& 0\\
      -\tau^{\dagger}& \mathbb{I}_n
    \end{pmatrix}
    \begin{pmatrix}
      e^{\xi} & 0\\
      0&  e^{-\xi^t  }  
    \end{pmatrix}
    \begin{pmatrix}
      \mathbb{I}_n& \tau\\
      0& \mathbb{I}_n
    \end{pmatrix}
     = 
     \begin{pmatrix}
      e^{\xi} & e^{\xi} \tau\\
      -\tau^\dag e^{\xi} & e^{-\xi^t} - \tau^\dag e^{\xi} \tau
    \end{pmatrix} \;.
\end{equation}
Therefore comparing Eqn.~\eqref{eq:matrepEta} and~\eqref{eq:matrepTau}, we find the connection between the matrices $\eta$ and $\tau$.
Explicitly one obtains the matrix relations
\begin{equation}
    e^{\gamma} = \sqrt{\mathbb{I}_n - z z^\dag} \;, \quad \tau = e^{-\gamma} z = (1 - z z^{\dagger})^{-1/2} z = z (1 - z^\dag z)^{-1/2} .
\end{equation}
These relations allow us to find the matrix elements of operators $\hat \Phi$ in terms of skew-symmetric matrices $\tau$ in a compact form.

\begin{equation}
\begin{aligned}
   & \Phi^{++}=\frac{i}{2}
       \left((\mathbb{I}+\tau^{\dagger} \tau)^{-1} \tau^{\dagger}+ \tau (\mathbb{I}+\tau^{\dagger} \tau)^{-1}+\tau(\mathbb{I}+\tau^{\dagger} \tau)^{-1} \tau^{\dagger}-\tau^{\dagger}  \tau(\mathbb{I}+\tau^{\dagger} \tau)^{-1}\right)\\
    &   \Phi^{+-}=\frac{1}{2}
       \left(-(\mathbb{I}+\tau^{\dagger} \tau)^{-1} \tau^{\dagger}+ \tau (\mathbb{I}+\tau^{\dagger} \tau)^{-1}-\tau(\mathbb{I}+\tau^{\dagger} \tau)^{-1} \tau^{\dagger}-\tau^{\dagger} \tau (\mathbb{I}+\tau^{\dagger} \tau)^{-1}+\mathbb{I}\right)\\
     &  \Phi^{-+}=\frac{1}{2}
       \left(-(\mathbb{I}+\tau^{\dagger} \tau)^{-1} \tau^{\dagger}+ \tau (\mathbb{I}+\tau^{\dagger} \tau)^{-1}+\tau(\mathbb{I}+\tau^{\dagger} \tau)^{-1} \tau^{\dagger}+\tau^{\dagger} \tau (\mathbb{I}+\tau^{\dagger} \tau)^{-1}-\mathbb{I}\right)\\
      & \Phi^{--}=-\frac{i}{2}
       \left((\mathbb{I}+\tau^{\dagger} \tau)^{-1} \tau^{\dagger}+ \tau (\mathbb{I}+\tau^{\dagger} \tau)^{-1}-\tau(\mathbb{I}+\tau^{\dagger} \tau)^{-1} \tau^{\dagger}+\tau^{\dagger}  \tau(\mathbb{I}+\tau^{\dagger} \tau)^{-1}\right)\;.
       \end{aligned}
       \end{equation}
 Using these expressions, one can easily recover the matrix elements of the Hamiltonian, as the matrices $\Phi$ serve as the fundamental building blocks of the Hamiltonian. Let us notice here that in order to recover cluster index one needs just to add an extra index $j$ to each $\tau \rightarrow \tau_j$.
\section{Details on the expansion of the action for small \texorpdfstring{$\tau_I$}{tau_I}}\label{App:tauexp}
In this appendix we perform the expansion of the kinetic term and the Hamilotnian discussed in the subsection \ref{ssec:integ_tau}, with respect to $\tau_I$. Remind that $\tau_I$ is the imaginary part of the matrix $\tau$. As we are going to show, monitoring part of the Hamiltonian has only second order in this expansion, while unitary part has also zero order. So the action has the following form:
\begin{equation}
    -\mathcal{S}=\int dt \left(N_F \tr (A \tau_I)+ J N_F^2 (\tr(B)+\tr(C\tau_IC\tau_I))+\Gamma N_F^2 \tr(D\tau_ID\tau_I) \right),
\end{equation}
where $A,B,C,D$ matrices that depend on the real part of $\tau$. Rescaling $\tau_I \rightarrow N_F \tau_I$ as we did with the parameter $y$ in the section \ref{sec:n2q2} we can neglect the quadratic order in the unitary part of the action, since it is subleading due to the scaling of the unitary  coupling $J \sim 1/N_F^{\alpha}$.
Therefore we proceed with the expansion of the kinetic term. For the first matrix under the trace in Eq.~\eqref{eq:kin} we get in the first order:
\begin{equation}
   (\mathbb{I}+\tau^{\dagger} \tau)^{-1}= (\mathbb{I}-( \tau_R -i \tau_I)(\tau_R+i \tau_I))^{-1}=((\mathbb{I}-\tau_R^2)^{-1}-i(\mathbb{I}-\tau_R^2)^{-1}[\tau_I, \tau_R](\mathbb{I}-\tau_R^2)^{-1}),
\end{equation}
where $[.,.]$ is a commutator, and
\begin{equation}
    \tau^{\dagger} \dot{\tau}-\dot{\tau}^{\dagger}\tau=-\tau_R \dot{\tau}_R+\dot \tau_R {\tau}_R-i\{\tau_R, \dot \tau_I\}+i\{\tau_I,  \dot\tau_R \},
\end{equation}
all together it gives:
\begin{equation}
\begin{split}
     &(\mathbb{I}+\tau^{\dagger} \tau)^{-1}(\tau^{\dagger} \dot{\tau}-\dot{\tau}^{\dagger}\tau)=\\&=((\mathbb{I}-\tau_R^2)^{-1}-i(\mathbb{I}-\tau_R^2)^{-1}[\tau_I, \tau_R](\mathbb{I}-\tau_R^2)^{-1}) ([\dot \tau_R,\tau_R]+i \{\tau_I,  \dot\tau_R \}-i \{\tau_R, \dot \tau_I\}),
     \end{split}
\end{equation}
and for the trace we naturally have the expansion
\begin{equation}\label{eq:kin_tau}
\begin{split}
    {\rm tr} (\mathbb{I}+\tau^{\dagger} \tau)^{-1}(\tau^{\dagger} \dot{\tau}-\dot{\tau}^{\dagger}\tau)=- i {\rm tr} (\mathbb{I}-\tau_R^2)^{-1} [\tau_I,\tau_R](\mathbb{I}-\tau_R^2)^{-1} [\dot \tau_R,\tau_R]+i {\rm tr}(\dot \tau_R (\mathbb{I}-\tau_R^2)^{-1} \tau_I)\\+i {\rm tr}( (\mathbb{I}-\tau_R^2)^{-1} \dot \tau_R \tau_I)-i {\rm tr}( (\mathbb{I}-\tau_R^2)^{-1} \tau_R \dot \tau_I)-i {\rm tr}( \tau_R (\mathbb{I}-\tau_R^2)^{-1}\dot \tau_I ).\end{split}\end{equation}
As in the case $n=2$ we can perform integration by parts for the last two terms
\begin{equation}
\begin{split}
    &-i \int dt ~~{\rm tr}( (\mathbb{I}-\tau_R^2)^{-1} \tau_R \dot \tau_I)-i \int dt ~~ {\rm tr}( \tau_R (\mathbb{I}-\tau_R^2)^{-1}\dot \tau_I  )=\\&=i \int dt ~~ {\rm tr}~\left( \tau_I \frac{\partial}{\partial t} (\mathbb{I}-\tau_R^2)^{-1} \tau_R \right)+ i \int dt ~~ {\rm tr}~\left( \tau_I \frac{\partial}{\partial t} \tau_R (1-\tau_R^2)^{-1}  \right)=\\&=
 i \int dt ~~ {\rm tr} \left( \tau_I(\mathbb{I}-\tau_R^2)^{-1}(\dot \tau_R \tau_R+\tau_R \dot \tau_R) \tau_R (\mathbb{I}-\tau_R^2)^{-1} \right) +  i \int dt ~~ {\rm tr} (\tau_I (\mathbb{I}-\tau_R^2)^{-1} \dot \tau_R)+\\&+
    i \int dt ~~ {\rm tr} (\tau_I \dot \tau_R (\mathbb{I}-\tau_R^2)^{-1} )+ i \int dt ~~ {\rm tr} \left( \tau_I (\mathbb{I}-\tau_R^2)^{-1} \tau_R (\dot \tau_R \tau_R+\tau_R \dot \tau_R)  (\mathbb{I}-\tau_R^2)^{-1} \right),
    \end{split}
\end{equation}
where we used $\frac{\partial }{\partial t} A^2= A' A+A A'$. Together with the first  terms from Eq.~\eqref{eq:kin_tau}
\begin{equation}
\begin{split}
    &-2 i \int dt~~ {\rm tr}~~ \frac{\tau_R}{\mathbb{I}-\tau_R^2} \tau_I \frac{\tau_R}{\mathbb{I}-\tau_R^2} \dot \tau_R+i \int d t~~{\rm tr}~~ \frac{1}{\mathbb{I}-\tau_R^2} \tau_I \frac{\tau^2_R}{1-\tau_R^2} \dot \tau_R+i \int d t~~{\rm tr}~~ \frac{\tau^2_R}{\mathbb{I}-\tau_R^2} \tau_I \frac{1}{\mathbb{I}-\tau_R^2} \dot \tau_R
    \\&+4 i \int dt ~~ {\rm tr} \left((\mathbb{I}-\tau_R^2)^{-1} \dot \tau_R (\mathbb{I}-\tau_R^2)^{-1}  \tau_I \right)-i\int dt~~ {\rm tr} \left((\mathbb{I}-\tau_R^2)^{-1} \tau_R^2 \dot \tau_R (\mathbb{I}-\tau_R^2)^{-1} \tau_I \right)-\\&-i\int dt~~ {\rm tr} \left((\mathbb{I}-\tau_R^2)^{-1}  \dot \tau_R \tau_R^2 (\mathbb{I}-\tau_R^2)^{-1} \tau_I \right)+2 i\int dt~~ {\rm tr} \left((\mathbb{I}-\tau_R^2)^{-1} \tau_R \dot \tau_R \tau_R (\mathbb{I}-\tau_R^2)^{-1} \tau_I \right)=\\&=
    i \int dt ~~{\rm tr}\left( 4 \dot \tau_R  \frac{1}{\mathbb{I}-\tau_R^2} \tau_I \frac{1}{\mathbb{I}-\tau_R^2}\right),
    \end{split}
\end{equation}
which gives us the expansion of the kinetic term, notice that it is in complete analogy with the case $n=2$.
For the Hamiltonian part, we first proceed with the expansion of the matrix $\Phi$. Starting from the expression we derived in the appendix \ref{App:matel} for $\Phi$  in terms of $\tau$,
we find the following expansion at linear order in $\tau_I$
\begin{equation}
\begin{split}
    \Phi^{++} =
    -\frac{1}{\mathbb{I}-\tau_R^2} \tau_I \frac{1}{\mathbb{I}-\tau_R^2}+\frac{1}{\mathbb{I}-\tau_R^2} \tau_I \frac{\tau_R}{\mathbb{I}-\tau_R^2}-\frac{\tau_R}{\mathbb{I}-\tau_R^2} \tau_I \frac{1}{\mathbb{I}-\tau_R^2}+\frac{\tau_R}{\mathbb{I}-\tau_R^2} \tau_I \frac{\tau_R}{\mathbb{I}-\tau_R^2},
    \end{split}
\end{equation}
\begin{equation}
\begin{split}
    \Phi^{--} 
    =
    \frac{1}{\mathbb{I}-\tau_R^2} \tau_I \frac{1}{\mathbb{I}-\tau_R^2}+\frac{1}{\mathbb{I}-\tau_R^2} \tau_I \frac{\tau_R}{\mathbb{I}-\tau_R^2}-\frac{\tau_R}{\mathbb{I}-\tau_R^2} \tau_I \frac{1}{\mathbb{I}-\tau_R^2}-\frac{\tau_R}{\mathbb{I}-\tau_R^2} \tau_I \frac{\tau_R}{\mathbb{I}-\tau_R^2}\;;
    \end{split}
\end{equation}
and 
\begin{equation}
    \Phi^{+-}=\frac{1}{2}\frac{\mathbb{I}-\tau_R}{\mathbb{I}+\tau_R}+O(\tau_I^2).
\end{equation}
So for the monitoring part of the Hamiltonian we arrive at 
\begin{equation}
    \mathcal{ H}_{\rm mon}= \frac{N_F^2 \Gamma }{4}\sum_{i=1}^L \left(2 {\rm tr} \left( \Phi^{++}_i(\Phi^{++}_i)^t \right) +2{\rm tr} \left( \Phi^{--}_i(\Phi^{--}_i)^t \right)\right).
\end{equation}
where we used $\tr \Phi \Phi^t=\frac{n}{4}={\rm tr} \Phi^{++}_i(\Phi^{++}_i)^t+{\rm tr} \Phi^{+-}_i(\Phi^{+-}_i)^t+{\rm tr} \Phi^{-+}_i(\Phi^{-+}_i)^t+{\rm tr} \Phi^{--}_i(\Phi^{--}_i)^t $, and  ${\rm tr} \Phi^{+-}_i(\Phi^{+-}_i)^t={\rm tr} \Phi^{-+}_i(\Phi^{-+}_i)^t$. And in terms of $\tau$ we find a full square under the trace: 
\begin{equation}
    \mathcal{ H}_{\rm mon}=- \Gamma N_f^2 \sum_{i=1}^L {\rm tr} \left(\frac{1}{(\mathbb{I}-\tau_{R,i}^2)^2} \tau_{I,i}-\frac{\tau_{R,i}^2}{(\mathbb{I}-\tau_{R,i}^2)^2} \tau_{I,i} \right)^2=- \Gamma N_f^2\sum_{i=1}^L {\rm tr}\left( \frac{1}{\mathbb{I}-\tau_{R,i}^2} \tau_{I,i}\frac{1}{\mathbb{I}-\tau_{R,i}^2} \tau_{I,i} \right).
\end{equation}
We also find the unitary part in the relevant zeroth order:

\begin{equation}
\begin{split}
    \mathcal{H}_{\rm uni}\Big|_{q_J=2} & = \frac{J}{4} N_F^2 \sum_{i=1}^L {\rm tr}\left(\frac{\mathbb{I} - \tau_{R,i}}{\mathbb{I} + \tau_{R,i}} \frac{\mathbb{I} - \tau^t_{R,i+1}}{\mathbb{I} + \tau^t_{R,i+1}}\right)+O(\tau^2_I)\\ 
    \mathcal{H}_{\rm uni}\Big|_{q_J=4} & = -\frac{J N_F^4}{64}\sum_{a,a'=1}^n\sum^L_{i=1}   \left( \left(\frac{\delta_{aa'} - \tau^{aa'}_{R,i}}{\delta_{aa'} + \tau^{aa'}_{R,i}}\right)^2 \left(\frac{\delta_{aa'} - \tau^{aa'}_{R,i+1}}{\delta_{aa'} + \tau_{R,i+1}^{aa'}}\right)^2 
        +\frac{1}{6}  \left(\frac{\delta_{aa'} - \tau^{aa'}_{R,i}}{\delta_{aa'} + \tau^{aa'}_{R,i}}\right)^4\right)+O(\tau^2_I).
    \end{split}
\end{equation}
In this appendix, we have presented the detailed derivation of the expansion of the kinetic term and the Hamiltonian with respect to $\tau_I$ modes. This expansion forms the basis for integrating out $\tau_I$ (see subsection \ref{ssec:integ_tau}), ultimately leading to the field theory descriptions for both interacting and non-interacting models, as discussed in the subsection \ref{ssec:NLSM}.

\section{Cayley transform}\label{App:Caley}In this appendix we derive the integration measure for the matrices formed by Cayle transformation of a skew-symmetric matrices.
Consider an antisymmetric real matrix $A$ of size $n$. The Cayley transform is defined as
\begin{equation}
    O = \frac{\mathbb{I} - A}{\mathbb{I} + A}.
\end{equation}
It is immediate to check that $O$ is an orthogonal matrix, i.e. $O^t O = \mathbb{I}$. Also, the transformation is an involution and therefore
\begin{equation}
    A = (\mathbb{I} - O)(\mathbb{I} + O)^{-1}.
\end{equation}
The matrix $O$ cannot have the eigenvalue $-1$ and therefore $\det O = 1$. So, the Cayley transform maps antisymmetric matrices to $SO(n)$.

We can use this transformation to express the Haar measure over $SO(n)$ by means of the simpler measure over the entries of $A$. Let us define the standard measure on real antisymmetric matrices as,
\begin{equation}
    dA = \prod_{1\leq i<j\leq n} dA_{ij}.
\end{equation}
We denote the Haar measure on $SO(n)$ as $dO_{\rm Haar}$. The Haar measure is invariant under left and right multiplication, $O \to W_1 O W_2$. We want to find a function $f$ such that
\begin{equation}
\label{eq:maphaar0}
    f(A) dA = dO_{\rm Haar}.
\end{equation}
To do so, we diagonalise $A$ writing $A = W X W^t$, with $W$ an orthogonal matrix and $X$ a diagonal matrix of eigenvalues. Since $A$ in antisymmetric, the spectrum is organized in pairs of imaginary numbers. 
We have to distinguish the two cases
\begin{align}
    &X = \mbox{diag}(+i x_1, -i x_1, +i x_2, -i x_2, \ldots, + i x_k, -i x_k) , \qquad n = 2k ,\\
    &X = \mbox{diag}(0, +i x_1, -i x_1, +i x_2, -i x_2, \ldots, + i x_k, -i x_k) , \qquad n = 2k + 1,
\end{align}
using standard measure, one can ttransform the measure $dA$ into a measure over the rotation $W$ and one over the eigenvalues. Following \cite{MEHTA1968449}, we obtain,
\begin{align}
    &dA = \frac{1}{Z_A} dW_{\rm Haar} \prod_{1\leq i<j \leq k} (x_i^2 - x_j^2)^2 \prod_{i=1}^k dx_i \qquad n = 2k \\
    &dA = \frac{1}{Z_A} dW_{\rm Haar} \prod_{1\leq i<j \leq k} (x_i^2 - x_j^2)^2 \prod_{i=1}^k x_i^2 dx_i \qquad n = 2k  + 1.
\end{align}
We can then apply the same procedure to the matrix $O$. In this case, the eigenvalues of $O$ come in complex conjugate pairs of the form $e^{\pm i \theta_k}$, with an additional $1$ in case of odd  and express the Haar measure as
\begin{align}
    &dO_{\rm Haar} = \frac{dW_{\rm Haar}}{Z_O} \prod_{1 \leq i < j \leq k} (\cos(\theta_i) - \cos(\theta_j))^2 \prod_{i=1}^k d\theta_k \qquad n = 2k \\
    &dO_{\rm Haar} = \frac{dW_{\rm Haar}}{Z_O} \prod_{1 \leq i < j \leq k} (\cos(\theta_i) - \cos(\theta_j))^2 \prod_{i=1}^k (1 - \cos(\theta_k)) d\theta_k \qquad n = 2k + 1.
\end{align}
We now use the Cayley transform to determine the function $f$ in Eq.~\eqref{eq:maphaar0} which produces the right correspondence.
We see that the Haar matrix over $W$ is common to both, so we can focus on the mapping of the eigenvalue distribution. 
Under the Cayley transformation, we have the mapping of the eigenvalues
\begin{equation}
    e^{i \theta} = \frac{1 + i x}{1 - i x} \quad \Rightarrow \quad \theta = 2 \arctan(x)  , \; x = \tan(\theta/2).
\end{equation}
With this transformation, we have
\begin{multline}
        \prod_{1 \leq i < j \leq k} (\cos(\theta_i) - \cos(\theta_j))^2 =     \prod_{1 \leq i < j \leq k} (\frac{2}{1 + x_i^2} - \frac{2}{1 + x_j^2})^2  =\\= 2^{k(k-1)}\prod_{1 \leq i<j} (x_i^2 - x_j^2)^2  \prod_{i=1}^k (1 + x_i^2)^{2(1-k)},
\end{multline}
also notice that this formula can be presented as a determinant:
\begin{equation}
    \prod_{i=1}^k (1 + x_i^2)^2 = \det(1 - A^2).
\end{equation}
So for the even number of replicas $n = 2k$, we arrive at
\begin{equation}
    dO_{\rm Haar} = \frac{Z_A}{Z_0} 2^{k^2} \prod_{i=1}^k (1 + x^2_i)^{1-2k} dA = \frac{Z_A}{Z_O} 2^{k^2} \det(1 - A^2)^{(1 - 2k)/2} dA.
\end{equation}For the odd number of replicas, we similarly  have
\begin{equation}
    \prod_{i=1}^k ( 1 - \cos(\theta_k)) = 2^{k} \prod_{i=1}^k \frac{x^2_i}{1 + x_i^2},
\end{equation}
which gives us the Haar measure on the $SO(n)$ group
\begin{equation}
    dO_{\rm Haar} = \frac{Z_A}{Z_O} 2^{k(k+1)} \det(1 - A^2)^{-k} dA.
\end{equation}
So we see that except for normalisation constants, we can set in both cases, $f(A) = \det(\mathbb{I} - A^2)^{(1 -n)/2}$. This result confirms Theorem 3 of \cite{toyama1948haar}. Therefore, in this appendix,1 we confirmed that the Caley transformation of a skew symmetric matrix gives a Haar measure on the special orthogonal group.
\section{Details on the measurement-only action}\label{App:mon}
In this appendix, we first provide the derivation of the expansion for the kinetic term and the expansion of the Hamiltonian part of the action. Here we start with the formula we derived in the section \ref{sec:kin}:

\begin{equation}
K[\tau,\tau^\ast] =  \frac{1}{2} \sum_{\alpha \beta} \left[\partial_{\tau_{ij}}(\ln \mathcal{N}) \dot \tau_{ij} - \partial_{\tau_{ij}^\ast}(\ln \mathcal{N}) \dot \tau^\ast_{ij} \right] d t.
\end{equation}
Now by computing the metric in $\tau + \Delta x \delta \tau$ at the first order
\begin{equation}
\begin{split}
    K[\tau + \Delta x \delta \tau, \tau + \Delta x \delta \tau^\ast] - &K[\tau, \tau^\ast] = \\ &\frac{\Delta x}{2} \big( \partial_{\tau_{ij},\tau_{kl}}(\ln \mathcal{N}) \dot \tau_{ij} \delta \tau_{kl} + \partial_{\tau_{ij},\tau_{kl}^\ast}(\ln \mathcal{N}) \dot \tau_{ij} \delta \tau_{kl}^\ast + \partial_{\tau_{ij}}(\ln \mathcal{N}) \delta \dot \tau_{ij} \\ 
    &- \partial_{\tau_{ij}^\ast \tau_{lk}} (\ln \mathcal{N}) \dot \tau^\ast_{ij} \delta \tau_{lk} - \partial_{\tau_{ij}^\ast \tau_{lk}^\ast} (\ln \mathcal{N}) \dot \tau^\ast_{ij} \delta \tau_{lk}^\ast - \partial_{\tau_{ij}^\ast}(\ln \mathcal{N}) \delta \dot \tau^\ast_{ij}  \big) + O(\delta \tau^2).
\end{split}
\end{equation}
By integrating by parts the terms containing $\delta \dot \tau$ we have: 
\begin{equation}
    \begin{split}
         \partial_{\tau_{ij}}(\ln \mathcal{N}) \delta \dot \tau_{ij} &=  - \partial_{\tau_{ij} \tau_{lk}}(\ln \mathcal{N}) \dot \tau_{lk}\delta \tau_{ij} - \partial_{\tau_{ij} \tau_{lk}^\ast}(\ln \mathcal{N}) \dot \tau_{lk}^\ast \delta \tau_{ij} \\
         \partial_{\tau_{ij}^\ast}(\ln \mathcal{N}) \delta \dot \tau_{ij}^\ast &=  - \partial_{\tau_{ij}^\ast \tau_{lk}}(\ln \mathcal{N}) \dot \tau_{lk} \delta \tau_{ij}^\ast - \partial_{\tau_{ij}^\ast \tau_{lk}^\ast}(\ln \mathcal{N}) \dot \tau_{lk}^\ast \delta \tau_{ij}^\ast, \\
    \end{split}
\end{equation}
so that we get 
\begin{equation}
\begin{split}
    K[\tau + \delta \tau, \tau + \delta \tau^\ast] - K[\tau, \tau^\ast] &=  \frac{\Delta x}{2} \big(   \partial_{\tau_{ij},\tau_{kl}^\ast}(\ln \mathcal{N}) \dot \tau_{ij} \delta \tau_{kl}^\ast - \partial_{\tau_{ij} \tau_{lk}^\ast}(\ln \mathcal{N}) \dot \tau_{lk}^\ast \delta \tau_{ij} \\ 
    &- \partial_{\tau_{ij}^\ast \tau_{lk}} (\ln \mathcal{N}) \dot \tau^\ast_{ij} \delta \tau_{lk} + \partial_{\tau_{ij}^\ast \tau_{lk}}(\ln \mathcal{N}) \dot \tau_{lk} \delta \tau_{ij}^\ast  \big) + O(\delta \tau^2) \\
    &= \mathtt{g}_{\alpha \beta} \dot \tau_\alpha \delta \tau_\beta^\ast - \mathtt{g}_{\beta\alpha} \dot \tau_\alpha^\ast  \delta \tau_\beta
\end{split}\;,
\end{equation} where we defined $\mathtt{g}_{\alpha \beta}= \frac{\partial^2 \ln \mathcal{N}}{\partial \tau_{\alpha } \partial \tau^*_{\beta}}$ is a metric tensor see Eq.~\eqref{eq:kahler}
(notice that $\mathtt{g}_{\beta \alpha} = \mathtt{g}_{\alpha \beta}^{*}$). Setting $\tau_{2i-1} = \tau + \Delta x\delta \tau$ and $\tau_{2i}= \tau -\Delta x \delta \tau$, one can expand the difference of kinetic terms in terms of the metric
\begin{equation}
    K_{2i-1} - K_{2i} = 2N_F \Delta x \mathtt{g}_{\alpha \beta}(\tau^{2i-1}) \dot \tau^{2i-1}_{\alpha} (\delta\tau^{2i-1}_{\beta})^*-c.c.
\end{equation}
Similarly, the interaction term expanded for small $\delta \tau$ up to second order must be invariant, and so we have:
\begin{equation}
    \sum_i H_{i ,i+1}=\sum_j H_{2j-1, 2j}+H_{2j,2j+1},
\end{equation}
where $H_{2j-1,2j}$ is a Hamiltonian defined inside each dimer and $H_{2j,2j+1}$ is a Hamiltonian defined in between different dimers. Therefore, for the interaction inside each dimer we write:
\begin{equation}
    H_{2i-1,2i}=- 4 N_F^2 \Delta x^2 \Gamma(1-\Delta) \  \mathtt{g}_{\alpha\beta}(\tau^{2i-1}) \delta \tau^{2i-1}_\alpha (\delta \tau^{2i-1}_\beta)^\ast,
\end{equation}
and for the interaction term between each dimer  we do the expansion:
\begin{equation}
    H_{2i,2i+1}= - 4 N_F^2 \Gamma(1+\Delta) \mathtt{g}_{\alpha \beta}(\tau^{2i-1})(\Delta x \delta \tau_{\alpha}^{2i-1}+\Delta x \partial_x \tau_{\alpha}^{2i-1})(\Delta x (\delta \tau_{\beta}^{2i-1})^*+\Delta x (\partial_x \tau_{\beta}^{2i-1})^*).
\end{equation}
Here we also provide the calculation of the Gaussian integral in $\delta \tau$. Notice that it can be presented as
\begin{equation}
    \int \delta \tau e^{-8 N_F^2 \Delta x \Gamma \int d^2 x \delta \tau_{\alpha} \mathtt{g}_{\alpha \beta} \delta \tau^*_{\beta}+\int d^2 x A_{\alpha} \delta \tau^*_{\alpha}+\int d^2 x \delta \tau_{\alpha} B_{\alpha}}\sim e^{\frac{A_{\alpha} \mathtt{g}^{-1}_{\alpha \beta }B_{\beta}}{8 N_F^2 \Delta x \Gamma}},
\end{equation}
where we assume summation over repeating indexes, here 
\begin{equation}
    A_{\alpha}=2 N_F \dot{\tau}_{\gamma} \mathtt{g}_{\gamma \alpha} -4 N_F^2\Gamma (1+\Delta) \Delta x \partial_x \tau_{\gamma} \mathtt{g}_{\gamma \alpha},
\end{equation}
\begin{equation}
    B_{\alpha}=-2 N_F \mathtt{g}_{\alpha \gamma} \dot{\tau}^*_{\gamma} -4 N_F^2\Gamma(1+\Delta) \Delta x \mathtt{g}_{\alpha \gamma} \partial_x \tau^*_{\gamma}.
\end{equation}

\section{Numerical simulations of SYK systems }\label{App:num}

\subsection{Time evolution with the stochastic Schrödinger equation}

The time discretized evolution corresponding to the SSE of Eq.~\eqref{eq:SSE} and the time-dependent random Hamiltonian Eq.~\eqref{eq_H_uni_plus_mon} is given by 
\begin{equation}\label{eq_discretized_SSE}
  \begin{aligned}
    \delta \ket{\psi} 
    = \sum_j \left\{\rule{0cm}{1cm}\right.  & i^{\frac{q_J}{2} - 1} \sqrt{J} \left[ 
    \sum_{{\boldsymbol{\mu}} {\boldsymbol{\nu}}}  \delta h^j_{ \boldsymbol{\mu}\boldsymbol{\nu}}(t) \prod^{q_J/2}_{k=1} \hchi_{j \mu_{k}} \prod_{l=1}^{q_J/2} \hchi_{j+1 \nu_{l}}    
    +  \sum_{{\boldsymbol{\tilde \mu}}}  \delta h^j_{ \boldsymbol{\tilde \mu}}(t) \prod^{q_J}_{k=1} \hchi_{j \tilde \mu_{k}}
    \right] 
    \\
    + & \ i^{\frac{q}{2}} \sqrt{\Gamma} \sum_{\boldsymbol{\tilde \nu}} \delta w_{\boldsymbol{\tilde{\nu}}}^j(t)   
    \left( \prod_{k=1}^{q}\hchi_{j\tilde \nu_k} - \left\langle \prod_{k=1}^{q}\hchi_{j\tilde \nu_k} \right\rangle_t \right)
    \\
    - & \left. \ i^q \frac{\Gamma \delta t}{2} \sum_{\boldsymbol{\tilde{\nu}}}  
    \left( \prod_{k=1}^{q}\hchi_{j\tilde \nu_k} - \left\langle \prod_{k=1}^{q}\hchi_{j\tilde \nu_k} \right\rangle_t \right)^2
    \right\} \ket{\psi}
    ,
    \\
  \end{aligned}
\end{equation}
with $\delta t \ll 1$.
The random variables $\delta h$ and $\delta w$ have zero mean and variance
\begin{subequations}
  \begin{equation}
    \mathbb{E}_{\rm G} \left[ \delta h_{\tmmu_1}^i \delta h_{\tmmu_2}^j \right]  = \delta_{ij} \delta_{\tmmu_1 \tmmu_2} \delta t , 
  \end{equation}
  \begin{equation}
    \mathbb{E}_{\rm G}\left[ \delta h_{\mmu_1\mnu_1}^i \delta h_{\mmu_2 \mnu_2}^j \right] = \delta_{ij} \delta_{\mnu_1 \mnu_2} \delta_{\mmu_1 \mmu_2} \delta t , 
  \end{equation}
  \begin{equation}
    \mathbb{E}_{\rm G} \left[ \delta w_{\mnu_1}^i \delta w_{\mnu_2}^j \right] = \delta_{ij} \delta_{\mnu_1 \mnu_2} \delta t .
  \end{equation}
\end{subequations}

As can be verified with a first-order expansion in $\delta t$, the evolution of the state is given by
\begin{equation}\label{eq:discrete_evol_step_op_cluster_mon}
  \begin{aligned}
    \ket{\psi(t + \delta t )} = &  
    \exp{  \sum_{j, \tmnu}  \left(  \delta w_{\tmnu}^j \sqrt{\Gamma} + i^{\frac{q}{2}} \Gamma \delta t \left\langle  \prod_{k=1}^{q}\hchi_{j\tilde \nu_k} \right\rangle_t   \right) i^{\frac{q}{2}} \prod_{k=1}^{q}\hchi_{j\tilde \nu_k}
    } \\ 
    \times & \exp{ i^{\frac{q_J}{2} - 1} \sqrt{J} \sum_j \left[  \sum_{{\boldsymbol{\mu}} {\boldsymbol{\nu}}}  \delta h^j_{ \boldsymbol{\mu}\boldsymbol{\nu}}(t) \prod^{q_J/2}_{k=1} \hchi_{j \mu_{k}} \prod_{l=1}^{q_J/2} \hchi_{j+1 \nu_{l}}    
    +  \sum_{{\boldsymbol{\tilde \mu}}}  \delta h^j_{ \boldsymbol{\tilde \mu}}(t) \prod^{q_J}_{k=1} \hchi_{j \tilde \mu_{k}} 
    \right] } \ket{\psi(t)}
  \end{aligned}
\end{equation}
up to a normalization factor. 

The discretized time evolution must be taken with a small enough $\delta t$, such that we see a convergence in the results in $\delta t$.
A proper $\delta t$ needs to be found empirically for each set of $q$, $q_J$, $J$, $\Gamma$ and $N_F$ (without loss of generality, we fix $\Gamma = 1$ to reduce the size of the parameter space).

\begin{figure}[t]
    \centering
    \includegraphics[width=0.6\columnwidth]{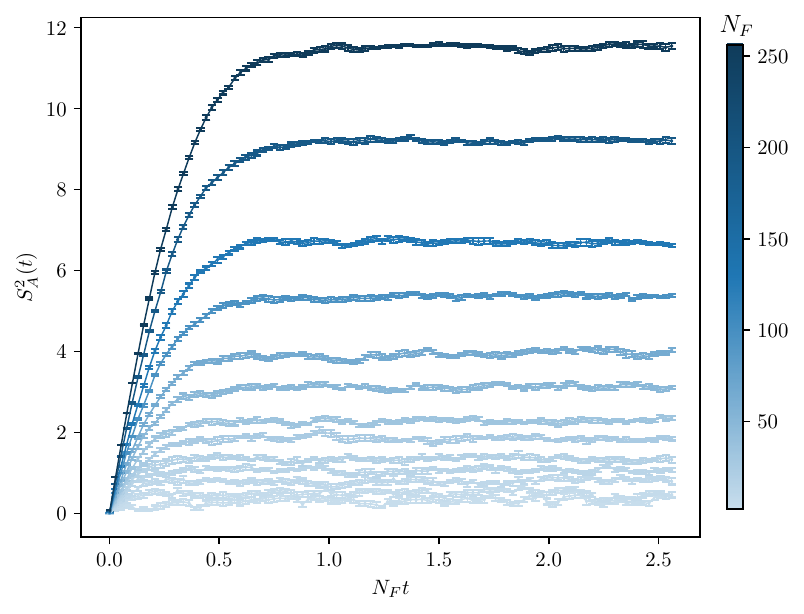}
    \caption{Time evolution of the Second Rényi entropy for a model with $q = q_J = 2$, $\Gamma = 1$ and $J = \sqrt{N_F}$.}
    \label{fig_entropy_saturation}
\end{figure}

We are interested in computing the entropies of Eq.~\eqref{eq_renyi_limit}.
Since the entropy saturates, as exemplified in Fig. \ref{fig_entropy_saturation}, we can instead compute time-averaged values over an appropriate window, reducing the total amount of need sampling. 
Note that this window also needs to be found empirically, for just changing $N_F$ can alter the timescale of saturation of the dynamics.

\subsection{The Gaussian case}

If all terms in the Hamiltonian are quadratic, meaning $q = q_J = 2$, then simulations can be greatly simplified since, due to Wick's theorem, all observables can be computed from the covariance matrix
\begin{equation}
    M_{jk} = \frac{i}{2} \Tr \Big( \rho [\hchi_j, \hchi_k] \Big) = i \left( \langle \hchi_j \hchi_k \rangle - \delta_{jk} \right) ,
\end{equation}
where we have introduced a flattening of the Majorana operator indices (cluster and flavour) which will be generally useful for numerics.


The unitary part of the discretized Stochastic Schrödinger's equation for the Gaussian case can generically be written in the form
\begin{equation}
    \delta \ket{\psi} = \sum_{ij}  \delta h_{ij} \hchi_i \hchi_j \ket{\psi} = -i \delta H \ket{\psi} ,
\end{equation}
where we have hidden all microscopic details and noise in the real skew-symmetric matrix $\delta h$, which can be found by comparing with Eq.~\eqref{eq_discretized_SSE}.
From the commutation relation
\begin{equation}
    [\delta H,\hchi_i] = -4 i \sum_{j} \delta h_{ij} \hchi_j ,
\end{equation}
using the Baker-Campbell-Hausdorff formula, we can show that the Majorana operators evolve in the Heisenberg picture as,
\begin{equation}
    \hchi_i \rightarrow  e^{-i \delta H} \hchi_i e^{i \delta H} = \sum_j \left[e^{4 \delta h}\right]_{ij} \hchi_j ,
\end{equation}
meaning the covariance matrix evolves as 
\begin{equation}
    M_{ij} \rightarrow O_{\delta h} M O_{\delta h}^t  ,
\end{equation}
where $O_{\delta h}$ is the orthogonal matrix given by $O_{\delta h} = e^{4\delta h}$.

The monitoring part of the evolution is given by Eq.~\eqref{eq:discrete_evol_step_op_cluster_mon} as, 
\begin{equation}
\begin{aligned}
    \ket{\psi} \rightarrow &
    \exp{  \sum_{j}  \sum_{\nu_1 < \nu_2} \left(  \delta w_{\nu_1 \nu_2}^j \sqrt{\Gamma} + \Gamma \delta t M_{(j\nu_1),(j\nu_2)}   \right) i \hchi_{j\nu_1} \hchi_{j\nu_2} }  \ket{\psi} \\ 
    = &
    \exp{ i \sum_{ij}    \delta w_{ij} \hchi_{i} \hchi_{j} }  \ket{\psi}
    = \mathcal{D} \ket{\psi} ,
\end{aligned}
\end{equation}
where we have hidden the microscopical details in the skew-symmetric real matrix $\delta w$.
Since $\mathcal{D}$ is not a unitary operator, the state also needs to be normalized as
\begin{equation}\label{eq_monitoring_evol_step}
  \ket{\psi} \rightarrow \frac{\mathcal{D}\ket{\psi}}{\|\mathcal{D} \ket{\psi}\|} .
\end{equation}
Following the approach of \cite{Fava:2023tgg}, we can note that any Gaussian state can be defined by a set of Dirac fermion annihilation operators (with respect to the state)
\begin{equation}
  d_\mu = \sum_jU_{\mu j} \hat{c}_j + V_{\mu j} \hat{c}_j^\dagger ,
\end{equation}
where $\hat{c}_j = \left(\hchi_{2j-1} + i\hchi_{2j} \right)/2$  and $\hat{c}_j^\dagger = \left(\hchi_{2j-1} - i\hchi_{2j} \right)/2$.
The operators $d_i$ satisfy the usual fermionic commutation relations if and only if
\begin{equation}\label{eq:an_op_commutation_relations}
  U U^\dagger + V V^\dagger = \mathds{1} \quad \land \quad V U^t + U V^t = 0 .
\end{equation}
After applying the monitoring step of Eq.~\eqref{eq_monitoring_evol_step}, the annihilation operators evolve as 
\begin{equation}\label{eq:an_op_UV_evol}
  d_\mu' =  \mathcal{D} d_\mu \mathcal{D}^{-1} \quad \Rightarrow \quad
  \begin{bmatrix}
    U' & V' \\
  \end{bmatrix}
  =
  \begin{bmatrix}
    U & V \\
  \end{bmatrix}
  \frac{1}{2}
  \begin{bmatrix}
    \mathds{1} & -i \mathds{1} \\
    \mathds{1} & i \mathds{1} \\
  \end{bmatrix}
  P e^{-i 4 \delta w} P^{-1}
  \begin{bmatrix}
    \mathds{1} & \mathds{1} \\
    i\mathds{1} & - i \mathds{1} \\
  \end{bmatrix}
  ,
\end{equation}
where $P$ is a permutation that reorders the Majorana operator flattened indices, such that the odd ones come before the even ones, since
\begin{equation}
    \mathcal{D} \hchi_i \mathcal{D}^{-1} = \sum_j [e^{-4i \delta w}]_{ij} \hchi_j .
\end{equation}

$U'$ and $V'$ do not generally respect the commutation relations of Eq.~\eqref{eq:an_op_commutation_relations}, but that can be fixed by applying the following QR decomposition
\begin{equation}\label{eq:QRUV}
  \begin{bmatrix}
    {U'}^t \\ 
    {V'}^t \\ 
  \end{bmatrix}
  = 
  Q
  \begin{bmatrix}
    R \\ 
    0 \\ 
  \end{bmatrix}
  =
  \begin{bmatrix}
    Q_{11} & Q_{12} \\
    Q_{21} & Q_{22} \\
  \end{bmatrix}
  \begin{bmatrix}
    R \\ 
    0 \\ 
  \end{bmatrix}
  = 
  \begin{bmatrix}
    Q_{11}R \\
    Q_{21}R \\
  \end{bmatrix}
  ,
\end{equation}
where $Q$ is a unitary matrix and $R$ is an upper triangular matrix.
We now define new valid annihilation operators 
\begin{equation}
  \tilde{d}_\mu = \sum_\nu \left[ {R^t}^{-1}\right]_\nu d'_\nu = \sum_j \tilde{U}_j c_j +  \tilde{V}_j c^\dagger_j , 
\end{equation}
with $\tilde{U} = Q_{11}^t$ and $\tilde{V} = Q_{21}^t$, with the correct commutation relations.

Finally, we can show that the covariance matrix can be found for any set of annihilation operators $d_\mu$.
For a Gaussian state $\ket{\psi} = \mathcal{O} \ket{0}$ such that $\mathcal{O}^\dagger \hchi_i \mathcal{O} = \sum_j O_{ij} \hchi_j$, the state is annihilated by $d_\mu = \mathcal{O} c_\mu \mathcal{O}^\dagger $, which is equivalent to the following relation
\begin{equation}
  O = 
  \frac{1}{2} 
  P^{-1}
  \begin{bmatrix}
    \mathds{1} & \mathds{1} \\
    -i\mathds{1} & i \mathds{1} \\
  \end{bmatrix}
  \begin{bmatrix}
    U^t & V^\dagger \\
    V^t & U^\dagger \\
  \end{bmatrix}
  \begin{bmatrix}
    \mathds{1} & i \mathds{1} \\
    \mathds{1} & -i \mathds{1} \\
  \end{bmatrix}
  P
  .
\end{equation}
The covariance matrix is given by $M = O M_0 O^t$, where $M_0$ is the vacuum state covariance matrix
\begin{equation}\label{eq:vacuum_covariance}
  M_0 = 
  \begin{bmatrix}
    0 & 1 & & & \\
    -1 & 0 & & & \\
    & & 0 & 1 & \\
    & & -1 & 0 & \\
    & & & & \ddots \\
  \end{bmatrix} .
\end{equation}

\subsection{Jordan-Wigner transformation to a spin chain}

If the model is not Gaussian, then we have to simulate the evolution of the full exponentially large many-body state. To avoid worrying about the fermionic antisymmetry of the wave function, it is useful to map the model to a spin chain.
For that, we use the following Jordan-Wigner transformation,
\begin{equation}
    \hchi_{2k-1} = \left(\prod_{k' <k} \hsigma^z_{k'} \right) \hsigma^x_k 
    \quad , \quad
    \hchi_{2k} = \left(\prod_{k' <k} \hsigma^z_{k'} \right) \hsigma^y_k  , 
\end{equation}
where $\hsigma^i$ are the usual Pauli matrices.
A string of Majorana operators with even length gets expanded as
\begin{equation}
    \hchi_{\mu_1} \hchi_{\mu_2} \dots \hchi_{\mu_q}  =
    \left( \prod_{k < k_1} \hsigma^z_k \right) \hsigma^{\alpha_1}_{k_1}
    \left( \prod_{k < k_2} \hsigma^z_k \right) \hsigma^{\alpha_2}_{k_2}
    \dots
    \left( \prod_{k < k_q} \hsigma^z_k \right) \hsigma^{\alpha_q}_{k_q}
    ,
\end{equation}
where we assumed sorted flattened indices $\mu_{i+1} > \mu_i$ and we defined
\begin{equation}
    \alpha_i(\mu_i) =  \left\{
    \begin{array}{ccc}
       x & \textnormal{if} & \mu_i \textnormal{ is odd}  \\
       y & \textnormal{if} & \mu_i \textnormal{ is even}  \\
    \end{array}
    \right.
    \quad , \quad
    k_i(\mu_i) =  \left\{
    \begin{array}{ccc}
       \frac{\mu_i +1}{2} & \textnormal{if} & \mu_i \textnormal{ is odd}  \\
       \frac{\mu_i}{2} & \textnormal{if} & \mu_i \textnormal{ is even}  \\
    \end{array}
    \right.  .
\end{equation}
Rearranging the products of $\hsigma^z$, the string of Majorana can be expressed as
\begin{equation}
    \hchi_{\mu_1} \hchi_{\mu_2} \dots \hchi_{\mu_q}  =
    \prod_{i \textnormal{ odd}}^{q-1} \left[ 
    \hsigma_{k_i}^{\alpha_i}
    \left( \prod_{k_i \leqslant k < k_{i+1}} \hsigma_k^z \right)
    \hsigma_{k_{i+1}}^{\alpha_{i+1}}
    \right]
    ,
\end{equation}
which allows for finding a matrix representation for all terms in Eq.~\eqref{eq:discrete_evol_step_op_cluster_mon}, to numerically construct the evolution operators.

\subsection{Trotterization}

Although we know how to express in a matrix representation the exponents in Eq.~\eqref{eq:discrete_evol_step_op_cluster_mon}, computing the exponential is still an expensive task, but that can be addressed with a Trotterization scheme.
Our evolution operators are of the general form 
\begin{equation}\label{eq:generict_evol_step_op}
    U = \exp{\sum_{i} \left( A_i \delta W_i + \delta t B_i \right) O_i}  , 
\end{equation}
where $O_i$ are generic operators, $\delta W_i$ are random variables with zero mean and variance $\delta t$, and $A_i, B_i \in \mathbb{C}$.
A first-order Trotterization scheme would instead correspond to
\begin{equation}
    U_1 =  \prod_{i=1}^N  \exp[\left( A_i \delta W_i + \delta t B_i \right) O_i] .
\end{equation}
A Taylor expansion of $U$ and $U_1$ allows for the computation of the average difference and fluctuations between the two evolution operators,
\begin{equation}
    \mathbb{E}[U - U_1] = \mathcal{O}(dt^2) \quad , \quad \mathbb{E}[\|U - U_1\|^2] = \mathcal{O}(\delta t^3) .
\end{equation}
This means that the lowest order terms vanish for both quantities, indicating that the approximation is reasonably valid.

We will now return to our original operators and apply the first-order Trotterization scheme,
\begin{equation}\label{eq_discretized_sse_trotterized}
  \begin{aligned}
    \ket{\psi(t + \delta t )} \approx &  \prod_{j, \tmnu} 
    \exp[    \left(  \delta w_{\tmnu}^j \sqrt{\Gamma} + i^{\frac{q}{2}} \Gamma \delta t \left\langle  \prod_{k=1}^{q}\hchi_{j\tilde \nu_k} \right\rangle_t   \right) i^{\frac{q}{2}} \prod_{k=1}^{q}\hchi_{j\tilde \nu_k} ] \\ 
    \times & \prod_{j\mmu\mnu} \exp[ i^{\frac{q_J}{2} - 1} \sqrt{J}  \delta h^j_{ \boldsymbol{\mu}\boldsymbol{\nu}}(t) \prod^{q_J/2}_{k=1} \hchi_{j \mu_{k}} \prod_{l=1}^{q_J/2} \hchi_{j+1 \nu_{l}} ] \\
    \times & \prod_{j\tmmu} \exp[ i^{\frac{q_J}{2} - 1} \sqrt{J}  \delta h^j_{ \boldsymbol{\tilde \mu}}(t) \prod^{q_J}_{k=1} \hchi_{j \tilde \mu_{k}} ] \ket{\psi(t)}
    .
  \end{aligned}
\end{equation}
Each exponential in Eq.~\eqref{eq_discretized_sse_trotterized} can be further simplified, taking into account that 
\begin{equation}
    (\hchi_{\mu_1} \hchi_{\mu_2} \dots \hchi_{\mu_q})^2 = (-1)^{\frac{q}{2}} , 
\end{equation}
with $\mu_{i} < \mu_{i+1}$, which allows to Taylor expand the exponentials  and find that
\begin{equation}
    \exp[ A ( \hchi_{\mu_1} \dots \hchi_{\mu_q})] 
    =  \left\{ 
    \begin{array}{lll}
        \cosh(A) \mathds{1} + \sinh(A) \hchi_{\mu_1} \dots \hchi_{\mu_q} & \textnormal{if} & \mod(q,4) = 0 \\
        \cos(A) \mathds{1} + \sin(A) \hchi_{\mu_1} \dots \hchi_{\mu_q} & \textnormal{if} & \mod(q,4) \neq 0
    \end{array}
    \right.
\end{equation},
for even $q$ and $A \in \mathbb{C}$.

\bibliographystyle{quantum}
\bibliography{biblio}
\end{document}